\documentclass[preprint,3p,12pt]{elsarticle}
\usepackage{mathrsfs}
\usepackage{amsmath}
\usepackage{stmaryrd}
\usepackage{bbding}
\usepackage{dcolumn}
\usepackage{graphicx}
\usepackage{amsfonts}
\usepackage{amssymb}
\usepackage{psfrag}
\usepackage{wrapfig}
\usepackage{subfigure}
\usepackage{makeidx}
\usepackage{bm}
\usepackage{epsf}
\usepackage{color}
\usepackage{epsfig}
\usepackage{setspace}
\usepackage{graphicx}
\usepackage{epstopdf}
\usepackage{psfrag}
\usepackage{subfigure}
\usepackage{multirow}
\usepackage{diagbox}
\usepackage{verbatim}
\newtheorem{rmk}{Remark}

\usepackage{makecell}
\usepackage{float}
\usepackage{esint}
\usepackage{comment}
\usepackage{movie15}
\usepackage{hyperref}

\newcommand{\mathsym}[1]{{}}
\newcommand{\unicode}[1]{{}}
\begin{document}

\title{A three-dimensional compact high-order gas-kinetic scheme on structured mesh}

\author[HKUST1]{Xing Ji}
\ead{xjiad@connect.ust.hk}

\author[HKUST2]{Fengxiang Zhao}
\ead{fzhaoac@connect.ust.hk}

\author[HKUST2]{Wei Shyy}
\ead{weishyy@ust.hk}

\author[HKUST1,HKUST2,HKUST3]{Kun Xu\corref{cor1}}
\ead{makxu@ust.hk}

\address[HKUST1]{Department of Mathematics, Hong Kong University of Science and Technology, Clear Water Bay, Kowloon, Hong Kong}
\address[HKUST2]{Department of Mechanical and Aerospace Engineering, Hong Kong University of Science and Technology, Clear Water Bay, Kowloon, Hong Kong}
\address[HKUST3]{Shenzhen Research Institute, Hong Kong University of Science and Technology, Shenzhen, China}
\cortext[cor1]{Corresponding author}

\begin{abstract}
In this paper, a third-order compact gas-kinetic scheme is firstly proposed for three-dimensional computation for the compressible Euler and Navier-Stokes solutions. As an extension of the previous compact gas-kinetic scheme (GKS) on two-dimensional structured/unstructured mesh \cite{ji2018compact,ji2020hweno}, the new scheme is based on three key ingredients: the time-accurate gas-kinetic evolution solution,
the Hermite weighted essentially non-oscillatory (HWENO) reconstruction, and the two-stage temporal discretization. The scheme achieves its compactness due to the time-dependent gas distribution function in GKS, which provides not only the fluxes but also the time accurate flow variables in the next time level at a cell interface. As a result, the cell averaged first-order spatial derivatives of flow variables can be obtained naturally through the Gauss's theorem. Then, a third-order compact reconstruction involving the cell averaged values and their first-order spatial derivatives can be achieved. The trilinear interpolation is used to treat possible non-coplanar elements on general hexahedral mesh.
The constrained least-square technique is applied to improve the accuracy in the smooth case. To deal with both smooth and discontinuous flows, a new HWENO reconstruction is designed in the current scheme by following the ideas in \cite{zhu2018new}. No identification of troubled cells is needed in the current scheme. In contrast to the Riemann solver-based method, the compact scheme can achieve a third-order temporal accuracy with the two-stage two-derivative temporal discretization, instead of the three-stage Runge-Kutta method. Overall, the proposed scheme inherits the high accuracy and efficiency of the previous ones in two-dimensional case. The desired third-order accuracy can be obtained with curved boundary.
The robustness of the scheme has been validated through many cases, including strong shocks in both inviscid and viscous flow computations.
Quantitative comparisons for both smooth and discontinuous cases show that the current third-order scheme can give competitive results against the fifth-order non-compact GKS under the same mesh. A large CFL number around 0.5 can be used in the present scheme.
	\end{abstract}


\maketitle
\section{Introduction}
Tremendous efforts have been paid on the
development of high-order computational fluid dynamics (CFD) methods for the compressible Euler and Navier-Stokes (N-S) equations in past decades. Representative methods include weighted essentially
non-oscillatory (WENO) methods \cite{shu2009weno-review}, discontinuous Galerkin (DG) methods \cite{shu2016weno-dg-review}, the flux reconstruction (FR) \cite{Huynh2007FR} or correction procedure via reconstruction (CPR) \cite{yu2014accuracy} methods, etc.
WENO-type reconstruction has been widely applied on structured grids,
which can keep good robustness with very high order of accuracy in space  \cite{balsara2016efficient}.
However, it is not a trivial task to extend the WENO approach to unstructured mesh or non-uniform mesh with the same high-order accuracy, since the stencil is too large.
Zhu et al. has proposed a class of new WENO schemes in the attempt for releasing the problems recently \cite{zhu2020new}.

On the other hand, high-order methods based on the compact stencil which only involves the target cell and its von Neumann neighbors are attractive because of their good mesh adaptability, high scalability, which becomes a hot topic in CFD research \cite{high-order-review}. Two main representatives of compact methods are the DG methods, which combine the finite volume framework and the finite element framework and the FR/CPR methods, which hybridize the finite difference and finite volume discretization originally.
These methods can achieve arbitrary spatial order of accuracy with only the targeted cell as the reconstruction stencil, which yields a great mesh adaptability.
Numerical results have demonstrated their power in large eddy simulation (LES) \cite{wang2017towards} and RANS simulation \cite{yang2019robust} for smooth flow.
However, in the flow simulations with strong discontinuities, these methods seem to lack robustness.
Many techniques have been used to limit the troubled internal degree of freedom \cite{qiu2016stability-limter-dg-book}.
At the same time, they have more restricted explicit time step for linear stability against the traditional high-order finite volume method.
New compact schemes have been proposed to improve the above shortcomings,
such as the multi-moment constrained finite volume scheme \cite{xie2017hybrid-compact-mcv-2d-euler},
$P_NP_M$\cite{yang2018reconstructed} scheme.
Most of these methods use Riemann solvers or approximate Riemann solvers for the flux evaluation, and the Runge-Kutta time-stepping methods are adopted for the
temporal accuracy.

In recent years,
a class of compact high-order gas-kinetic scheme (HGKS) has been developed \cite{pan2016unstructuredcompact}.
With the adoption of the two-stage time discretization \cite{li2016twostage,du2018hermite},
and the Hermite WENO (HWENO) reconstruction \cite{qiu2004hermite},
a fourth-order compact gas-kinetic scheme (GKS) on two-dimensional structured mesh is constructed \cite{ji2018compact}.
It has higher resolution than the conventional non-compact fourth-order GKS \cite{pan2018two} while performs competitive robustness as the second-order scheme.
Most importantly, it allows a CFL number around 0.5 in the computation.
The success of the above scheme lies in the gas-kinetic framework.
In comparison with traditional Riemann solver based high-order CFD
methods, it includes the following distinguishable points:
(i) GKS is based on an analytical integral solution of the BGK equation, which can recover the N-S equations from the Chapman-Enskog expansion \cite{GKS-2001}.
The time-dependent gas distribution function at a cell interface provides a multiple
scale flow evolution from the kinetic particle transport to the
hydrodynamic wave propagation, which bridges the approaches seamlessly between the
kinetic flux vector splitting and the central difference Lax-Wendroff
type discretization.
(ii) Both inviscid and viscous fluxes are
obtained from the moments of a single gas distribution function, which has special advantages in the construction of scheme on unstructured mesh.
(iii) The GKS is a multi-dimensional scheme
\cite{xu2005multidimensional-implicitGKS}, where both normal and tangential derivatives of
flow variables around a cell interface participate in the time
evolution of the gas distribution function.
(iv) Besides fluxes, the time-dependent gas distribution function also provides
time-evolving flow variables at each Gaussian-point on a cell interface.
By using these time-accurate solutions, the first-order spatial derivatives within a cell can be calculated exactly through the Green-Gauss theorem.
Then, the compact HWENO reconstruction can be adopted.
It makes fundamental differences against DG/CPR methods,
in which the slopes can be wrongly evaluated by weak formulations or differential equations around discontinuities.
Therefore, how to treat discontinuities becomes a big issue in these schemes.
In GKS, the same large time step as the purely high-order finite volume scheme can be used and the same robustness can be kept in capturing discontinuous solution.
(v) The multi-stage multi-derivative (MSMD) time discretization can be adopted
due to the existence of the time-derivative of the flux function in HGKS.
For the most of high-order methods based on the time-independent Riemann solvers, the Runge-Kutta (RK) time discretization is used for improving temporal accuracy.
There are well-developed stability theories,
such as strong stability preserving (SSP) property \cite{RK-advantage2} for the traditional RK methods.
However, the Nth-order accuracy in RK methods requires no less than N stages.
For a classical fourth-order RK method, four stages are needed.
It is true that the GKS flux solver is more expensive than the time-independent Riemann solvers.
But, with the incorporation of the MSMD method, the fourth-order time accuracy can be achieved with only two stages in HGKS \cite{Pan2016twostage}, which becomes efficient in comparison with Riemann-solver based RK methods \cite{ji2020performance}.

The compact GKS \cite{ji2018compact} has been extended to eighth-order spatial accuracy in rectangular mesh \cite{zhao2019compact},
which behaves a spectral-like resolution at large wavenumber
and has been validated in the aeroacoustic problems
involving both shock-shock interactions and linear acoustic waves \cite{zhao2019acoustic}.
The extension on triangular mesh \cite{ji2020hweno} demonstrates excellent robustness in the test cases with strong shocks, such as the hypersonic flow passing a cylinder up to Mach number  $20$.
Thus, it becomes nature to extend the compact HGKS to three-dimensional compressible flow computations.

In this paper, a compact third-order scheme on three-dimensional multi-block structured mesh will be proposed, following the compact GKS framework and the MSMD temporal discretization in \cite{ji2018family,ji2020hweno}.
To handle mesh with general hexahedron, unlike the direction-by-direction reconstruction strategy  in \cite{ji2018family,zhao2019compact}, a second-order multi-dimensional polynomial is directly constructed in a least-squares sense.
For structured mesh with curved boundary, the vortexes of a hexahedral cell can be easily non-coplanar.
The geometric information, such as the locations and normal directions of the Gaussian points at a cell interface cannot be determined.
Therefore, the trilinear/bilinear interpolation is used so that these parameters for cells and cell interfaces are uniquely determined with the satisfaction of global mesh volume conservation.
In this way, the scheme will keep the formal order of accuracy under non-uniform or curvilinear meshes.
At the same time, a HWENO-type reconstruction following the similar idea in \cite{zhu2018new} with compact stencils is designed.
Unlike the traditional WENO reconstruction \cite{hu1999weighted,zhu2009hermite}, which requires a large number of sub-stencils and can easily have non-positive weights on irregular mesh \cite{zhao2017weighted},
the HWENO reconstruction here is to construct a whole weighted polynomial in a cell rather than function value at each Gaussian point.
The linear weights can be chosen to be any positive number as long as the summation goes to one. The number of sub-stencils is less than the conventional WENO method.
and the scheme keeps the expected order of accuracy in smooth region.
As a result, the new compact reconstruction provides an accurate and reliable initial condition for the gas-kinetic evolution.
The new scheme inherits the advantages of the previous compact schemes \cite{ji2018compact,ji2020hweno}, which are efficient, accurate and robust.
In comparison with the third-order Runge-Kutta (RK) time-stepping method, it achieves a third-order accuracy in time with one middle stage only.
A CFL number of 0.5 can be taken safely in both smooth and discontinuous cases,
while the CFL number is on the order of 0.2 for a third-order DG.
To appreciate its high accuracy, quantitative comparisons are given between the proposed scheme and the non-compact HGKS \cite{ji2020performance}, which shows that the formal one has superiority in the implicit LES simulations.
Stringent tests including hypersonic flow passing through a sphere validate the robustness of the current compact scheme in 3D structured mesh.

This paper is organized as follows.
The basic framework for the three-dimensional compact high-order GKS is presented in section 2.
In section 3, the general formulation for the two-stage high-order temporal discretization is introduced.
In section 4, the compact third-order HWENO reconstruction on hexahedral mesh is presented.
Numerical experiments including inviscid and viscous test cases are given in section 5 to validate the proposed scheme.
The last section is the conclusion.

\section{Compact finite volume gas-kinetic scheme}

The three-dimensional gas-kinetic BGK equation \cite{BGK} can be written as
\begin{equation}\label{bgk}
f_t+\textbf{u}\cdot\nabla f=\frac{g-f}{\tau},
\end{equation}
where $f$ is the gas distribution function,
$g$ is the corresponding equilibrium state,
and $\tau$ is the collision time.
$f=f(\textbf{x},t,\textbf{u},\xi)$,
where $\textbf{x}$ is location in physical space,
t is time and $\textbf{u}$ is particle velocity in phase space.

The collision term satisfies the following compatibility condition
\begin{equation}\label{compatibility}
\int \frac{g-f}{\tau} \pmb{\psi} \text{d}\Xi=0,
\end{equation}
where $\pmb{\psi}=(1,\textbf{u},\displaystyle \frac{1}{2}(\textbf{u}^2+\xi^2))^T$,
$\text{d}\Xi=\text{d}u_1\text{d}u_2\text{d}u_3\text{d}\xi_1...\text{d}\xi_{K}$,
$K$ is the number of internal degree of freedom, i.e.
$K=(5-3\gamma)/(\gamma-1)$ in three-dimensional case, and $\gamma$
is the specific heat ratio.

Based on the Chapman-Enskog expansion for BGK equation
\cite{xu2014direct}, the gas distribution function in the continuum
regime can be expanded as
\begin{align*}
f=g-\tau D_{\textbf{u}}g+\tau D_{\textbf{u}}(\tau
D_{\textbf{u}})g-\tau D_{\textbf{u}}[\tau D_{\textbf{u}}(\tau
D_{\textbf{u}})g]+...,
\end{align*}
where $D_{\textbf{u}}={\partial}/{\partial t}+\textbf{u}\cdot
\nabla$.
By truncating on different orders of $\tau$, the
corresponding macroscopic equations can be derived.
If the zeroth-order truncation is taken, i.e., $f=g$,
the Euler equations can be recovered by multiplying $\pmb{\psi}$ on Eq.\eqref{bgk} and integrating it over the phase space,
\begin{equation*}\label{euler-conservation}
\begin{split}
\textbf{W}_t+ \nabla \cdot \textbf{F}(\textbf{W})=0.
\end{split}
\end{equation*}
If the first-order truncated distribution function is applied, i.e.,
\begin{align} \label{ce-ns}
f=g-\tau (\textbf{u} \cdot \nabla g + g_t),
\end{align}
the N-S equations can be obtained,
\begin{equation*}\label{ns-conservation}
\begin{split}
\textbf{W}_t+ \nabla \cdot \textbf{F}(\textbf{W},\nabla \textbf{W} )=0,
\end{split}
\end{equation*}
with $\tau = \mu / p$ and a fixed Prandtl number $Pr=1$.

Taking moments of the time-dependent distribution function $f$, the conservative variables can be obtained
\begin{align}\label{point}
\textbf{W}(\textbf{x},t)=\int \pmb{\psi} f(\textbf{x},t,\textbf{u},\xi)\text{d}\Xi.
\end{align}
so as the macroscopic fluxes $\textbf{F} (\textbf{W}(\textbf{x},t))$,
\begin{equation}\label{f-to-flux}
\textbf{F}(\textbf{x},t)=
\int \textbf{u} \pmb{\psi} f(\textbf{x},t,\textbf{u},\xi)\text{d}\Xi.
\end{equation}

\begin{rmk}
	It is well known that the cell-averaged conservative variables can be updated through the interface fluxes under the finite volume framework.
	Beside the fluxes, Eq.\eqref{point} provides additional information, which can be updated at the next time level.
	It is the key part in constructing the compact GKS, which will be introduced in detail in subsection \ref{slope-section}.
	An obvious prerequisite is that the $\textbf{W}(\textbf{x},t)$ must be time-accurate, while the time-independent Riemann solution cannot provide
such a time accurate evolution solution \end{rmk}

\subsection{Finite volume scheme on general structured mesh}

For a polyhedron cell $\Omega_i$ in 3-D case, the boundary can be expressed as
\begin{equation*}
\partial \Omega_i=\bigcup_{p=1}^{N_f}\Gamma_{ip},
\end{equation*}
where $N_f$ is the number of cell interfaces for cell $\Omega_i$, e.g., $N_f=4$
for tetrahedron and $N_f=6$ for cuboid or general hexahedron.

The increment of the cell averaged conservative flow variables in a finite control volume i in a time interval $[t_n,t_{n+1}]$  can be expressed as
\begin{equation}\label{fv-3d-general}
\textbf{W}^{n+1}_i \left| \Omega_i \right|=\textbf{W}^n_i \left| \Omega_i \right|-\sum_{p=1}^{N_f}\int_{\Gamma_{ip}}\int_{t_n}^{t_{n+1}}
\textbf{F}(\textbf{x},t)\cdot\textbf{n}_p \text{d}t\text{d}s,
\end{equation}
with
\begin{equation}\label{f-to-flux-in-normal-direction}
\textbf{F}(\textbf{x},t)\cdot \textbf{n}_p=\int\pmb{\psi}  f(\textbf{x},t,\textbf{u},\xi) \textbf{u}\cdot \textbf{n}_p \text{d}\Xi,
\end{equation}
where $\textbf{W}_{i}$ is the cell averaged value over cell $\Omega_i$, $\left|
\Omega_i \right|$ is the volume of $\Omega_i$, $\textbf{F}$ is the interface flux, and $\textbf{n}_p=(n_1,n_2,n_3)^T$ is the unit vector representing the outer normal direction of $\Gamma_{ip}$.
The semi-discretized form of finite volume scheme can be written as
\begin{equation}\label{semidiscrete}
\frac{\text{d} \textbf{W}_{i}}{\text{d}t}=\mathcal{L}(\textbf{W}_i)=-\frac{1}{\left| \Omega_i \right|} \sum_{p=1}^{N_f} \int_{\Gamma_{ip}}
\textbf{F}(\textbf{W})\cdot\textbf{n}_p \text{d}s,
\end{equation}

Numerical quadratures can be adopted to give a high-order spatial approximation for $\textbf{F}_{ip}(t)$ or $\widetilde{\textbf{F}}_{ip}(t)$,
where Eq.\eqref{fv-3d-general} can be rewritten  as
\begin{equation}\label{fv-3d-general-quadrature}
\textbf{W}^{n+1}_i \left| \Omega_i \right|
=\textbf{W}^n_i \left| \Omega_i \right|-\sum_{p=1}^{N_f}  \left|\Gamma_{ip}\right| \sum_{k=1}^{M} \omega_k\int_{t_n}^{t_{n+1}}
\textbf{F}(\textbf{x}_{p,k},t)\cdot\textbf{n}_p \text{d}t.
\end{equation}

\begin{rmk}
	The bilinear interpolation is used to describe a given quadrilateral interface $\Gamma_{ip}$ with coplanar or non-coplanar vertexes,
	\begin{align*}
	\textbf{X} (\xi, \eta)= \sum_{l=1}^4 \textbf{x}_l \phi_l (\xi, \eta),
	\end{align*}
	where $(\xi, \eta) \in [-1/2, 1/2]^2$, $\textbf{x}_l$ is the locations of the mth vertex
	and $\phi_l$ is the base function as follows
	\begin{equation*}
	\begin{aligned}
	&\phi_{1}=\frac{1}{4}(1-2 \xi)(1-2 \eta),
	&\phi_{2}=\frac{1}{4}(1-2 \xi)(1-2 \eta),\\
	&\phi_{3}=\frac{1}{4}(1-2 \xi)(1+2 \eta),
	&\phi_{4}=\frac{1}{4}(1-2 \xi)(1+2 \eta),
	\end{aligned}
	\end{equation*}

The flux across $\Gamma_{ip}$ in Eq.\eqref{fv-3d-general} can be rewritten as
\begin{equation*}\label{flux-para-coord}
\begin{split}
\textbf{F}_{ip}(t)
=\int_{\Gamma_{ip}} \textbf{F}(\textbf{x},t)\cdot\textbf{n}_p \text{d}s
=\int_{-1/2}^{1/2}\int_{-1/2}^{1/2} \textbf{F}(\textbf{W}(\textbf{X}(\xi, \eta))) \cdot\textbf{n}_p \left|\frac{\partial(x, y, z)}{\partial(\xi, \eta)}\right| \text{d}\xi \text{d}\eta.
\end{split}
\end{equation*}
To meet the requirement of a third-order spatial accuracy,
the above equation can be approximated through Gaussian quadrature as
\begin{equation*}\label{flux-para-coord-gauss}
\textbf{F}_{ip}(t)
=\sum_{m, n=1}^{2} \omega_{m, n} \textbf{F}_{m, n}(t) \cdot (\textbf{n}_p )_{m, n}\left|\frac{\partial(x, y, z)}{\partial(\xi, \eta)}\right|_{m, n} \Delta \xi \Delta \eta,
\end{equation*}
where $\Delta \xi = \Delta \eta =1$, and the local normal direction
$(\textbf{n}_p )_{m, n}=\left(\boldsymbol{X}_{\xi} \times \boldsymbol{X}_{\eta}\right) /\left\|\boldsymbol{X}_{\xi} \times \boldsymbol{X}_{\eta}\right\|$.
The standard Gaussian points are
\begin{equation*}
\begin{split}
&(\xi,\eta)_{1,1}=(-\frac{1}{2\sqrt{3}}\Delta \xi,-\frac{1}{2\sqrt{3}}\Delta \eta),
(\xi,\eta)_{1,2}=(-\frac{1}{2\sqrt{3}}\Delta \xi,\frac{1}{2\sqrt{3}}\Delta \eta),\\
&(\xi,\eta)_{2,1}=(\frac{1}{2\sqrt{3}}\Delta \xi,-\frac{1}{2\sqrt{3}}\Delta \eta),
(\xi,\eta)_{2,2}=(\frac{1}{2\sqrt{3}}\Delta \xi,\frac{1}{2\sqrt{3}}\Delta \eta),
\end{split}
\end{equation*}
with $\omega_{1,1}=\omega_{1,2}=\omega_{2,1}=\omega_{2,2}=\frac{1}{4}$.

Compared with Eq.\eqref{fv-3d-general-quadrature}, we have
\begin{equation*}
\omega_{k} = \omega_{m, n} \left|\frac{\partial(x, y, z)}{\partial(\xi, \eta)}\right|_{m, n},~
\textbf{x}_{p,k} = \textbf{x}((\xi,\eta)_{m,n}),~
\textbf{n}_{p,k} =(\textbf{n}_{p})_{m, n},
\end{equation*}
with $k = 2m + n$, $m=1,2$, $n = 1,2$.
	
\end{rmk}

According to the coordinate transformation, the local coordinate for
the cell interface $\Gamma_{ip}$ is expressed as
$(\widetilde{x}_1,\widetilde{x}_2,\widetilde{x}_3)^T=(0, \widetilde{x}_2,, \widetilde{x}_3)^T$, where
$(\widetilde{x}_2,, \widetilde{x}_3)^T \in \Gamma_{ip}$, and the
velocities in the local coordinate are given by
\begin{equation}\label{rotate-3d}
\begin{split}
\begin{cases}
&\tilde{u_1}=u_1 n_1 + u_2 n_2 + u_3 n_3 , \\
&\tilde{u_2}=-u_1 n_2 + u_2 (n_1+\frac{n_3^2}{1+n_1}) - u_3 \frac{n_2n_3}{1+n_1} , \\[3pt]
&\tilde{u_3}=-u_1 n_3 - u_2 \frac{n_2n_3}{1+n_1}  + u_3 (1-\frac{n_3^2}{1+n_1}),
\end{cases}
~~~ n_1\neq -1.
\end{split}
\end{equation}

And the macroscopic conservative variables in the local coordinate are given as
\begin{align*}
\widetilde{\textbf{W}}(\widetilde{\textbf{x}},t)=\textbf{T}\textbf{W}(\textbf{x},t) ,
\end{align*}
where $\textbf{T}$ is the rotation matrix
\begin{equation}\label{rotate-matrix-3d}
\textbf{T}= \left(
\begin{array}{ccccc}
1 &0 & 0 & 0 & 0 \\
0 & n_1 &n_2 & n_3 &0 \\
0 &-n_2 &n_1+\frac{n_3^2}{1+n_1} & -\frac{n_2n_3}{1+n_1} & 0 \\[3pt]
0 &-n_3 &-\frac{n_2n_3}{1+n_1} & 1-\frac{n_3^2}{1+n_1} &0  \\
0 &0 & 0 & 0 & 1 \\
\end{array}
\right),~~~ n_1\neq -1.
\end{equation}
Note that when $n_1=-1$, the Eq.\eqref{rotate-3d} changes to
$(\tilde{u_1},\tilde{u_2},\tilde{u_3})^T=(-u_1,-u_2,u_3)^T$ and
the matrix \eqref{rotate-matrix-3d} is replaced by a diagonal matrix
$\Lambda=\text{diag}(1,-1,-1,1,1)$.

For the gas distribution function in the local coordinate,
$\widetilde{f}(\widetilde{\textbf{x}},t,\widetilde{\textbf{u}},\xi)=f(\textbf{x},t,\textbf{u},\xi)$
and $|\text{d}\textbf{u}|=|\text{d}\widetilde{\textbf{u}}|$, then the numerical fluxes
can be transformed as
\begin{align}\label{flux-global-local}
\textbf{F}(\textbf{x},t)=\int\pmb{\psi} f(\textbf{x},t,\textbf{u},\xi) \textbf{u} \cdot \textbf{n}_p
\text{d} \textbf{u} \text{d} \xi
=\int\pmb{\psi}\widetilde{f}(\widetilde{\textbf{x}},t,\widetilde{\textbf{u}},\xi)
\widetilde{u}_1 \text{d} \widetilde{\textbf{u}} \text{d}\xi .
\end{align}
In the computation, the fluxes in the local coordinate are obtained first by taking moments of the gas distribution function in the local coordinate
\begin{align}\label{flux-local}
\widetilde{\textbf{F}}(\widetilde{\textbf{x}},t)=\int \widetilde{\pmb{\psi}} \widetilde{f}(\widetilde{\textbf{x}},t,\widetilde{\textbf{u}},\xi)
\widetilde{u}_1 \text{d} \widetilde{\textbf{u}} \text{d}\xi,
\end{align}
where
$\widetilde{\pmb{\psi}}=(1,\widetilde{\textbf{u}},\displaystyle\frac{1}{2}(\widetilde{\textbf{u}}^2+\xi^2))^T$.
According to Eq.\eqref{rotate-3d}, Eq.\eqref{flux-global-local} and Eq.\eqref{flux-local}, the fluxes
in the global coordinate can be expressed as a combination of the
fluxes in the local coordinate
\begin{align}\label{flux-local-to-global}
\textbf{F}(\textbf{W}(\textbf{x},t))\cdot\textbf{n}=
\textbf{T}^{-1}\widetilde{\textbf{F}}(\widetilde{\textbf{W}}(\widetilde{\textbf{x}},t)).
\end{align}

\subsection{Gas-kinetic solver}

In order to construct the numerical
fluxes at $\textbf{x}=(0,0,0)^T$, the integral solution of BGK equation Eq.\eqref{bgk} is used
\begin{equation}\label{integral1}
f(\textbf{x},t,\textbf{u},\xi)=\frac{1}{\tau}\int_0^t g(\textbf{x}',t',\textbf{u},\xi)e^{-(t-t')/\tau}\text{d}t'
+e^{-t/\tau}f_0(\textbf{x}-\textbf{u}t,\textbf{u},\xi),
\end{equation}
where $\textbf{x}=\textbf{x}'+\textbf{u}(t-t')$ is the trajectory of
particle. $f_0$ is the initial gas distribution function, $g$ is the corresponding
equilibrium state.
The integral solution basically states a physical process from the particle free transport in $f_0$ in the kinetic scale
to the hydrodynamic flow evolution in the integral of $g$ term.
The flow evolution at the cell interface depends on the ratio of time step
to the  local particle collision time $\Delta t/\tau$.

To construct a time evolution solution of gas distribution function at a cell interface,
the following notations are introduced first
\begin{align*}
a_{x_i} \equiv  (\partial g/\partial x_i)/g=g_{x_i}/g,
A \equiv (\partial g/\partial t)/g=g_t/g,
\end{align*}
where $g$ is the equilibrium state.  The variables $(a_{x_i}, A)$, denoted by $s$,
depend on particle velocity in the form of
\cite{GKS-2001}
\begin{align*}
s=s_j\psi_j =s_{1}+s_{2}u_1+s_{3}u_2+s_{4}u_3
+s_{5}\displaystyle \frac{1}{2}(u_1^2+u_2^2+u_3^2+\xi^2).
\end{align*}
For the kinetic part of the integral solution Eq.\eqref{integral1},
the initial gas distribution function can be constructed as
\begin{equation*}
f_0=f_0^l(\textbf{x},\textbf{u})\mathbb{H} (x_1)+f_0^r(\textbf{x},\textbf{u})(1- \mathbb{H}(x_1)),
\end{equation*}
where $\mathbb{H}(x_1)$ is the Heaviside function. Here $f_0^l$ and $f_0^r$ are the
initial gas distribution functions on both sides of a cell
interface, which have one to one correspondence with the initially
reconstructed macroscopic variables. The first-order
Taylor expansion for the gas distribution function in space around
$\textbf{x}=\textbf{0}$ is expressed as
\begin{align}\label{flux-3d-1}
f_0^k(\textbf{x})=f_G^k(\textbf{0})+\frac{\partial f_G^k}{\partial x_i}(\textbf{0})x_i
=f_G^k(\textbf{0})+\frac{\partial f_G^k}{\partial x_1}(\textbf{0})x_1
+\frac{\partial f_G^k}{\partial x_2}(\textbf{0})x_2
+\frac{\partial f_G^k}{\partial x_3}(\textbf{0})x_3,
\end{align}
for $k=l,r$.
According to Eq.\eqref{ce-ns}, $f_{G}^k$ has the form
\begin{align}\label{flux-3d-2}
f_{G}^k(\textbf{0})=g^k(\textbf{0})-\tau(u_ig_{x_i}^{k}(\textbf{0})+g_t^k(\textbf{0})),
\end{align}
where $g^k$ are the equilibrium states with the form of a Maxwell distribution.
$g^k$ can be fully determined from  the
reconstructed macroscopic variables $\textbf{W}
^l, \textbf{W}
^r$ at the left and right sides of a cell interface as
\begin{align}\label{get-glr}
\int\pmb{\psi} g^{l}\text{d}\Xi=\textbf{W}
^l,\int\pmb{\psi} g^{r}\text{d}\Xi=\textbf{W}
^r.
\end{align}
Substituting Eq.\eqref{flux-3d-1} and Eq.\eqref{flux-3d-2} into Eq.\eqref{integral1},
the kinetic part for the integral solution can be written as
\begin{equation}\label{dis1}
\begin{aligned}
e^{-t/\tau}f_0^k(-\textbf{u}t,\textbf{u},\xi)
=e^{-t/\tau}g^k[1-\tau(a_{x_i}^{k}u_i+A^k)-ta^{k}_{x_i}u_i],
\end{aligned}
\end{equation}
where the coefficients $a_{x_1}^{k},...,A^k, k=l,r$ are defined according
to the expansion of $g^{k}$.
After determining the kinetic part
$f_0$, the equilibrium state $g$ in the integral solution
Eq.\eqref{integral1} can be expanded in space and time as follows
\begin{align}\label{equli}
g(\textbf{x},t)= g^{c}(\textbf{0},0)+\frac{\partial  g^{c}}{\partial x_i}(\textbf{0},0)x_i+\frac{\partial  g^{c}}{\partial t}(\textbf{0},0)t,
\end{align}
where $ g^{c}$ is the Maxwellian equilibrium state located at an interface.
Similarly, $\textbf{W}^c$ are the macroscopic flow variables for the determination of the
equilibrium state $ g^{c}$
\begin{align}\label{compatibility2}
\int\pmb{\psi} g^{c}\text{d}\Xi=\textbf{W}^c.
\end{align}
Substituting Eq.\eqref{equli} into Eq.\eqref{integral1}, the hydrodynamic part for the integral solution
can be written as
\begin{equation}\label{dis2}
\begin{aligned}
\frac{1}{\tau}\int_0^t
g&(\textbf{x}',t',\textbf{u},\xi)e^{-(t-t')/\tau}\text{d}t'
=C_1 g^{c}+C_2 a_{x_i}^{c} u_i g^{c} +C_3 A^{c} g^{c} ,
\end{aligned}
\end{equation}
where the coefficients
$a_{x_i}^{c},A^{c}$ are
defined from the expansion of the equilibrium state $ g^{c}$. The
coefficients $C_m, m=1,2,3$ in Eq.\eqref{dis2}
are given by
\begin{align*}
C_1=1-&e^{-t/\tau}, C_2=(t+\tau)e^{-t/\tau}-\tau, C_3=t-\tau+\tau e^{-t/\tau}.
\end{align*}
The coefficients in Eq.\eqref{dis1} and Eq.\eqref{dis2}
can be determined by the spatial derivatives of macroscopic flow
variables and the compatibility condition as follows
\begin{align}\label{co}
&\langle a_{x_1}\rangle =\frac{\partial \textbf{W} }{\partial x_1}=\textbf{W}_{x_1},
\langle a_{x_2}\rangle =\frac{\partial \textbf{W} }{\partial x_2}=\textbf{W}_{x_2},
\langle a_{x_3}\rangle =\frac{\partial \textbf{W} }{\partial x_3}=\textbf{W}_{x_3},\nonumber\\
&\langle A+a_{x_1}u_1+a_{x_2}u_2+a_{x_3}u_3\rangle=0,
\end{align}
where $\left\langle ... \right\rangle$ are the moments of a gas distribution function defined by
\begin{align}\label{co-moment}
\langle (...) \rangle  = \int \pmb{\psi} (...) g \text{d} \Xi .
\end{align}
The details for calculation of each microscopic term from macroscopic quantities can refer \cite{ji2019high}.
Then the second-order time dependent gas distribution function is given as
\begin{align*}
f(\textbf{0},t,\textbf{u},\xi)
=&(1-e^{-t/\tau}) g^{c}+[(t+\tau)e^{-t/\tau}-\tau]a_{x_i}^{c}u_i g^{c}\nonumber
+(t-\tau+\tau e^{-t/\tau})A^{c}  g^{c}\nonumber\\
+&e^{-t/\tau}g^l[1-(\tau+t)a_{x_i}^{l}u_i-\tau A^l]H(u_1)\nonumber\\
+&e^{-t/\tau}g^r[1-(\tau+t)a_{x_i}^{r}u_i-\tau A^r] (1-H(u_1)).
\end{align*}
Finally in order to properly capture the un-resolved shock structure, additional numerical dissipation is needed.
The physical collision time $\tau$ in the exponential function part can be replaced by a numerical collision time $\tau_n$, which will be defined  later,
\begin{align}\label{2nd-flux}
f(\textbf{0},t,\textbf{u},\xi)
=&(1-e^{-t/\tau_n}) g^{c}+[(t+\tau)e^{-t/\tau_n}-\tau]a_{x_i}^{c}u_i g^{c}\nonumber
+(t-\tau+\tau e^{-t/\tau_n})A^{c}  g^{c}\nonumber\\
+&e^{-t/\tau_n}g^l[1-(\tau+t)a_{x_i}^{l}u_i-\tau A^l]H(u_1)\nonumber\\
+&e^{-t/\tau_n}g^r[1-(\tau+t)a_{x_i}^{r}u_i-\tau A^r] (1-H(u_1)).
\end{align}

\subsection{Direct evolution of the cell averaged first-order spatial derivatives} \label{slope-section}

Distinguished from the approximate Riemann solver with a constant state at a cell interface,
the gas-kinetic scheme provides a time evolution solution.
Recall Eq.(\ref{point}), the conservative variables at the Gaussian point  $\textbf{x}_{p,k}$ can be updated by the moments with $\pmb{\psi}$,
\begin{align}\label{point-interface}
\textbf{W}_{p,k}(t^{n+1})=\int \pmb{\psi} f^n(\textbf{x}_{p,k},t^{n+1},\textbf{u},\xi) \text{d}\Xi,~ k=1,...,M
\end{align}

According to the Gauss's theorem, the cell-averaged first-order derivatives within each element at $t^{n+1}$ can be given by
\begin{equation}\label{gauss-formula}
\begin{aligned}
\overline{W}_x^{n+1}
&=\int_{V} \nabla \cdot (\overline{W}(t^{n+1}),0,0) \text{d}V
=\frac{1}{ \Delta V}\int_{\partial V} (1,0,0) \cdot \textbf{n}  \overline{W}(t^{n+1}) \text{d}S \\
&=\frac{1}{ \Delta V}\int_{\partial V} \overline{W}(t^{n+1}) n_1   \text{d}S
 =\frac{1}{ \Delta V} \sum_{p=1}^{N_f}\sum_{k=1}^{M} \omega_{p,k} W^{n+1}_{p,k} (n_1)_{p,k} \Delta S_p,
 \\
\overline{W}_y^{n+1}
&=\int_{V} \nabla \cdot (0,\overline{W}(t^{n+1}),0) \text{d}V
= \frac{1}{ \Delta V}\int_{\partial V} (0,1,0) \cdot \textbf{n}  \overline{W}(t^{n+1}) \text{d}S \\
& = \frac{1}{ \Delta V}\int_{\partial V} \overline{W}(t^{n+1}) n_2   \text{d}S
 =\frac{1}{ \Delta V} \sum_{p=1}^{N_f}\sum_{k=1}^{M} \omega_{p,k} W^{n+1}_{p,k} (n_2)_{p,k} \Delta S_p,
\\
\overline{W}_z^{n+1}
&=\int_{V} \nabla \cdot (0,0,\overline{W}(t^{n+1})) \text{d}V
=\frac{1}{ \Delta V}\int_{\partial V} (0,0,1) \cdot \textbf{n}  \overline{W}(t^{n+1}) \text{d}S \\
&=\frac{1}{ \Delta V}\int_{\partial V} \overline{W}(t^{n+1}) n_3 \text{d}S
 =\frac{1}{ \Delta V} \sum_{p=1}^{N_f}\sum_{k=1}^{M} \omega_{p,k} W^{n+1}_{p,k} (n_3)_{p,k} \Delta S_p,
\end{aligned}
\end{equation}
where $\textbf{n}_{m,k}=((n_{1})_{m,k},(n_{2})_{m,k},(n_{3})_{m,k})$ is the outer unit normal direction at each Gaussian point $\textbf{x}_{p,k}$.

\section{Two-stage temporal discretization}

The two-stage fourth-order (S2O4) temporal discretization which has
been adopted in the previous compact schemes on two-dimensional cases is implemented here \cite{ji2018compact,ji2020hweno}.
Following the definition of Eq.\eqref{semidiscrete},
a fourth-order time-accurate solution for cell-averaged conservative flow variables $\textbf{W}_i$ is updated by
\begin{equation}\label{s2o4}
\begin{aligned}
\textbf{W}_i^*&=\textbf{W}_i^n+\frac{1}{2}\Delta t\mathcal
{L}(\textbf{W}_i^n)+\frac{1}{8}\Delta t^2\frac{\partial}{\partial
	t}\mathcal{L}(\textbf{W}_i^n), \\
\textbf{W}_i^{n+1}&=\textbf{W}_i^n+\Delta t\mathcal
{L}(\textbf{W}_i^n)+\frac{1}{6}\Delta t^2\big(\frac{\partial}{\partial
	t}\mathcal{L}(\textbf{W}_i^n)+2\frac{\partial}{\partial
	t}\mathcal{L}(\textbf{W}_i^*)\big),
\end{aligned}
\end{equation}
where
$\mathcal{L}(\textbf{W}_i^n)$ and $\frac{\partial}{\partial t}\mathcal{L}(\textbf{W}_i^n)$ are
\begin{equation} \label{flux-operator}
\begin{aligned}
\mathcal{L}(\textbf{W}_i^n)&= -\frac{1}{\left| \Omega_i \right|} \sum_{p=1}^{N_f}
\sum_{k=1}^{M} \omega_{p,k} \textbf{F}(\textbf{x}_{p,k},t_n)\cdot \textbf{n}_{p,k},\\
\frac{\partial}{\partial t}\mathcal{L}(\textbf{W}_i^n)&= -\frac{1}{\left| \Omega_i \right|} \sum_{p=1}^{N_f}\sum_{k=1}^{M} \omega_{p,k}
\partial_t \textbf{F}(\textbf{x}_{p,k},t_n)\cdot \textbf{n}_{p,k}, \\
\frac{\partial}{\partial t}\mathcal{L}(\textbf{W}_{i}^*)&=-\frac{1}{\left| \Omega_i \right|} \sum_{p=1}^{N_f}\sum_{k=1}^{M} \omega_{p,k}
\partial_t \textbf{F}(\textbf{x}_{p,k},t_*)\cdot \textbf{n}_{p,k}.
\end{aligned}
\end{equation}
The proof for fourth-order accuracy can be found in \cite{li2016twostage}.

In order to obtain the numerical fluxes $\textbf{F}_{p,k}$ and their time derivatives  $\partial_t \textbf{F}_{p,k}$ at $t_n$ and $t_*=t_n + \Delta t/2$,
the time accurate flux function, as shown in Eq.\eqref{2nd-flux}, can be approximated as a linear function of time within a time interval.
Let's first
introduce the following notation,
\begin{align*}
\mathbb{F}_{p,k}(\textbf{W}^n,\delta)=\int_{t_n}^{t_n+\delta} \textbf{F}_{p,k}(\textbf{W}^n,t)\text{d}t.
\end{align*}
For convenience, assume $t_n=0$,
the flux in the time interval $[t_n, t_n+\Delta t]$ is expanded as
the following linear form
\begin{align*}
\textbf{F}_{p,k}(\textbf{W}^n,t)=\textbf{F}_{p,k}^n+ t \partial_t \textbf{F}_{p,k}^n  .
\end{align*}
The coefficients $\textbf{F}_{p,k}^n$ and $\partial_t\textbf{F}_{p,k}^n$ can be
fully determined by
\begin{align*}
\textbf{F}_{p,k}(\textbf{W}^n,t_n)\Delta t&+\frac{1}{2}\partial_t
\textbf{F}_{p,k}(\textbf{W}^n,t_n)\Delta t^2 =\mathbb{F}_{p,k}(\textbf{W}^n,\Delta t) , \\
\frac{1}{2}\textbf{F}_{p,k}(\textbf{W}^n,t_n)\Delta t&+\frac{1}{8}\partial_t
\textbf{F}_{p,k}(\textbf{W}^n,t_n)\Delta t^2 =\mathbb{F}_{p,k}(\textbf{W}^n,\Delta t/2).
\end{align*}
By solving the linear system, we have
\begin{equation}\label{linear-system}
\begin{aligned}
\textbf{F}_{p,k}(\textbf{W}^n,t_n)&=(4\mathbb{F}_{p,k}(\textbf{W}^n,\Delta t/2)-\mathbb{F}_{p,k}(\textbf{W}^n,\Delta t))/\Delta t,\\
\partial_t \textbf{F}_{p,k}(\textbf{W}^n,t_n)&=4(\mathbb{F}_{p,k}(\textbf{W}^n,\Delta t)-2\mathbb{F}_{p,k}(\textbf{W}^n,\Delta t/2))/\Delta
t^2.
\end{aligned}
\end{equation}
Finally, with Eq.\eqref{flux-operator}, and Eq.\eqref{linear-system},
$W_{i}^{n+1}$ at $\displaystyle t_{n+1}$ can be updated by Eq.\eqref{s2o4}.

\begin{rmk}
	The use of S2O4 time integration in GKS is due to the following reasons.
	Firstly, the fourth-order two-derivative method with two-stage is unique and is efficient in comparison with the RK method.
	Secondly, the accuracy and robustness of S2O4 time integration has been validated by numerical tests for both non-compact and compact schemes \cite{Pan2016twostage,ji2018family,ji2018compact,pan2018two,ji2020performance},
	and has been extended to compressible multi-component flow \cite{pan2017two-multicomponent},
	hypersonic non-equilibrium multi-temperature flow \cite{cao2018physical},
	and direction simulation of compressible homogeneous turbulence \cite{cao2019three}.
	The S2O4 method becomes a building block of a family of HGKS for time integration.
\end{rmk}

Similar to the two-stage temporal discretization in the flux evaluation,
the time dependent gas distribution function at a cell interface is updated as
\begin{equation}\label{step-du}
\begin{split}
&f^*=f^n+\frac{1}{2}\Delta tf_t^n,\\
&f^{n+1}=f^n+\Delta tf_{t}^*,
\end{split}
\end{equation}
where a second-order evolution model is used for the update of gas distribution function on the cell interface and for the
evaluation of flow variables.

In order to construct the first-order time derivative of the gas distribution function,
the distribution function in Eq.(\ref{2nd-flux}) is approximated by the linear function
\begin{align*}
f(t)=f(\textbf{x}_{p,k},t,\textbf{u},\xi)=f^n+
f_{t}^n(t-t^n).
\end{align*}
According to the gas-distribution function at
$t=0$ and $\Delta t$
\begin{align*}
f^n&=f(0),\\
f^n&+f_{t}^n\Delta t=f(\Delta t),
\end{align*}
the coefficients $f^n, f_{t}^n$ can be
determined by
\begin{align*}
f^n&=f(0),\\
f^n_t&=(f(\Delta t)-f(0))/\Delta t.\\
\end{align*}
Thus, $f^*$ and $f^{n+1}$ are fully determined at the cell interface for the evaluation of macroscopic flow variables.
This temporal evolution for the interface value is similar to the one used in GRP solver \cite{du2018hermite} on Cartesian grid.
There is no rigorous proof for its temporal accuracy on general mesh so far.
However, numerical tests demonstrate that the proposed compact scheme
achieves a third-order temporal accuracy.

\section{Compact HWENO reconstruction}

In this section,  the discontinuous values and their first-order derivatives of flow variables at each Gaussian point of a cell interface
will be constructed by a newly designed compact HWENO-type reconstruction.
Then, based on such an initial condition the time-dependent gas distribution function in Eq.\ref{2nd-flux} can be fully determined.

As a starting point of WENO reconstruction, a linear reconstruction
will be presented first. For
a piecewise smooth function $Q( \textbf{x} )$ (Q can be conservative or characteristic variables) over cell $\Omega_{0}$, a
polynomial $P^r(\textbf{x})$ with degrees $r$ can be constructed to
approximate $Q(\textbf{x})$ as follows
\begin{equation*}
P^r(\textbf{x})=Q(\textbf{x})+O(\Delta h^{r+1}),
\end{equation*}
where $\Delta h \sim |\Omega_{0}|^{\frac{1}{3}}$ is the equivalent cell size.
In order to achieve a third-order accuracy and satisfy conservative property,
the following quadratic polynomial over cell $\Omega_{0}$ is obtained
\begin{equation}\label{p2-def}
P^2(\textbf{x})= \overline{Q}_{0}+\sum_{|k|=1}^2a_kp^k(\textbf{x}),
\end{equation}
where $\overline{Q}_{0}$ is the cell averaged value of $Q(\textbf{x})$ over cell $\Omega_{0}$, $k=(k_1,k_2,k_3)$, $|k| = k_1+k_2+k_3$.
The $p^k(\textbf{x})$ are basis functions, which are given by
\begin{align}\label{base}
\displaystyle p^k(\textbf{x})=x_1^{k_1}x_2^{k_2}x_3^{k_3}-\frac{1}{\left| \Omega_{0} \right|}\displaystyle\iiint_{\Omega_{0}}x_1^{k_1}x_2^{k_2}x_3^{k_3} \text{d}V.
\end{align}

\begin{rmk}
The trilinear interpolation is used to describe a given hexahedron $\Omega_{0}$ with coplanar or non-coplanar vertexes,
\begin{align*}
\textbf{X} (\xi, \eta, \zeta)= \sum_{i=1}^8 \textbf{x}_i \phi_i (\xi, \eta, \zeta),
\end{align*}
where $(\xi, \eta, \zeta) \in [-1/2, 1/2]^3$, $\textbf{x}_i$ is the locations of the ith vertex
and $\phi_i$ is the base function as follows
\begin{equation}\begin{aligned}
&\phi_{1}=\frac{1}{8}(1-2 \xi)(1-2 \eta)(1-2 \zeta),\\
&\phi_{2}=\frac{1}{8}(1-2 \xi)(1-2 \eta)(1+2 \zeta),\\
&\phi_{3}=\frac{1}{8}(1-2 \xi)(1+2 \eta)(1-2 \zeta),\\
&\phi_{4}=\frac{1}{8}(1-2 \xi)(1+2 \eta)(1+2 \zeta),\\
&\phi_{5}=\frac{1}{8}(1+2 \xi)(1-2 \eta)(1-2 \zeta),\\
&\phi_{6}=\frac{1}{8}(1+2 \xi)(1-2 \eta)(1+2 \zeta),\\
&\phi_{7}=\frac{1}{8}(1+2 \xi)(1+2 \eta)(1-2 \zeta),\\
&\phi_{8}=\frac{1}{8}(1+2 \xi)(1+2 \eta)(1+2 \zeta).
\end{aligned}\end{equation}
Then the integration of monomial in Eq.\eqref{base} can be given by
\begin{equation} \label{base-int-change}
\begin{array}{l}
\iiint_{\Omega} x^{k_1} y^{k_2} z^{k_3} \mathrm{d} x \mathrm{d} y \mathrm{d} z
=\iiint_{\Omega} x^{k_1} y^{k_2} z^{k_3} (\xi,\eta,\zeta)
\left|\frac{\partial(x, y, z)}{\partial(\xi, \eta, \zeta)}\right|
\mathrm{d} \xi \mathrm{d} \eta \mathrm{d} \zeta.
\end{array}\end{equation}
It can be evaluated numerically as
\begin{equation}\begin{array}{l}
 \iiint_{\Omega} x^{k_1} y^{k_2} z^{k_3} \mathrm{d} x \mathrm{d} y \mathrm{d} z
 =\sum_{l, m, n=1}^{M} \omega_{l,m,n} x^{k_1} y^{k_2} z^{k_3} \left(\xi_{l}, \eta_{m}, \zeta_{n}\right)\left|\frac{\partial(x, y, z)}{\partial(\xi, \eta, \zeta)}\right|_{\left(\xi_{l}, \eta_{m}, \zeta_{n}\right)} \Delta \xi \Delta \eta \Delta \zeta,
 \end{array}\end{equation}
where $\omega_{l,m,n}$ is the quadrature weight for the Gaussian point $\left(\xi_{l}, \eta_{m}, \zeta_{n}\right)$.
 For the current third-order scheme,
 a $2 \times 2 \times 2$ Gaussian quadrature with fourth-order spatial accuracy is used with $\omega_{l,m,n}= (\frac{1}{2})^3$ and $\left(\xi_{l}, \eta_{m}, \zeta_{n}\right) = (\pm \frac{\sqrt{3}}{6}, \pm \frac{\sqrt{3}}{6}, \pm \frac{\sqrt{3}}{6}) $.
\end{rmk}

\subsection{Large stencil and sub-stencils}

\begin{figure}[!h]
	\centering
	\includegraphics[height=0.4\textwidth]{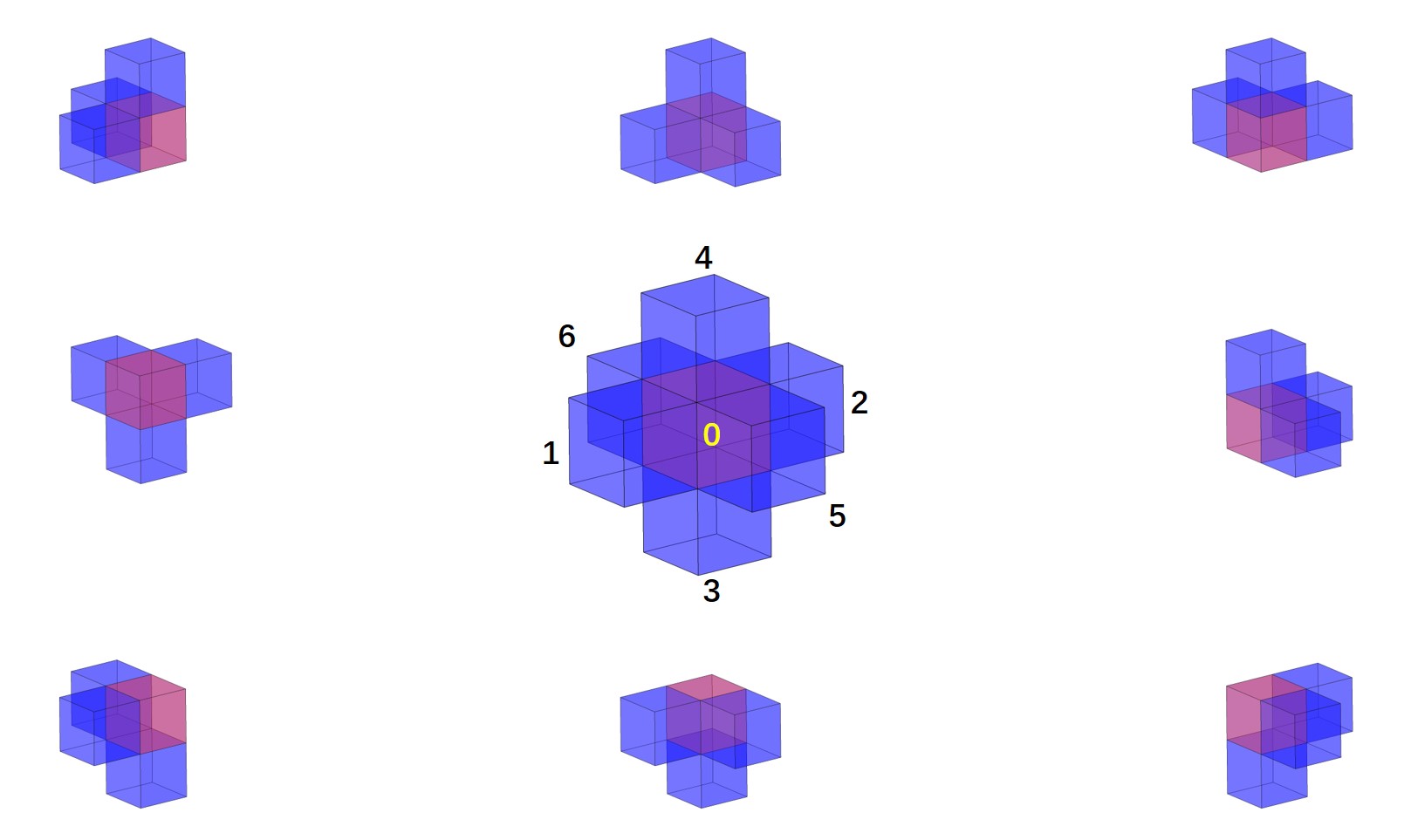}
	\caption{The compact stencils of cell $\Omega_{0}$ for HWENO-type reconstruction. Center: the large stencil. Others: the eight sub-stencils.}
	\label{stencil}
\end{figure}

In order to reconstruct the quadratic polynomial $P^2(\textbf{x})$ on $\Omega_{0}$,
the large stencil for reconstruction includes $\Omega_{0}$ and all its von-Neumann neighbors, $\Omega_{m}, m=1,...,6$
where the averages of $Q(\textbf{x})$ and averaged derivatives of $Q(\textbf{x})$ over each cell are known.

The following values are used to obtain $P^2(\textbf{x})$,

\begin{itemize}
	\item cell averages $\overline{Q}$ for cell 0, 1, 2, 3, 4, 5, 6,
	\item cell averages of the $x$-direction partial derivative $\overline{Q}_x$ for cell 1, 2, 3, 4, 5, 6,
	\item cell averages of the $y$-direction partial derivative $\overline{Q}_y$ for cell 1, 2, 3, 4, 5, 6,
		\item cell averages of the $z$-direction partial derivative $\overline{Q}_z$ for cell 1, 2, 3, 4, 5, 6.
\end{itemize}

The polynomial $P^2(x,y)$ is required to exactly satisfy
\begin{align} \label{large-stenci-condition-1}
\iiint_{\Omega_{m}}P^2(x,y)\text{d}V=\overline{Q}_{m}\left| \Omega_{m}\right|,
\end{align}
where $\overline{Q}_m$ is the cell averaged value over $\Omega_{m},~ m=1,...,6$.
Then we require  the following condition satisfied in a least-squares sense
\begin{equation}\label{large-stenci-condition-2}
\begin{split}
\iiint_{\Omega_{m}}
\frac{\partial}{\partial x_1} P^2(\textbf{x})\text{d}V=(\overline{Q}_{x_1})_m|\Omega_{m}|,\\
\iiint_{\Omega_{m}}
\frac{\partial}{\partial x_2} P^2(\textbf{x})\text{d}V=(\overline{Q}_{x_2})_m|\Omega_{m}|,\\
\iiint_{\Omega_{m}}
\frac{\partial}{\partial x_3} P^2(\textbf{x})\text{d}V=(\overline{Q}_{x_3})_m|\Omega_{m}|,
\end{split}	
\end{equation}
where  $\overline{Q}_{x_i}, i=1,2,3$  are the cell averaged directional derivatives over $\Omega_{m}$ in a global coordinate,
respectively.
On a regular mesh, the system has $24$ independent equations.
The constrained least-square method is used to solve the above linear system \cite{li2014efficient}.

\begin{rmk} \label{rmk-reconstruction}
	The constraints introduced in Eq.\eqref{large-stenci-condition-1} can improve the linear stability of the reconstruction and reduce the numerical errors.
	The information from the cell-averaged values takes up a high proportion using the technique.
	Under uniform mesh, the coefficients in Eq.\eqref{p2-def} for cell $\Omega_{i,j,k}$ are given in a concise and elegant form
\begin{equation}
\begin{split}
	&a_{1,0,0} = \frac{\overline{Q}_{i+1,j,k}-\overline{Q}_{i-1,j,k}}{2 h},~
	a_{0,1,0} = \frac{\overline{Q}_{i,j+1,k}-\overline{Q}_{i,j-1,k}}{2 h},~	
	a_{0,0,1} = \frac{\overline{Q}_{i,j,k+1}-\overline{Q}_{i,j,k-1}}{2 h},~\\
	&a_{2,0,0} = \frac{\overline{Q}_{i+1,j,k}+\overline{Q}_{i-1,j,k}-2 \overline{Q}_{i,j,k}}{2 h^2},~
    a_{0,2,0} = \frac{\overline{Q}_{i,j+1,k}+\overline{Q}_{i,j+1,k}-2 \overline{Q}_{i,j,k}}{2 h^2},~\\	
    &a_{0,0,2} = \frac{\overline{Q}_{i,j,k+1}+\overline{Q}_{i,j+1,k-1}-2 \overline{Q}_{i,j,k}}{2 h^2},~\\			
   &  a_{1,1,0} = \frac{(\overline{Q}_{x_1})_{i,j+1,k}-(\overline{Q}_{x_1})_{i,j-1,k} + (\overline{Q}_{x_2})_{i+1,j,k}-(\overline{Q}_{x_2})_{i-1,j,k}}{4 h},\\
    &  a_{0,1,1} = \frac{(\overline{Q}_{x_2})_{i,j,k+1}-(\overline{Q}_{x_2})_{i,j,k-1} + (\overline{Q}_{x_3})_{i,j+1,k}-(\overline{Q}_{x_3})_{i,j-1,k}}{4 h},  \\
    & a_{1,0,1} = \frac{(\overline{Q}_{x_1})_{i,j,k+1}-(\overline{Q}_{x_1})_{i,j,k-1} + (\overline{Q}_{x_3})_{i+1,j,k}-(\overline{Q}_{x_3})_{i-1,j,k}}{4 h},
\end{split}	
\end{equation}
The information from the derivatives only shows in the cross terms.
In addition, it will reduce to a third-order scheme involving the cell-averaged values only in the 1-D case.
\end{rmk}

In order to deal with discontinuity, eight sub-stencils $S_{j},
j=1,...,8$ are selected from the large one given in
Fig.\ref{stencil}. And the following cell averaged values for each sub-stencil are
used to get the linear polynomial $P^1_j(\textbf{x})$,
\begin{align*}
P_{1}^1 ~&\text{on}~ S_1=\{\bar{Q}_0,\bar{Q}_1,\bar{Q}_3,\bar{Q}_5 \},
~~~P_{2}^1 ~\text{on}~ S_2=\{\bar{Q}_0,\bar{Q}_3,\bar{Q}_5,\bar{Q}_2 \},\\
P_{3}^1 ~&\text{on}~ S_3=\{\bar{Q}_0,\bar{Q}_5,\bar{Q}_2 ,\bar{Q}_4 \},
~~~P_{4}^1 ~\text{on}~ S_4=\{\bar{Q}_0,\bar{Q}_4,\bar{Q}_2,\bar{Q}_6 \},\\
P_{5}^1 ~&\text{on}~ S_5=\{\bar{Q}_0,\bar{Q}_2,\bar{Q}_6,\bar{Q}_3 \},
~~~P_{6}^1 ~\text{on}~ S_6=\{\bar{Q}_0,\bar{Q}_6,\bar{Q}_3 ,\bar{Q}_1 \},\\
P_{7}^1 ~&\text{on}~ S_7=\{\bar{Q}_0,\bar{Q}_6,\bar{Q}_4,\bar{Q}_1 \},
~~~P_{7}^1 ~\text{on}~ S_7=\{\bar{Q}_0,\bar{Q}_4,\bar{Q}_1,\bar{Q}_5 \},
\end{align*}
There is always one sub-stencil in smooth region with the appearance of discontinuity
near any one of the interfaces of the target cell.
The method in \cite{zhao2017weighted} can be used to obtain $P^1_j(x,y)$, which avoids the singularity
caused by mesh irregularity,
and the linear polynomial is expressed as
\begin{equation}\label{linear-def}
P^1_j(\textbf{x})=\bar{Q}_{0}+\sum_{|k|=1}^1a_{j,k}p^k(\textbf{x}).
\end{equation}
Note that the choice of the large and sub-stencils is not unique.

\subsection{Define the values of linear weights}
$P^2(\textbf{x})$ is written as
\begin{align*}
P^2(\textbf{x})=d_0[\frac{1}{d_0} P^2(\textbf{x}) - \sum_{j=1}^{8} \frac{d_{j}}{d_0} P_{j}^1(\textbf{x})]+\sum_{j=1}^{8} d_{j} P_{j}^1(\textbf{x}) ,
\end{align*}
where the linear weights are chosen as $\gamma_0=0.92, \gamma_j= 0.01, j=1,...,8$ without special statement according to \cite{zhu2018new,ji2020hweno}.

\subsection{Compute the non-linear weights}

The smoothness indicators $\beta_{j}, j=0,...,8$ are defined as
\begin{equation*}
\beta_j=\sum_{|\alpha|=1}^{r_j}|\Omega|^{ \frac{2}{3}|\alpha|-1}\iiint_{\Omega}\big(D^{\alpha}P_j(\textbf{x})\big)^2
 \text{d} V,
\end{equation*}
where $\alpha$ is a multi-index and $D$ is the derivative operator, $r_0=2$, $r_j=1, j=1,...,8$.
The smoothness indicators in Taylor series at $(x_0,y_0)$ have the order
\begin{align*}
\beta_0&=O\{|\Omega_0|^{\frac{2}{3}}[1+O(|\Omega_0|^{\frac{2}{3}})]\}=O(|\Omega_0|)^{\frac{2}{3}} = O(h^2),\\
\beta_j&=O\{|\Omega_0|^{\frac{2}{3}}[1+O(|\Omega_0|^{\frac{1}{3}})]\}=O(|\Omega_0|)^{\frac{2}{3}} = O(h^2),j=1,...,8.
\end{align*}
By using a similar technique \cite{zhu2018new}, a global smoothness indicator $\sigma$ can be defined
\begin{equation*}
\sigma = (\frac{1}{8}\sum_1^8|\beta_0-\beta_j|)^2 = O(|\Omega_0|^2) = O(h^6),
\end{equation*}
then the corresponding non-linear weights are given by
\begin{equation}\label{non-linear-weight}
\begin{split}
&\omega_{j}=d_{j}(1+\frac{\sigma}{\epsilon+\beta_{j}}),~j = 0,...,8, \\
&\delta_{j}=\frac{\omega_{j}}{\sum_{l=0}^{8}\omega_{l}} = d_{j} + O(h^4),
\end{split}
\end{equation}
where $\epsilon$ takes $10^{-8}$ to avoid zero in the denominator.

The final reconstruction polynomial for the approximation of $Q(\textbf{x})$ yields
\begin{equation} \label{final_hweno_expression}
R(\textbf{x})=\delta_0[\frac{1}{d_0} P^2(\textbf{x}) - \sum_{j=1}^{8} \frac{d_{j}}{d_0} P_{j}^1(\textbf{x})]+\sum_{j=1}^{8} \delta_{j} P_{j}^1(\textbf{x}).
\end{equation}
As a result, the non-linear reconstruction achieves a third-order accuracy  $R(\textbf{x})=Q(\textbf{x})+O(h^3)$.
If any of these values yield negative density or pressure,  the first-order reconstruction is used instead.
So all the desired quantities at Gaussian points can be fully determined as
\begin{align*}
&Q^{l,r}_{p,k}=R^{l,r}(\textbf{x}_{p,k}), ~(Q^{l,r}_{x_i})_{p,k}= \frac{\partial R^{l,r}}{\partial {x_i}}(\textbf{x}_{p,k}).
\end{align*}

\subsection{Reconstruction of the equilibrium state}
The reconstructions for the non-equilibrium states  have the uniform order and can be used to get the equilibrium state directly,
such as $ g^{c},g_{x_i}^{c}$ by a suitable average of $g^{l,r},g_{x_i}^{l,r}$.
The simplest way is to use the arithmetic average, but it is only applicable for smooth flow.
To be consistent with the construction of  $ g^{c}$, we make an analogy of the kinetic-based weighting method for $g_{x_i}^{c}$, which are given by
\begin{align}\label{new-equ-part}
&\int\pmb{\psi} g^{c}\text{d}\Xi=\textbf{W}^c=\int_{u>0}\pmb{\psi}
g^{l}\text{d}\Xi+\int_{u<0}\pmb{\psi} g^{r}\text{d}\Xi, \nonumber \\
&\int\pmb{\psi} g^{c}_{x_i}\text{d}\Xi=\textbf{W}_{x_i}^c=\int_{u>0}\pmb{\psi}
g_{x_i}^{l}\text{d}\Xi+\int_{u<0}\pmb{\psi} g_{x_i}^{r}\text{d}\Xi.
\end{align}
This method has been used in an early version of second-order GKS \cite{GKS-lecture} and validated in the non-compact WENO5-GKS \cite{ji2020performance}.
In this way, all components of the microscopic slopes in Eq.\eqref{2nd-flux} have been fully obtained.

\section{Numerical tests}

In this section, numerical tests will be presented to validate the compact high-order GKS. For the
inviscid flow, the collision time $\tau_n$ is defined by
\begin{align*}
\tau_n=\epsilon \Delta t+C\displaystyle|\frac{p_l-p_r}{p_l+p_r}|\Delta
t,
\end{align*}
where $\varepsilon=0.01$ and $C=1$. For the viscous flow, the collision time is related to the viscosity coefficient,
\begin{align*}
\tau_n=\frac{\mu}{p}+C \displaystyle|\frac{p_l-p_r}{p_l+p_r}|\Delta t,
\end{align*}
where $p_l$ and $p_r$ denote the pressure on the left and right
sides of the cell interface, $\mu$ is the dynamic viscosity coefficient, and
$p$ is the pressure at the cell interface. In  smooth flow region,
it reduces to $\tau_n=\tau=\mu/p$. The ratio of specific heats takes
$\gamma=1.4$.
The inclusion of the pressure jump term is to enlarge the collision time in the discontinuous region, where the numerical cell size is not enough to resolve the shock structure.
It increases  the non-equilibrium transport mechanism in the flux function to mimic the physical process in the shock layer.

All reconstructions will be performed on the characteristic variables.
Ghost cells are mainly adopted in the current scheme for boundary treatment.
After we obtain the inner state at the boundary, a ghost state can be assigned according to boundary condition, and the corresponding gas distribution function in Eq.\eqref{2nd-flux} can be determined.
A high-order boundary reconstruction is only applied to the test of subsonic flow passing through a circular cylinder.
Explorations on the construction of accurate and stable condition on curved boundaries will continue.
The time step is determined by
\begin{align*}
\Delta t = C_{CFL} \mbox{Min} ( \frac{ \Delta r_i}{||\textbf{U}_i||+(a_s)_i}, \frac{ (\Delta r_i)^2}{3\nu _i}),
\end{align*}
where $C_{CFL}$ is the CFL number, and $||\textbf{U}_i||$, $(a_s)_i$, and $\nu _i= (\mu /\rho) _i$ are the magnitude of velocities, sound speed, and kinematic viscosity coefficient for cell i. The $\Delta r_i$ is taken as
\begin{align*}
\Delta r_i = \frac{|\Omega_i|}{\mbox{Max}|\Gamma_{ip}|}.
\end{align*}
The CFL number is set as  $0.5$ if no specifies in the test cases.

\subsection{Accuracy test}
\noindent{\sl{(a) 3-D sinusoidal wave propagation }}

The advection of density perturbation is tested, and the initial
condition is given as follows
\begin{align*}
\rho(x,y,z)=1+0.2\sin(\pi (x+y+z)),\ \ \ \textbf{U}(x,y,z)=(1,1,1),  \ \ \  p(x,y,z)=1,
\end{align*}
within a cubic domain $[0, 2]\times[0, 2]\times[0, 2]$.
In the computation, a series of uniform meshes with $N^3$ cells are used.
With the periodic boundary condition in all directions, the analytic
solution is
\begin{align*}
\rho(x,y,z,t)=1+0.2\sin(\pi(x+y+z-t)),\ \ \ \textbf{U}(x,y,z)=(1,1,1),\ \ \  p(x,y,z,t)=1.
\end{align*}
The collision time $\tau=0$ is set since the flow is smooth and inviscid.
The $CFL=0.5$ is used for computation.
The $L^1$, $L^2$ and $L^{\infty}$ errors and the corresponding orders with linear and non-linear Z-type weights at $t=2$ are given in Tab.\ref{3d-accuracy-linear} and Tab.\ref{3d-accuracy-weno-1}. The expected accuracy is confirmed.

\begin{table}[htp]
	\small
	\begin{center}
		\def\temptablewidth{1\textwidth}
		{\rule{\temptablewidth}{1pt}}
		\begin{tabular*}{\temptablewidth}{@{\extracolsep{\fill}}c|cc|cc|cc}
			
			mesh number & $L^1$ error & Order & $L^2$ error & Order& $L^{\infty}$ error & Order  \\
			\hline
$5^3$ & 8.591164e-02 & ~ & 9.529661e-02 & ~ & 1.327066e-01 & ~ \\
$10^3$ & 2.201313e-02 & 1.96 & 2.442492e-02 & 1.96 & 3.422233e-02 & 1.96 \\
$20^3$ & 3.084179e-03 & 2.84 & 3.432916e-03 & 2.83 & 5.054260e-03 & 2.76 \\
$40^3$ & 3.949479e-04 & 2.97 & 4.378248e-04 & 2.97 & 6.582257e-04 & 2.94 \\
$80^3$ & 4.954332e-05 & 2.99 & 5.490190e-05 & 3.00 & 8.289161e-05 & 2.99 \\ 				
		\end{tabular*}
		{\rule{\temptablewidth}{0.1pt}}
	\end{center}
	\vspace{-4mm} \caption{\label{3d-accuracy-linear} Accuracy test for the 3D sin-wave
		propagation by the linear third-order compact reconstruction. $CFL=0.5$.  }
\end{table}

\begin{table}[htp]
	\small
	\begin{center}
		\def\temptablewidth{1\textwidth}
		{\rule{\temptablewidth}{1pt}}
		\begin{tabular*}{\temptablewidth}{@{\extracolsep{\fill}}c|cc|cc|cc}
			
			mesh number & $L^1$ error & Order & $L^2$ error & Order& $L^{\infty}$ error & Order  \\
			\hline
$5^3$ & 8.428434e-02 & ~ & 9.425905e-02 & ~ & 1.299919e-01 & ~ \\
$10^3$ & 2.532893e-02 & 1.73 & 2.806704e-02 & 1.75 & 4.319655e-02 & 1.59 \\
$20^3$ & 3.113958e-03 & 3.02 & 3.595147e-03 & 2.96 & 6.419271e-03 & 2.75 \\
$40^3$ & 3.949729e-04 & 2.98 & 4.378852e-04 & 3.04 & 6.576984e-04 & 3.29 \\
$80^3$ & 4.954332e-05 & 2.99 & 5.490190e-05 & 3.00 & 8.289162e-05 & 2.99 \\ 			
		\end{tabular*}
		{\rule{\temptablewidth}{0.1pt}}
	\end{center}
	\vspace{-4mm} \caption{\label{3d-accuracy-weno-1} Accuracy test for the 3D sin-wave
		propagation by the non-linear third-order compact HWENO reconstruction with $d_0=0.92$, $d_i=0.01,~ i=1,...,8$. $CFL=0.5$.  }
\end{table}

\noindent{\sl{(b) Subsonic flow past a circular cylinder }}

This 2-D test has been widely used to test the spatial accuracy for a high-order scheme with curved wall boundary \cite{krivodonova2006high,luo2008computation,wang2016compact}.

\begin{figure}[htp]	
	\centering
	\includegraphics[width=0.4\textwidth]
	{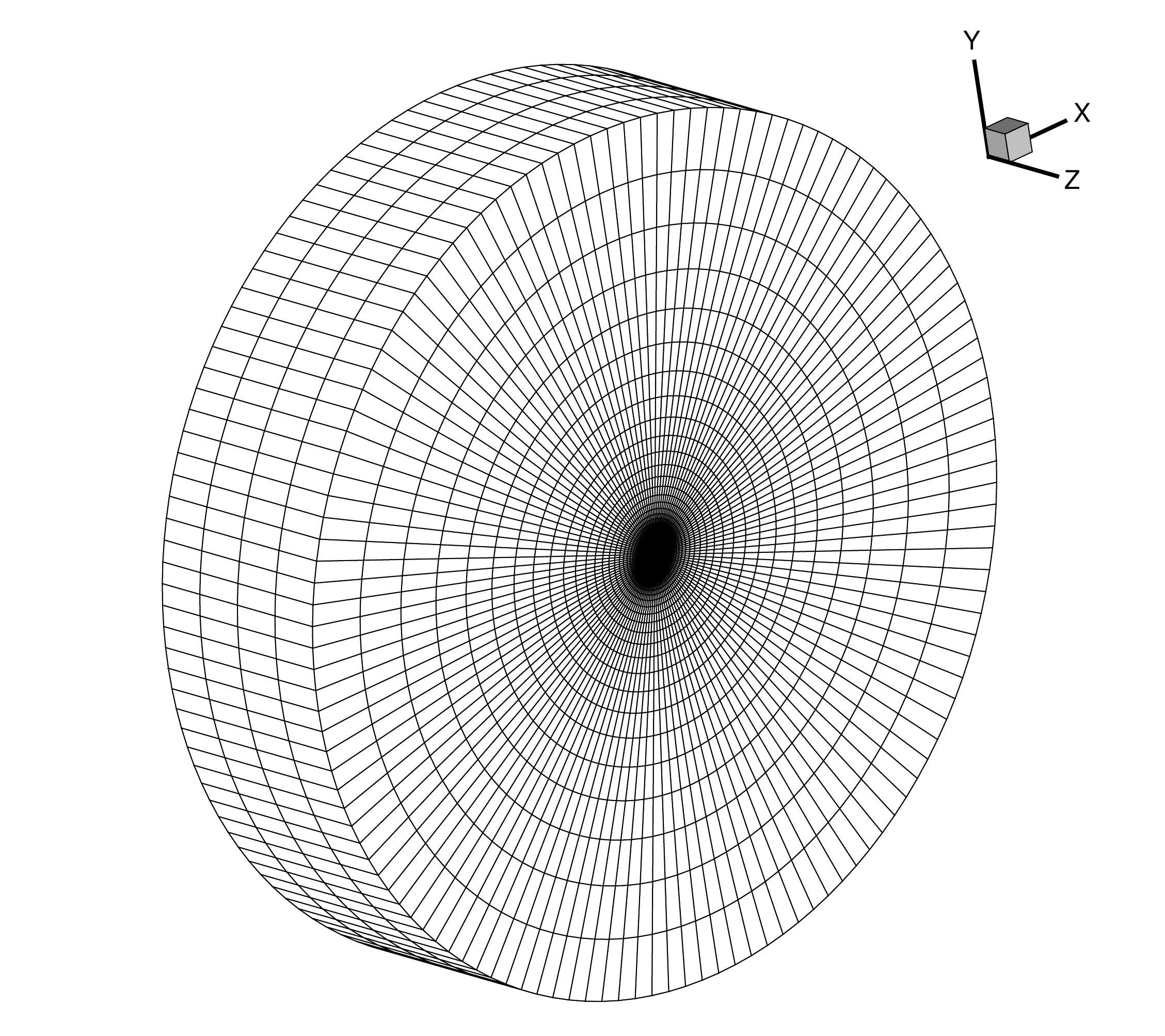}
	\includegraphics[width=0.4\textwidth]
	{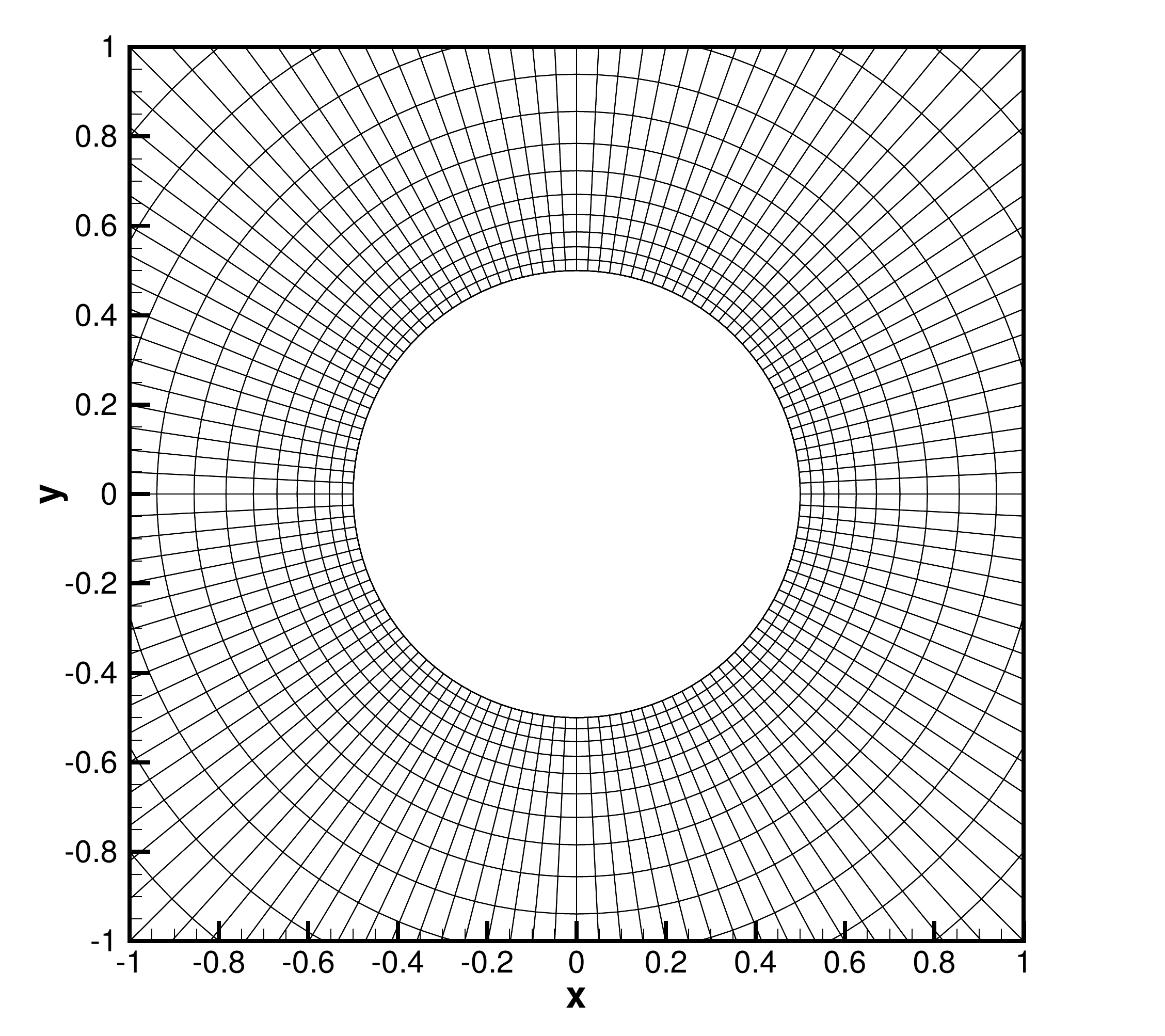}
	\caption{The mesh sample  with $128 \times 32\times 4 $ cells for inviscid flow passing through a 3-D circular cylinder.}
	\label{accuracy-cylinder-mesh}
\end{figure}

A circular cylinder is put in the center of the computational domain with a radius of $r_0=0.5$.
The  concentric computational domain is bounded by a circle $r_{out}=20$.
Four successively refined meshes with $16 \times 4 \times 4$, $32 \times 8 \times 4$, $64 \times 16 \times 4$, and $128 \times 32 \times 4$ cells are given according to \cite{krivodonova2006high}.
Mesh distributions are shown in Fig.\ref{accuracy-cylinder-mesh}.

\begin{figure}[htp]	
	\centering
	\includegraphics[width=0.48\textwidth]
	{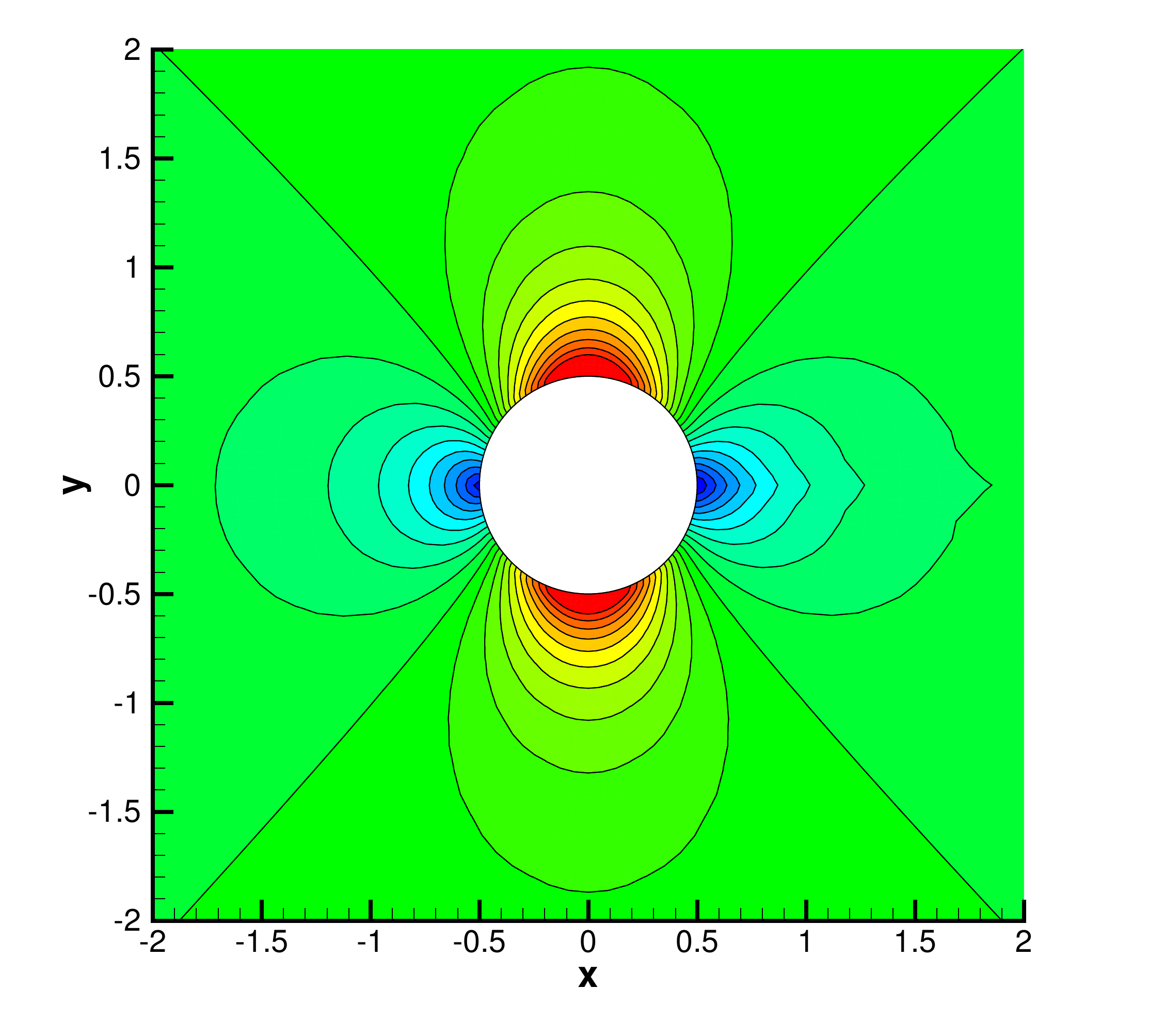}
	\includegraphics[width=0.48\textwidth]
	{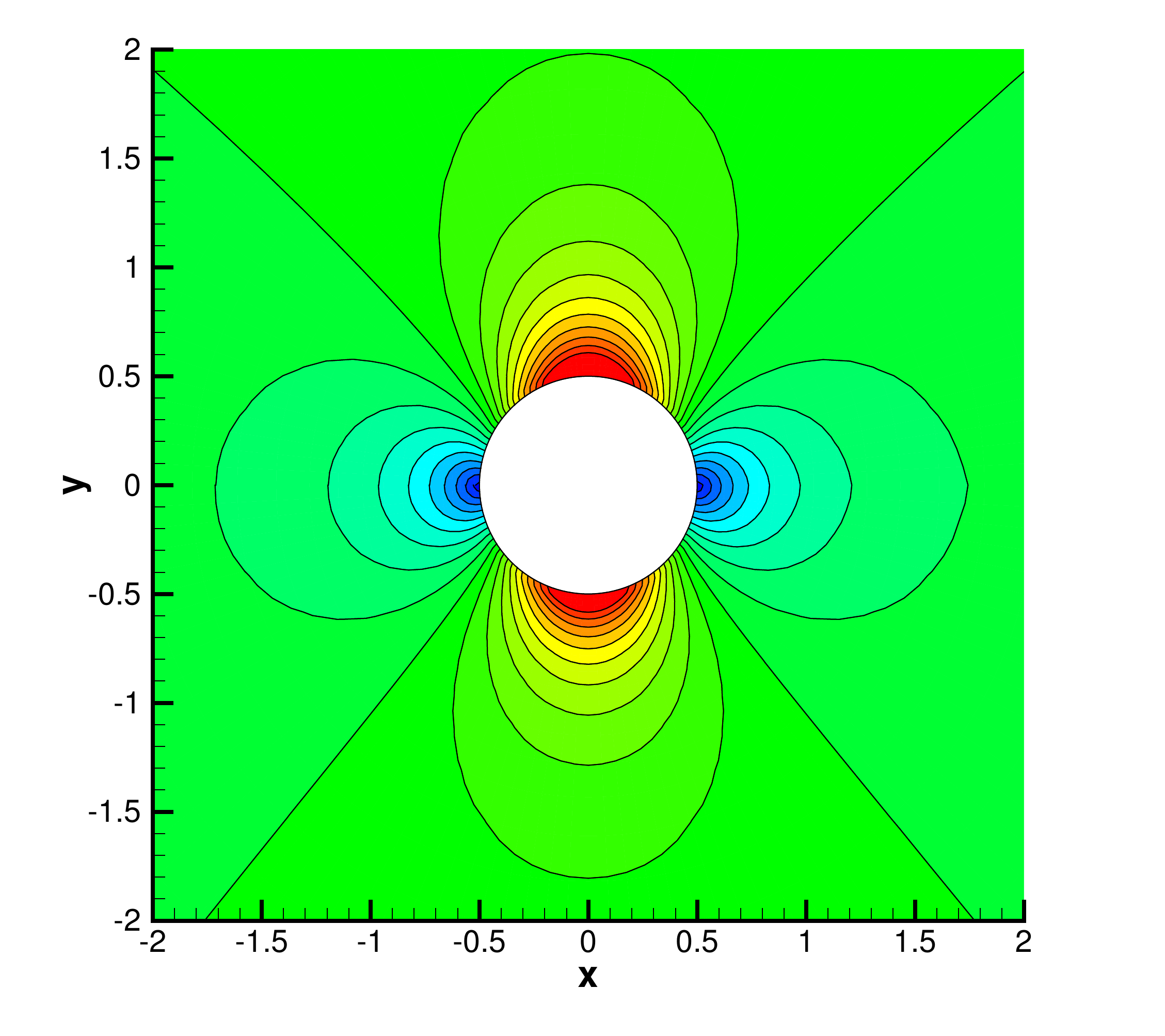}
	\caption{Subsonic flow passing through a circular cylinder: 20 equidistant Mach contours from $0.038$ to $0.76$. Mesh $128 \times 32 \times 4$. Left: the second-order one-sided boundary reconstruction. Right: the third-order compact boundary reconstruction. The curved boundary treatment is applied in both cases.}
	\label{accuracy-cylinder-ma}
\end{figure}

The reflective boundary condition is imposed on the wall of the cylinder.
The far-field boundary condition is set around the outside of the domain, which has a free stream condition
\begin{equation*}
\begin{split}
(\rho,U,V,W,p)_{\infty} =(1,0.38,0,0,\frac{1}{\gamma}),
\end{split}
\end{equation*}
with $\gamma=1.4$.
The periodic boundary condition is given in the Z-direction.
It describes a subsonic inviscid flow at $Ma_{\infty}=0.38$ passing through a cylinder.
Ideally, the flow is isentropic with
\begin{equation*}
\begin{split}
S(x,y,t)=S_{\infty}.
\end{split}
\end{equation*}
Thus, an entropy error, defined as
\begin{equation*}
\begin{split}
\epsilon_{s}=\frac{S-S_{\infty}}{S_{\infty}}=\frac{p}{p_{\infty}}(\frac{\rho_{\infty}}{\rho})^{\gamma}-1,
\end{split}
\end{equation*}
is used for measuring the error of the numerical solution.
The simulation is initialized with the free stream value.
The error is recorded when the flow gets to a steady state.
To achieve a third-order accuracy on the cylinder wall, a one-side compact stencil with six cells $0,1,...,5$ is used to reconstruct a smooth polynomial within a boundary cell 0. This stencil includes 21 data, i.e.,
\begin{itemize}
	\item cell averages $\bar{W}$ for cell 0, 1, 2, 3, 4, 5,
	\item cell averages of the $x$-direction partial derivative $\bar{W}_x$ for cell 0, 1, 2, 3, 4, 5,
	\item cell averages of the $y$-direction partial derivative $\bar{W}_y$ for cell 0, 1, 2, 3, 4, 5,
	\item cell averages of the $z$-direction partial derivative $\bar{W}_z$ for cell 0, 1, 2, 3, 4, 5,	
\end{itemize}
and a quadratic polynomial can determined in a least-squares sense.
Moreover, the curved boundary modification proposed in \cite{krivodonova2006high}  is adopted, by adjusting the normal directions on the boundary at Gaussian point.
A third-order convergence rate is achieved through the above treatment, as shown in Tab.\ref{accuracy-cylinder}.
The numerical result with low-order boundary reconstruction has a more visible wake than that with the high-order one, as shown in Fig.\ref{accuracy-cylinder-ma}.

\begin{table}[htp]
	\small
	\begin{center}
		\def\temptablewidth{0.9\textwidth}
		{\rule{\temptablewidth}{1pt}}
		\begin{tabular*}{\temptablewidth}{@{\extracolsep{\fill}}c|cc|cc}
			~&Second-order&~&Third-order&~\\
			\hline
			mesh number  & $L^2$ error & Order& $L^2$ error & Order  \\
			\hline
			$64\times 16$ & 1.15e-03 &~ &1.40e-4 & ~  \\
			$128\times 32$ & 5.02e-5 & 4.51& 1.01e-5 & 3.79 \\
			$256\times 64$ & 8.75e-6 & 2.52& 1.19e-6 & 3.21 				
		\end{tabular*}
		{\rule{\temptablewidth}{0.1pt}}
	\end{center}
	\vspace{-4mm} \caption{\label{accuracy-cylinder} Accuracy test for the subsonic flow past a circular cylinder by different boundary reconstructions. The curved boundary treatment is applied in  both cases.}
\end{table}

\subsection{Subsonic viscous flow passing through a sphere: Re=118}

This test case is used to test the capability of the proposed method in resolving low-speed viscous flow.
The Reynolds number based on the diameter of the sphere is 118.
In such case, the drag coefficient $C_D=1$ according to the experimental work \cite{taneda1956experimental}.

\begin{figure}[htbp]	
	\centering
	\includegraphics[width=0.4\textwidth]
	{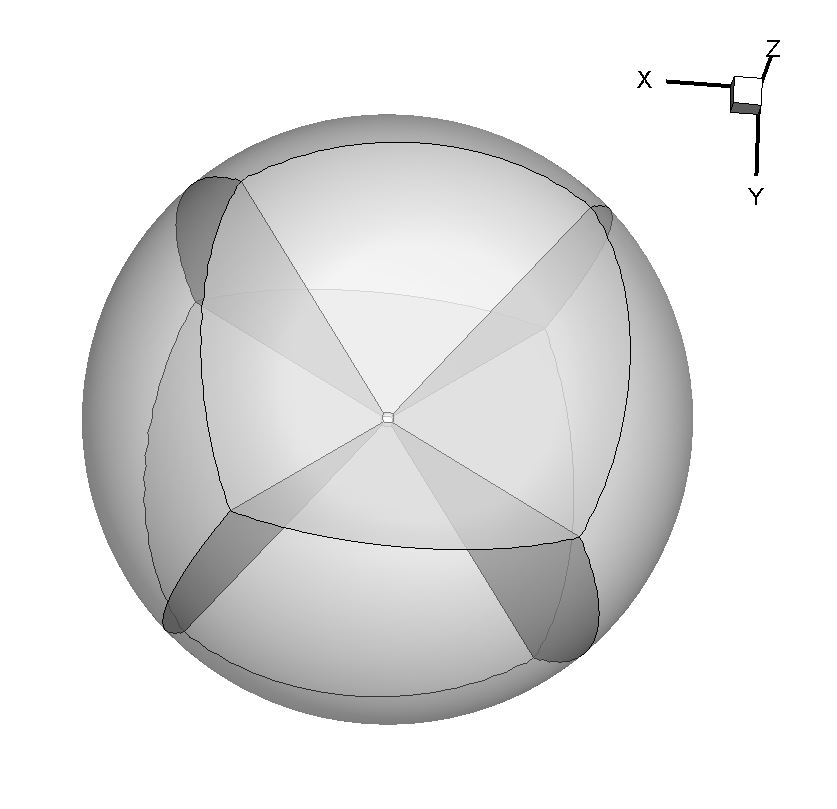}
	\includegraphics[width=0.4\textwidth]{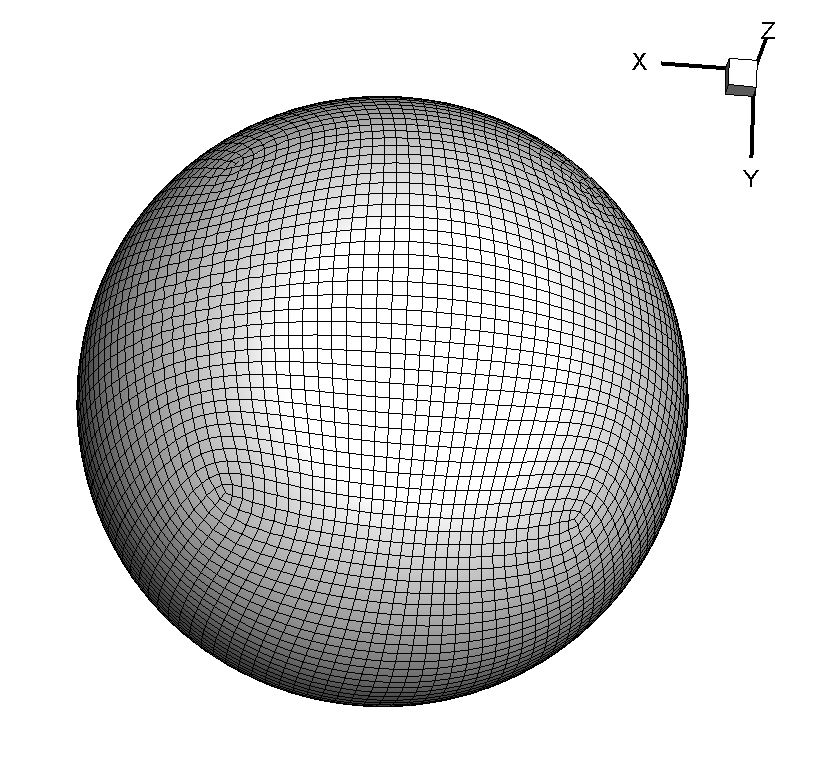}
	\caption{The mesh for viscous flow passing through a sphere with $32\times 32 \times 32 \times 6 $ cells.}
	\label{sphere-mesh}
\end{figure}

\begin{figure}[htbp]	
	\centering
	\includegraphics[width=0.7\textwidth]
	{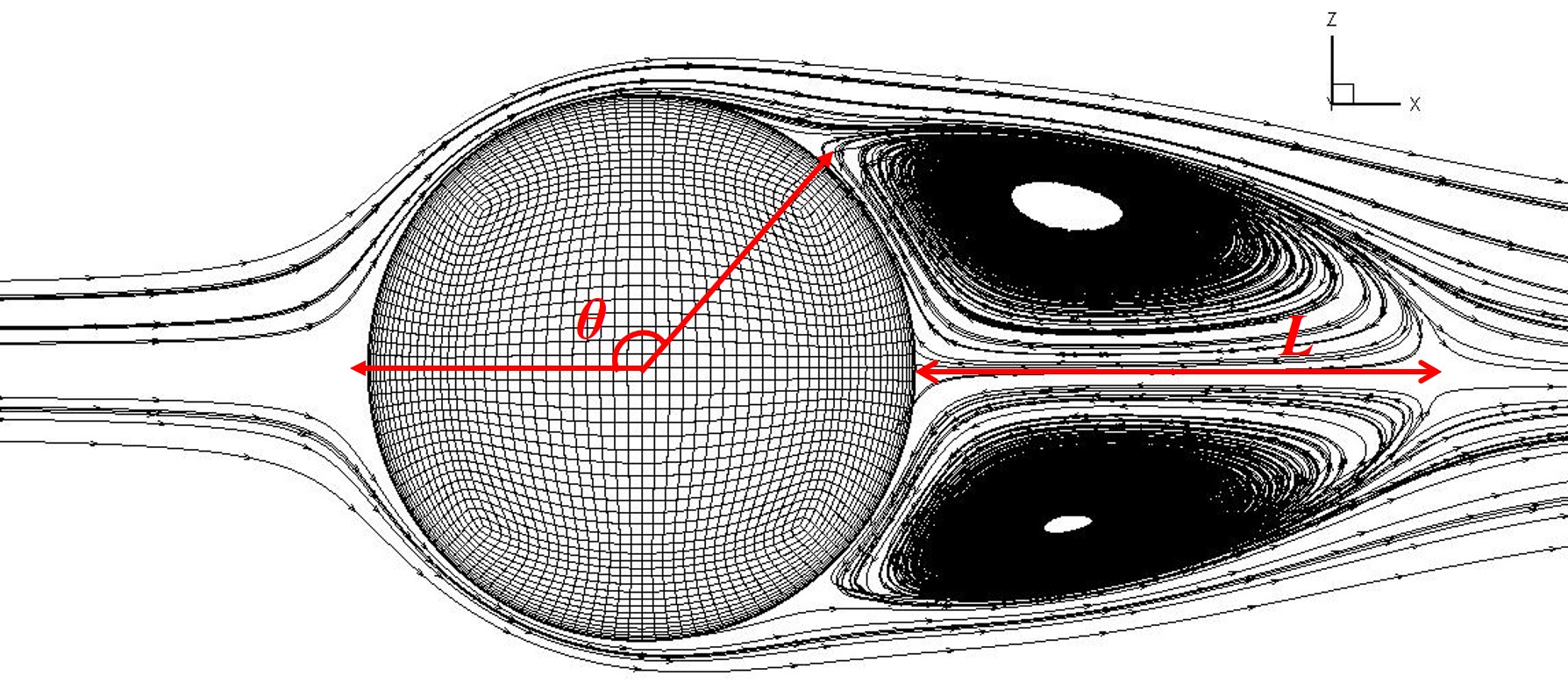}	
	\caption{The definition of the separation angle $\theta$ and the closed wake length $L$.}
	\label{re118-sphere-sketch}
\end{figure}

\begin{figure}[htbp]	
	\centering
	\includegraphics[width=0.49\textwidth]
	{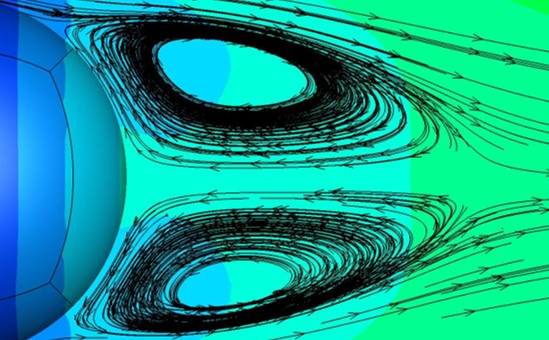}\\
	\includegraphics[width=0.49\textwidth]
	{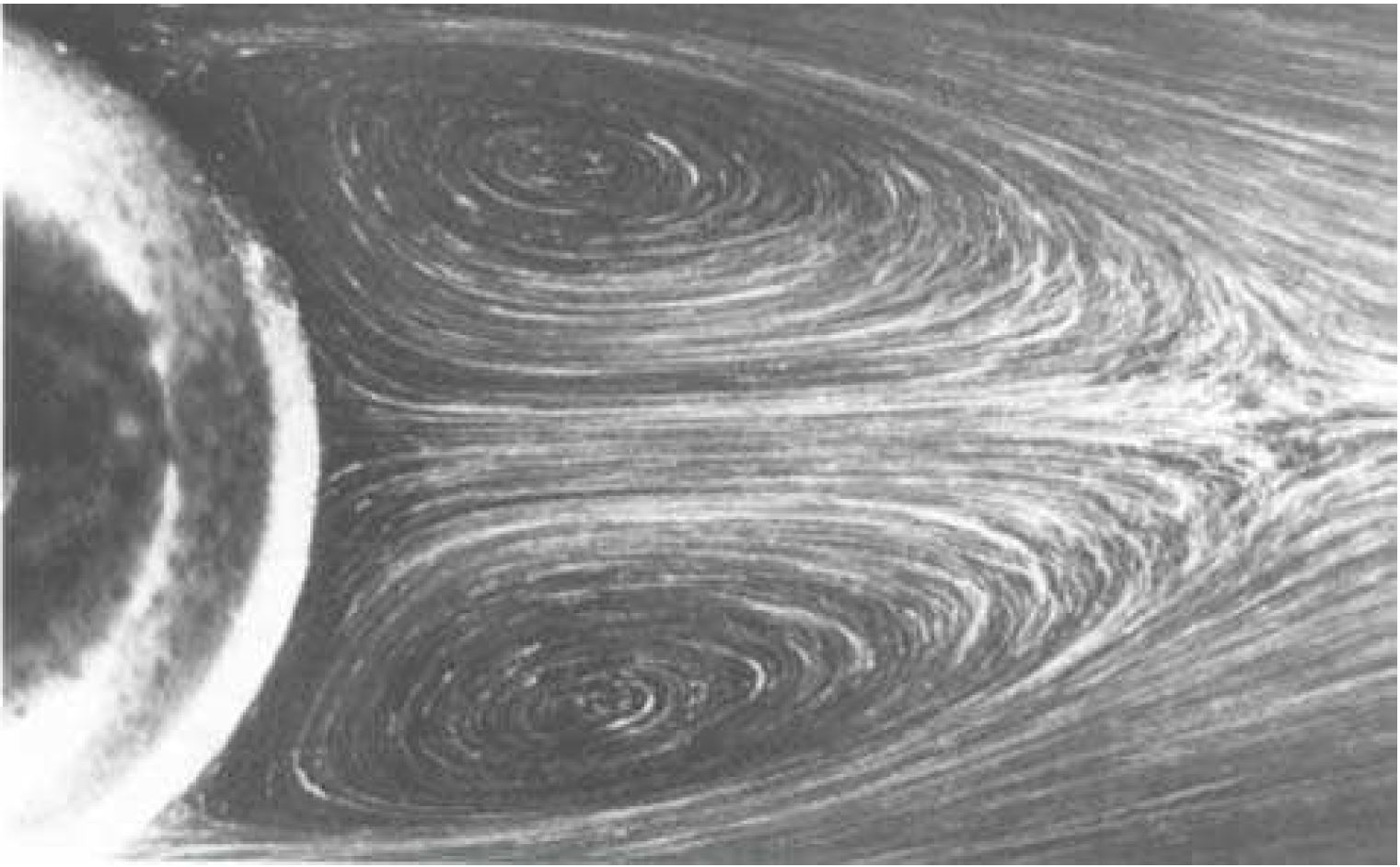}
	\caption{Comparison of streamlines between the experiment \cite{taneda1956experimental} and computation.}
	\label{re-118-sphere-stream}
\end{figure}

The far-field boundary condition is set around the outside of the domain, which has a free stream condition
\begin{equation*}
\begin{split}
(\rho,U,V,W,p)_{\infty} =(1,0.2535,0,0,\frac{1}{\gamma}),
\end{split}
\end{equation*}
with $\gamma=1.4$, $Ma_{\infty}=0.2535$.
The non-slip adiabatic boundary condition is imposed on the surface of the sphere.
The structured grids with six blocks are used in the computation, as shown in Fig.\ref{sphere-mesh}.
The diameter of the sphere is $D=1$ and the first grid off the wall is about $2.45 \times 10^{-2} D$.
The height of the grid grows from the wall with a constant ratio and stop at $20D$ in the radial direction.
The computational streamlines are compared with the experimental streamlines \cite{taneda1956experimental}, as shown in Fig.\ref{re-118-sphere-stream}.
The shape of the steady separation bubble agrees well with each other.
The quantitative results are given in Tab.\ref{re-118-sphere}, including the drag coefficient $C_D$,  the separation angle $\theta$, and the closed wake length $L$.
The drag coefficient are defined as
\begin{equation*}
\begin{split}
C_D=\frac{F_D}{\frac{1}{2} \rho_{\infty} U_{\infty}^2 S},
\end{split}
\end{equation*}
where $S= \frac{1}{4}\pi D^2$. The definition of $\theta$ and $L$ is given in Fig.\ref{re118-sphere-sketch}.
With similar DOFs, the current compact scheme has the closest drag coefficient to the experiment data.
The closed wake length has a visible difference with the experiment, but very close to the DG's result \cite{cheng2017parallel}.
It is probably due to the compressible effect since the experiment is conducted in the low-speed water tank.

\begin{table}[htbp]
	\small
	\begin{center}
		\def\temptablewidth{1.0\textwidth}
		{\rule{\temptablewidth}{1pt}}
		\begin{tabular*}{\temptablewidth}{@{\extracolsep{\fill}}c|c|c|c|c}
			Scheme & DOF & Cd  & $\theta$  &L\\
			\hline
			Experiment \cite{taneda1956experimental}	&-- & 1.0  & 151 & 1.07 \\ 	
			Current & 524,288 & 1.009  & 125.1 & 0.95\\
			Implicit third-order DDG \cite{cheng2017parallel} & 1,608,680 & 1.016 & 123.7 & 0.96\\
			Implicit fourth-order VFV \cite{wang2017thesis}  & 458,915 & 1.014 & --& --\\
		    Implicit third-order AMR-VFV \cite{pan2018high} &621,440   & 1.016 &--& --\\
			Fourth-order FR \cite{sun2007sd} & --& -- & 123.6 & 1.04\\	
		\end{tabular*}
		{\rule{\temptablewidth}{0.1pt}}
	\end{center}
	\vspace{-4mm} \caption{\label{re-118-sphere} Quantitative comparisons among different compact schemes  for the viscous flow past a sphere.}
\end{table}

\subsection{Taylor-Green vortex}

The implicit large eddy simulation (ILES) of a three-dimensional Taylor-Green vortex \cite{debonis2013solutions} is conducted to validate the new compact GKS for nearly incompressible viscous flow.
The initial flow field  is given by
\begin{align*}
&U_1=V_0\sin(\frac{x}{L})\cos(\frac{y}{L})\cos(\frac{z}{L}),\\
&U_2=-V_0\cos(\frac{x}{L})\sin(\frac{y}{L})\cos(\frac{z}{L}),\\
&U_3=0,\\
&p=p_0+\frac{\rho_0V_0^2}{16}(\cos(\frac{2x}{L})+\cos(\frac{2y}{L}))(\cos(\frac{2z}{L})+2),
\end{align*}
within a periodic cubic box $-\pi L\leq x, y, z\leq \pi L$.
The density distribution is given by keeping a constant temperature.
In the computation, $L=1, V_0=1, \rho_0=1$, and the Mach number takes $M_0=V_0/a_s=0.1$,
where $a_s$ is the sound speed. The characteristic convective time $t_c = L/V_0$.
The specific heat ratio $\gamma=1.4$ and the Prandtl number is $Pr=1$. Two Reynolds number $Re=280$ and $1600$ are studied here.
The linear weights of reconstruction and the smooth flux function are adopted in this case.
The equilibrium state  is obtained by the arithmetic average of the non-equilibrium states to further reduce the numerical dissipations.
The CFL is set as 0.3.

\begin{figure}[htbp]	
	\centering
	\subfigure[Re=280, t=5]{
		\label{tg-re280-qc-t5}
		\includegraphics[width=0.48\textwidth]
		{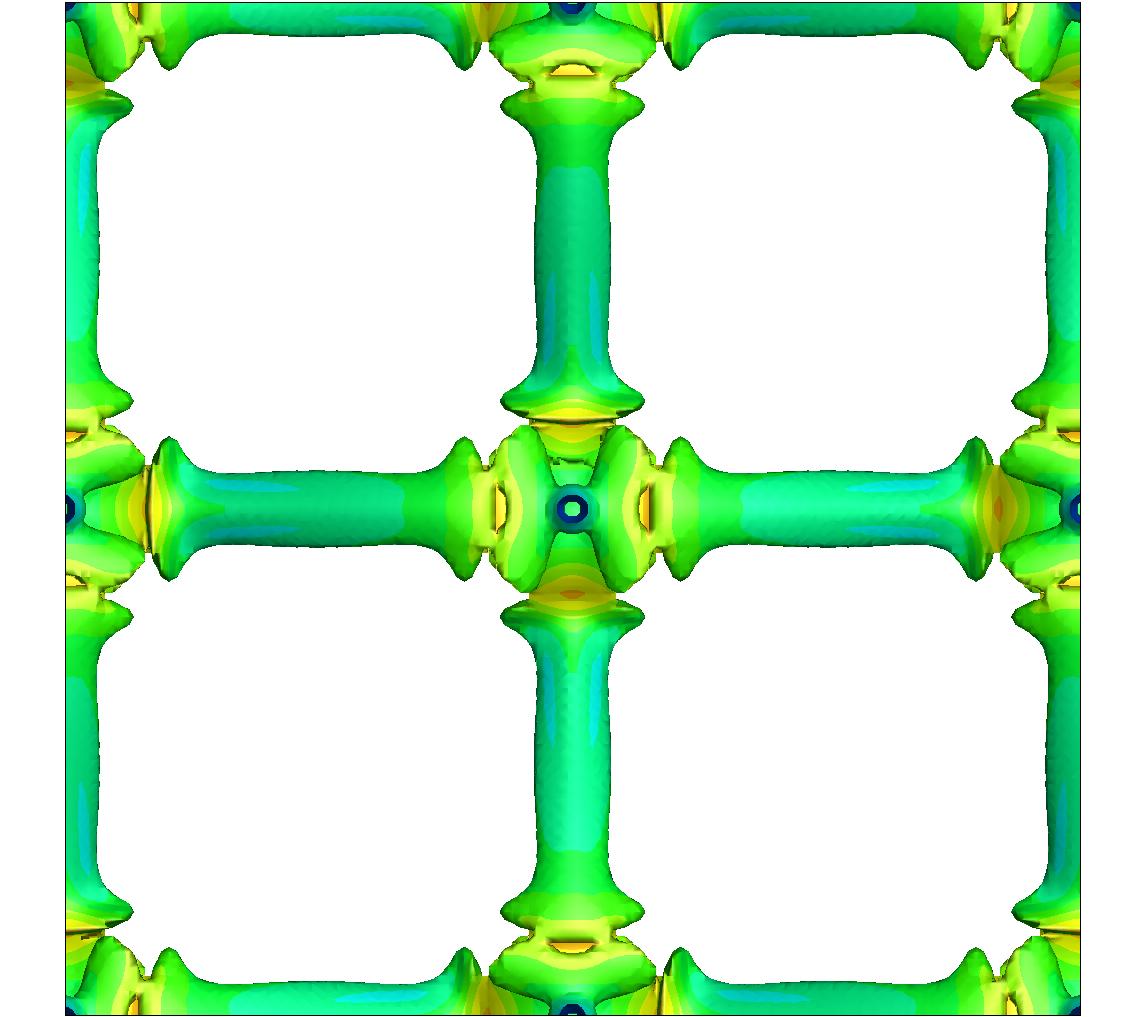}}
	\subfigure[Re=280, t=10]{
		\label{tg-re280-qc-t10}
		\includegraphics[width=0.48\textwidth]
		{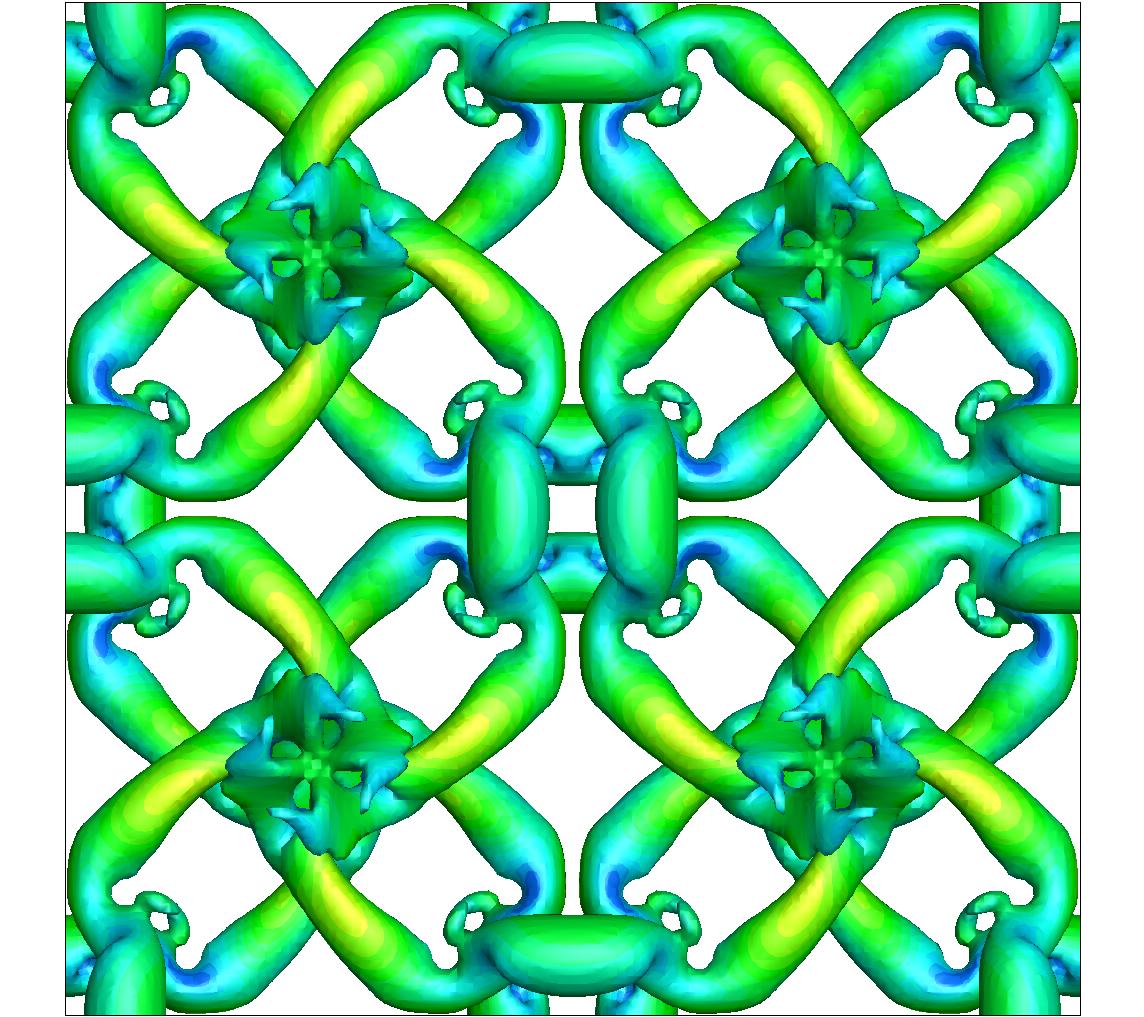}}
	\caption{Taylor-Green vortex: the iso-surfaces of Q criterion colored by Mach number at
		time t = 5, 10 for Re = 280. Cell number = $128^3$. The x-y plane is shown.}
	\label{tg-re280-contour}
\end{figure}

\begin{figure}[htbp]	
	\centering
	\subfigure[Re=1600, t=5]{
		\label{tg-re1600-qc-t5}
		\includegraphics[width=0.48\textwidth]
		{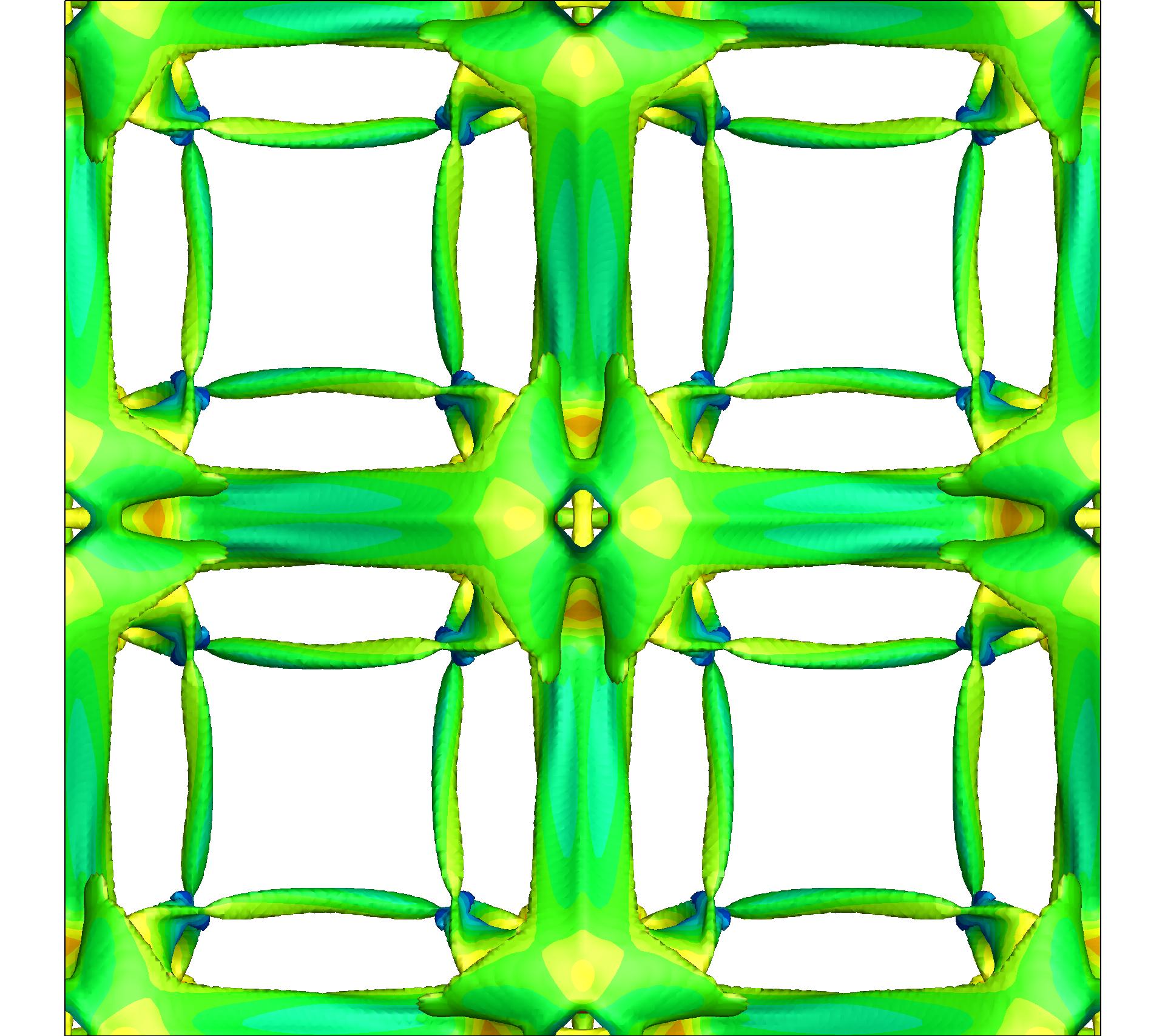}}
	\subfigure[Re=1600, t=10]{
		\label{tg-re1600-qc-t10}
		\includegraphics[width=0.48\textwidth]
		{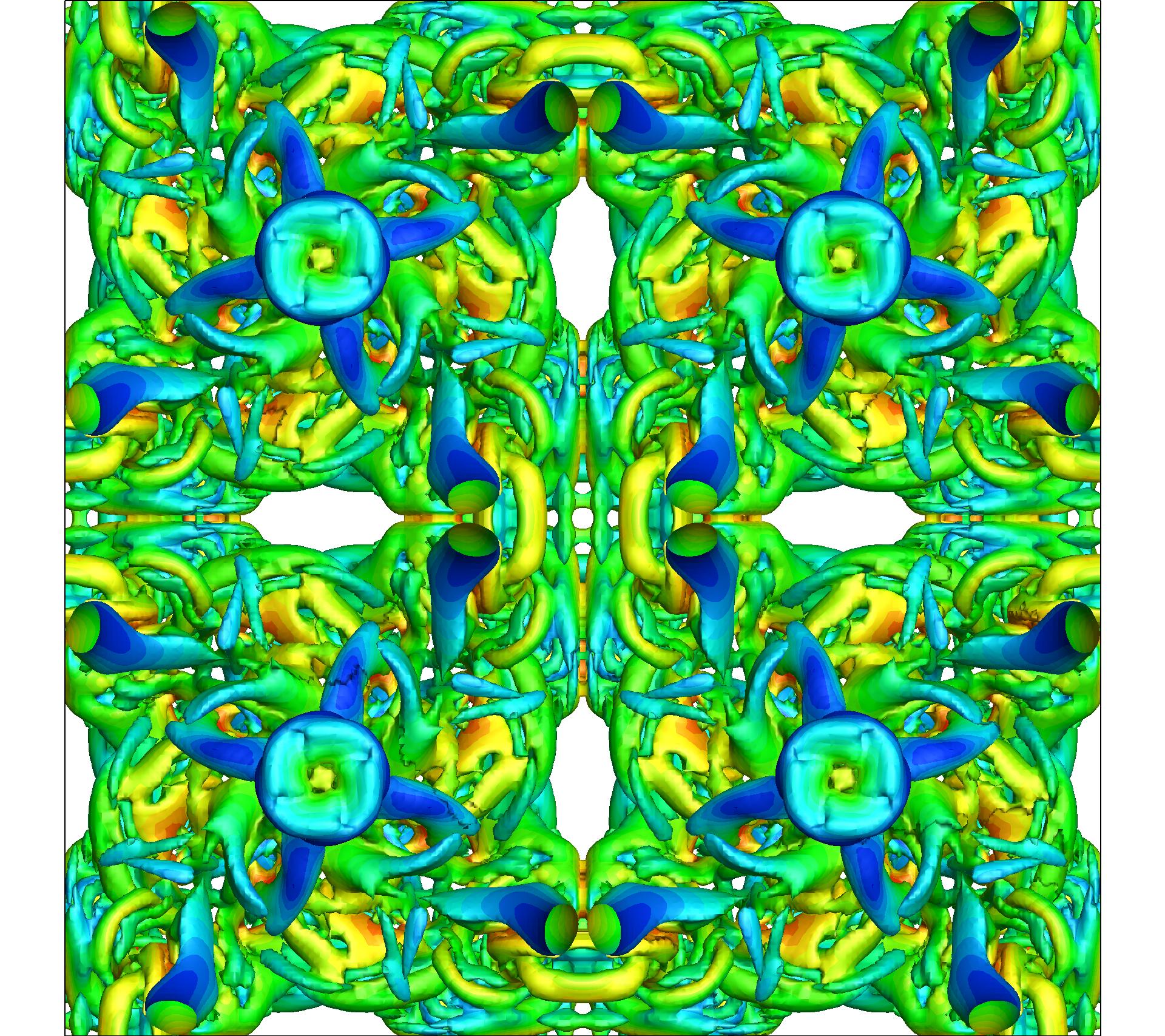}}
	\caption{Taylor-Green vortex: the iso-surfaces of Q criterion colored by Mach number at
		time t = 5, 10 for Re = 1600. Cell number = $196^3$. The x-y plane is shown.}
	\label{tg-re1600-contour}
\end{figure}

Two quantities are investigated in the current study as the flow evolves in time.
The first one is the volume-averaged kinetic energy
\begin{align*}
E_k=\frac{1}{\rho_0\Omega}\int_\Omega\frac{1}{2}\rho\textbf{U}\cdot\textbf{U}\text{d}\Omega,
\end{align*}
where $\Omega$ is the volume of the computational domain.

\begin{align*}
\varepsilon_k=-\frac{\text{d}E_k}{\text{d}t} = - \frac{\text{d}E_k}{\text{d}t} ,
\end{align*}
By using the data $E_k(t_i), t_i \in [0, t_{stop}]$,
the dissipation rate of the kinetic energy is given by a second-order interpolation
\begin{align*}
\varepsilon_k (t_i) = - \frac{E_k (t_{i+1})-E_k (t_{i-1})}{ t_{i+1}-t_{i-1}} .
\end{align*}
The numerical results are compared with the reference DNS data in \cite{debonis2013solutions} and the non-compact fifth-order GKS \cite{ji2020performance}.
The iso-surfaces of $Q$ criterion colored by Mach number at $t=5$ and $10$ for both Reynolds number are shown in Fig.\ref{tg-re280-contour} and Fig.\ref{tg-re1600-contour}.
With the time increment, the vortex structures become denser and smaller. The case with higher Reynolds number has richer structures, which requires high resolution for a numerical scheme.

\begin{figure}[htbp]	
	\centering
	\subfigure[$E_k$]{
		\label{tg-re280-ke}
		\includegraphics[width=0.32\textwidth]
		{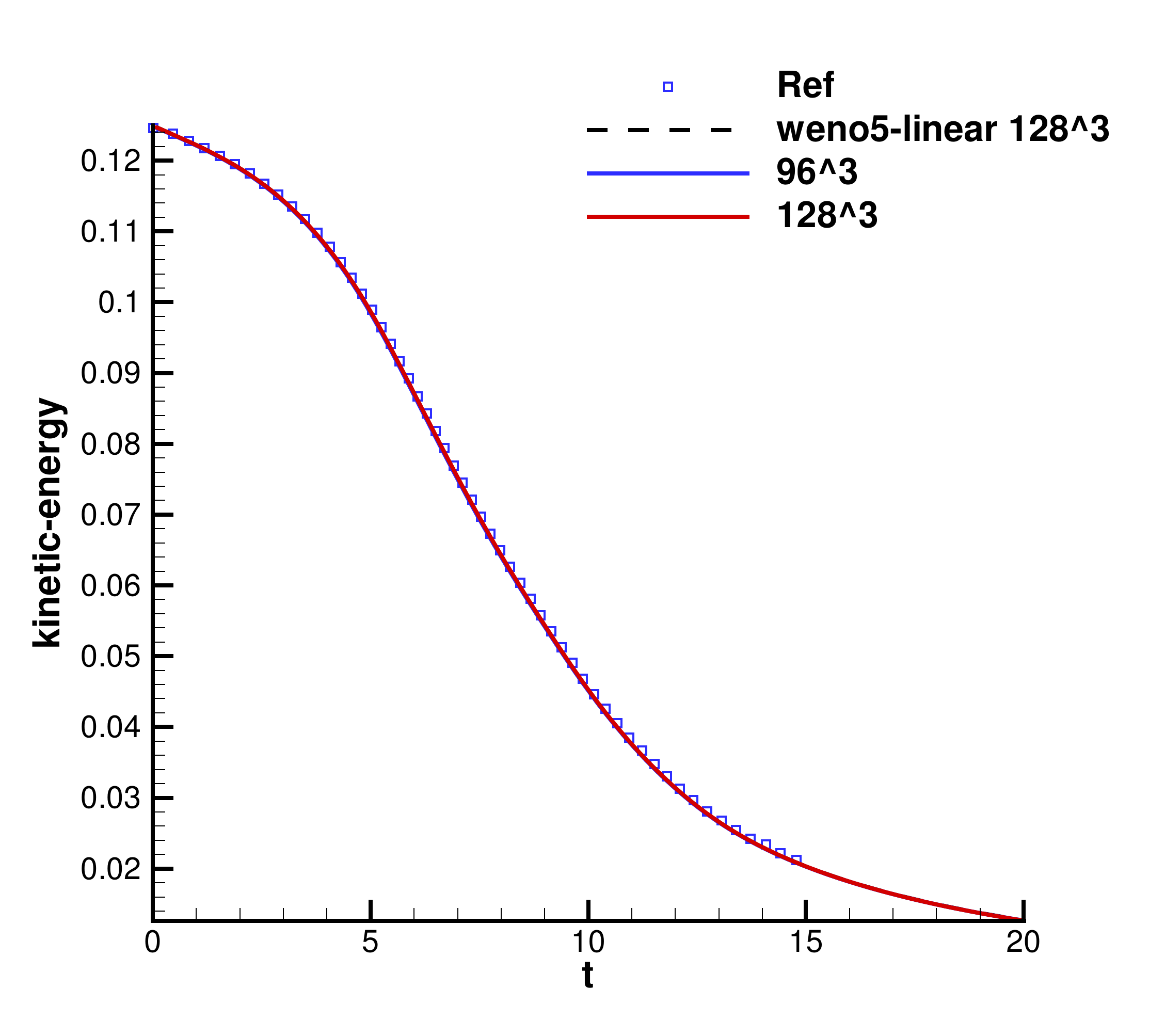}}
	\subfigure[$\varepsilon_k$]{
		\label{tg-re280-dk}
		\includegraphics[width=0.32\textwidth]
		{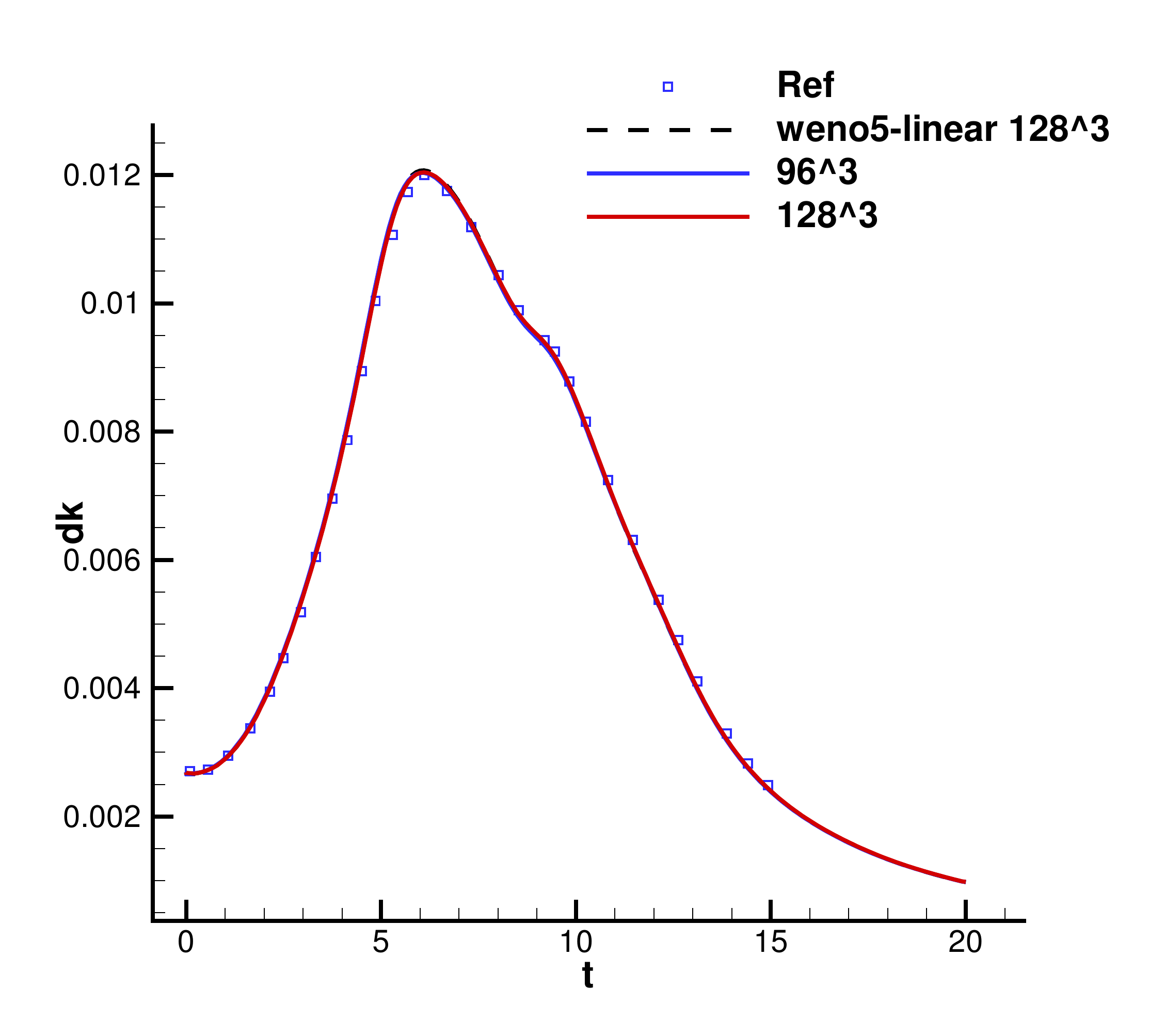}}
		\subfigure[Local enrlargement of $\varepsilon_k$]{
		\label{tg-re280-dk-local}
		\includegraphics[width=0.32\textwidth]
		{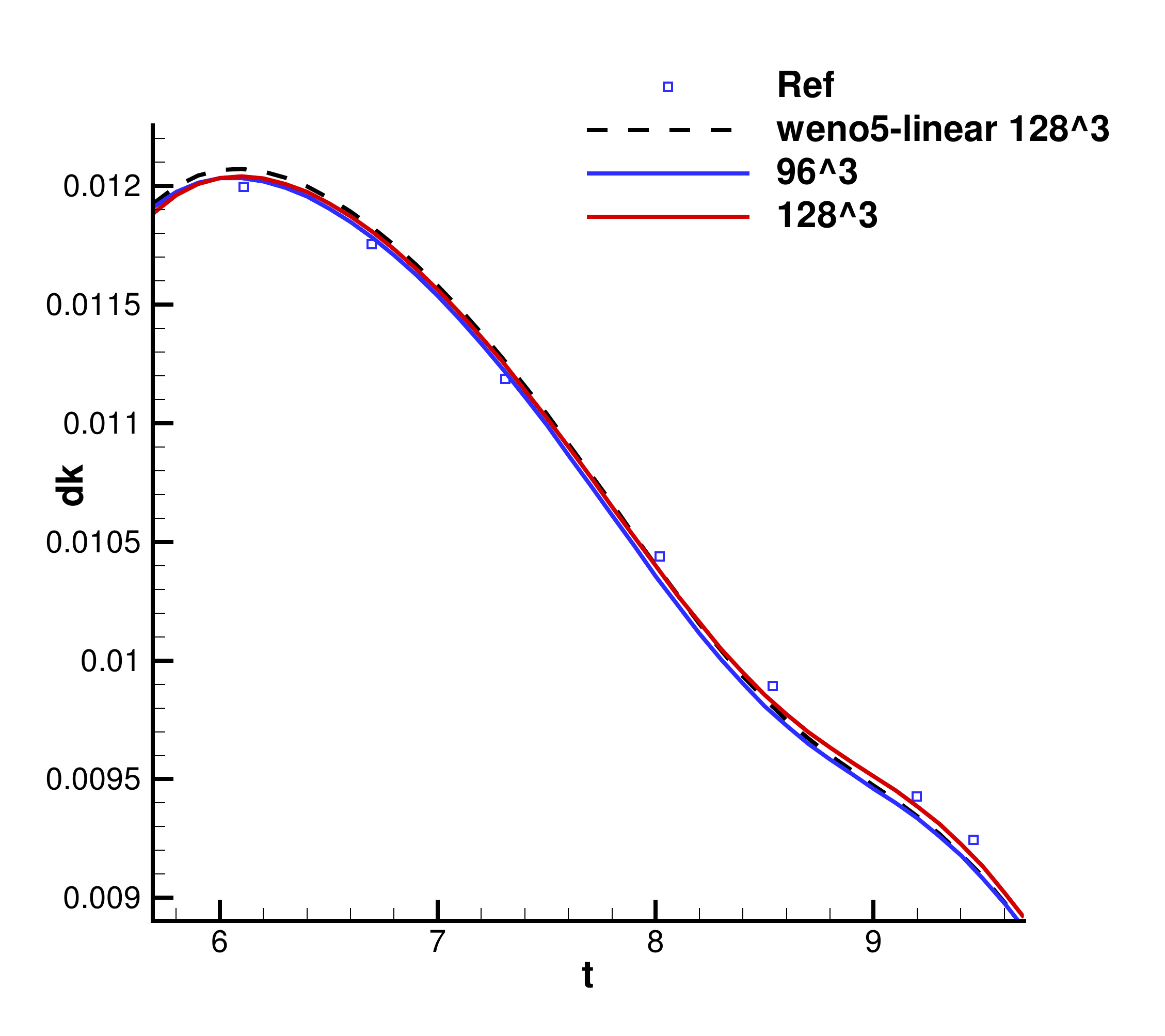}}
	\caption{Taylor-Green vortex: Re=280. The time history of kinetic energy and its dissipation rate. }
	\label{tg-re280}
\end{figure}

\begin{figure}[htbp]	
	\centering
	\subfigure[$E_k$]{
		\label{tg-re1600-ke}
		\includegraphics[width=0.4\textwidth]
		{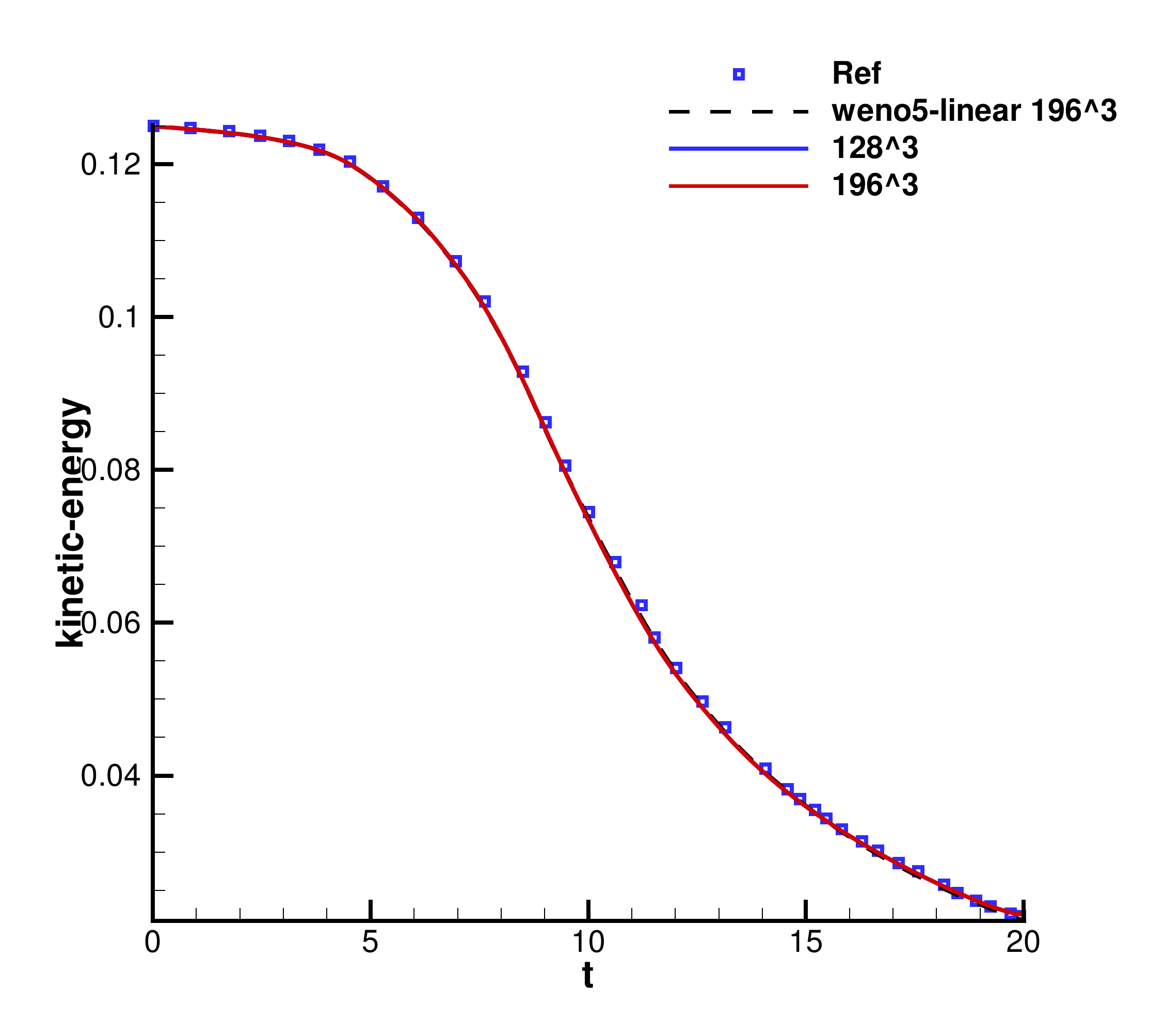}}
	\subfigure[$\varepsilon_k$]{
		\label{tg-re1600-dk}
		\includegraphics[width=0.4\textwidth]
		{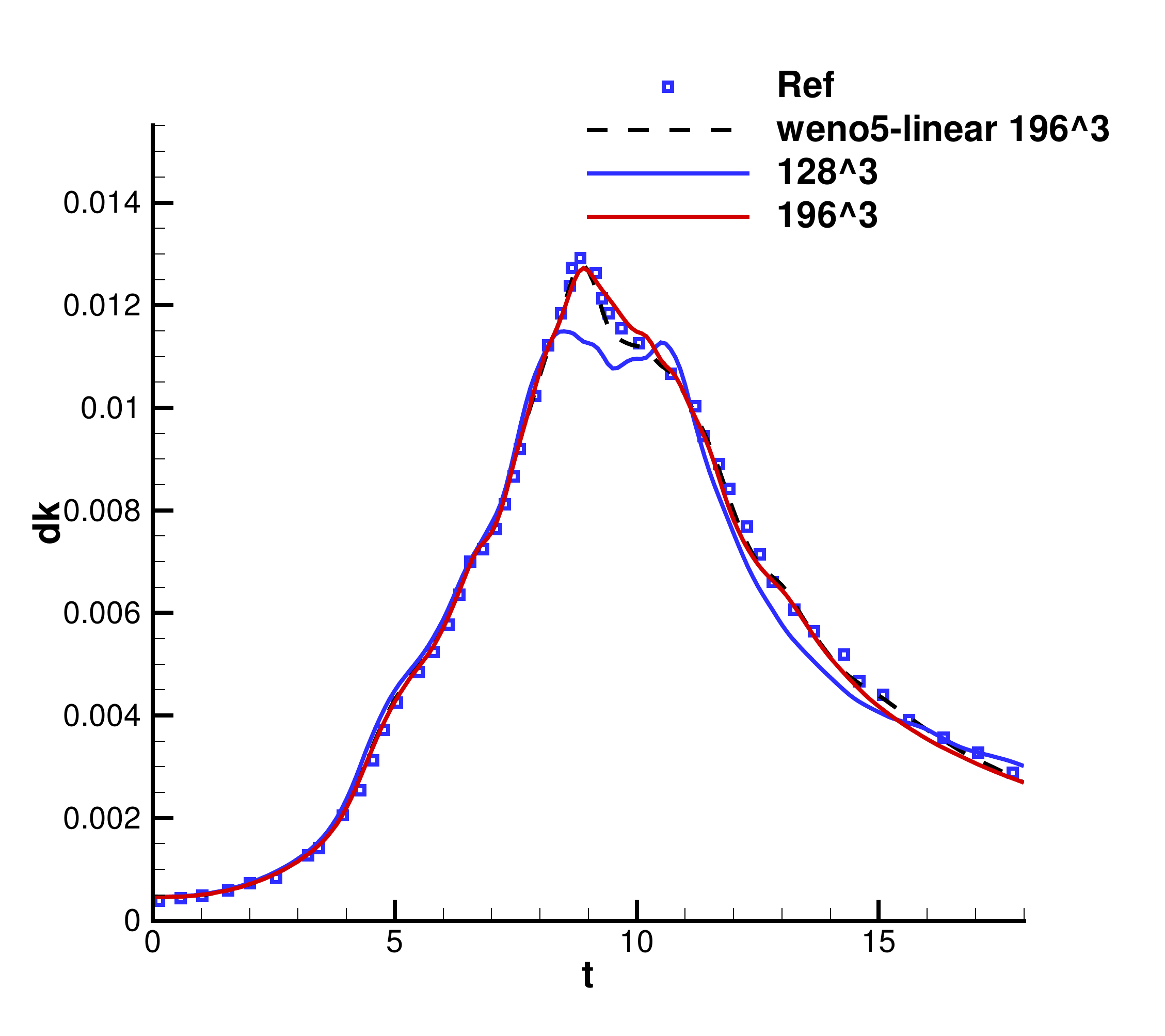}}
	\caption{Taylor-Green vortex: Re=1600. The time history of kinetic energy and its dissipation rate. }
	\label{tg-re1600}
\end{figure}

\begin{figure}[htbp]	
	\centering
	\subfigure[$E_k$]{
		\label{tg-re1600-dk-1}
		\includegraphics[width=0.32\textwidth]
		{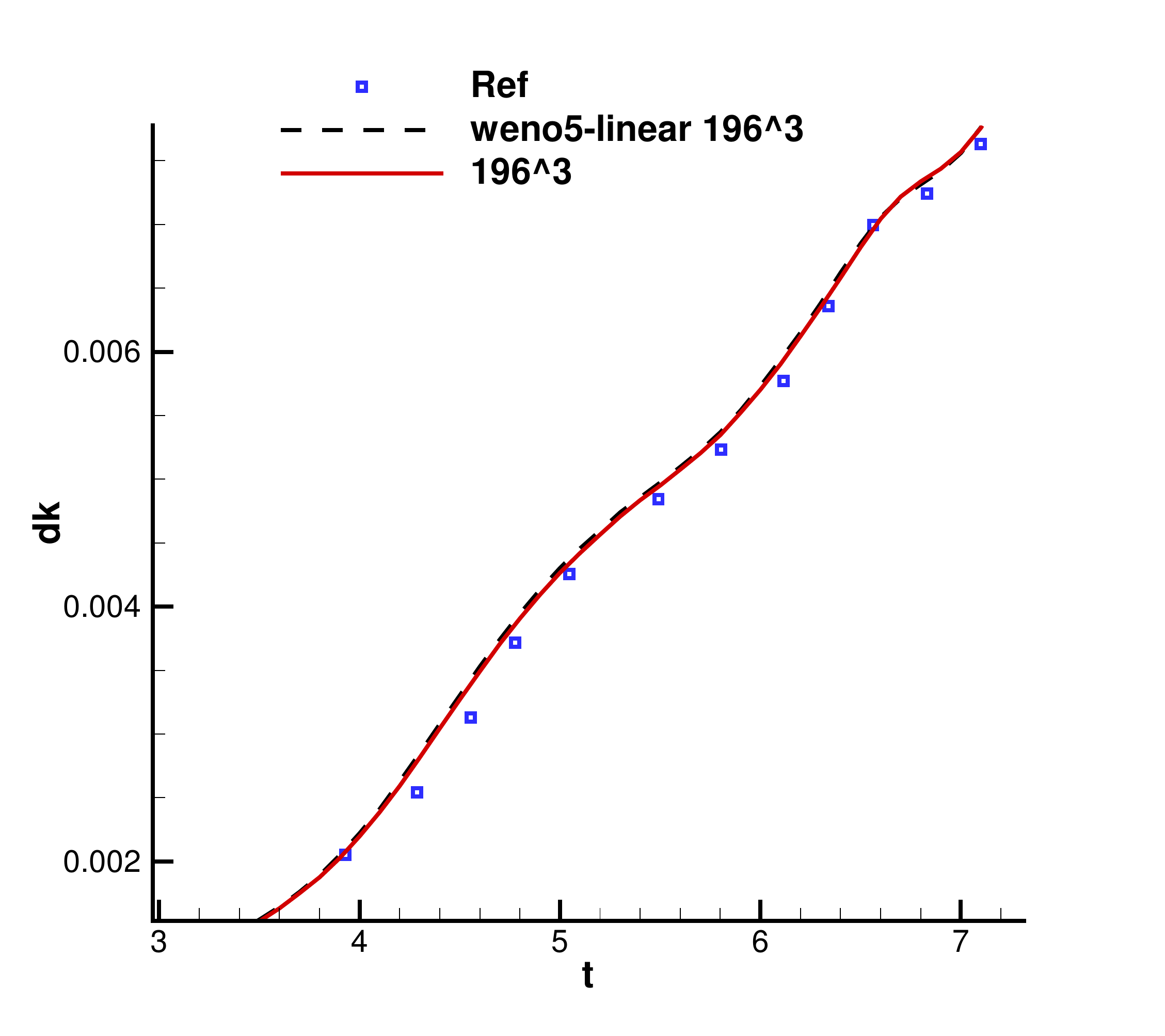}}
	\subfigure[$\varepsilon_k$]{
		\label{tg-re1600-dk-2}
		\includegraphics[width=0.32\textwidth]
		{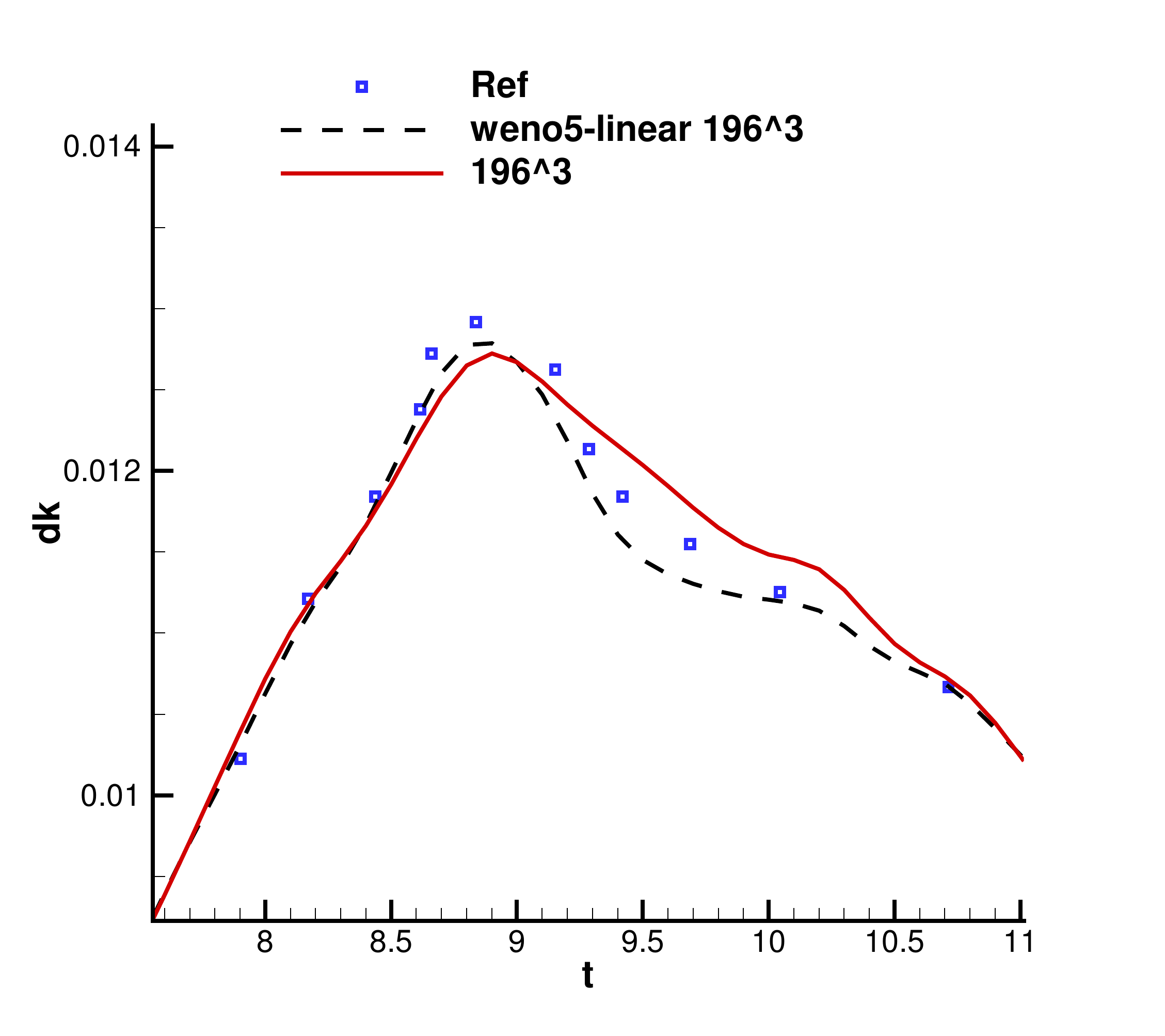}}
		\subfigure[$\varepsilon_k$]{
		\label{tg-re1600-dk-3}
		\includegraphics[width=0.32\textwidth]
		{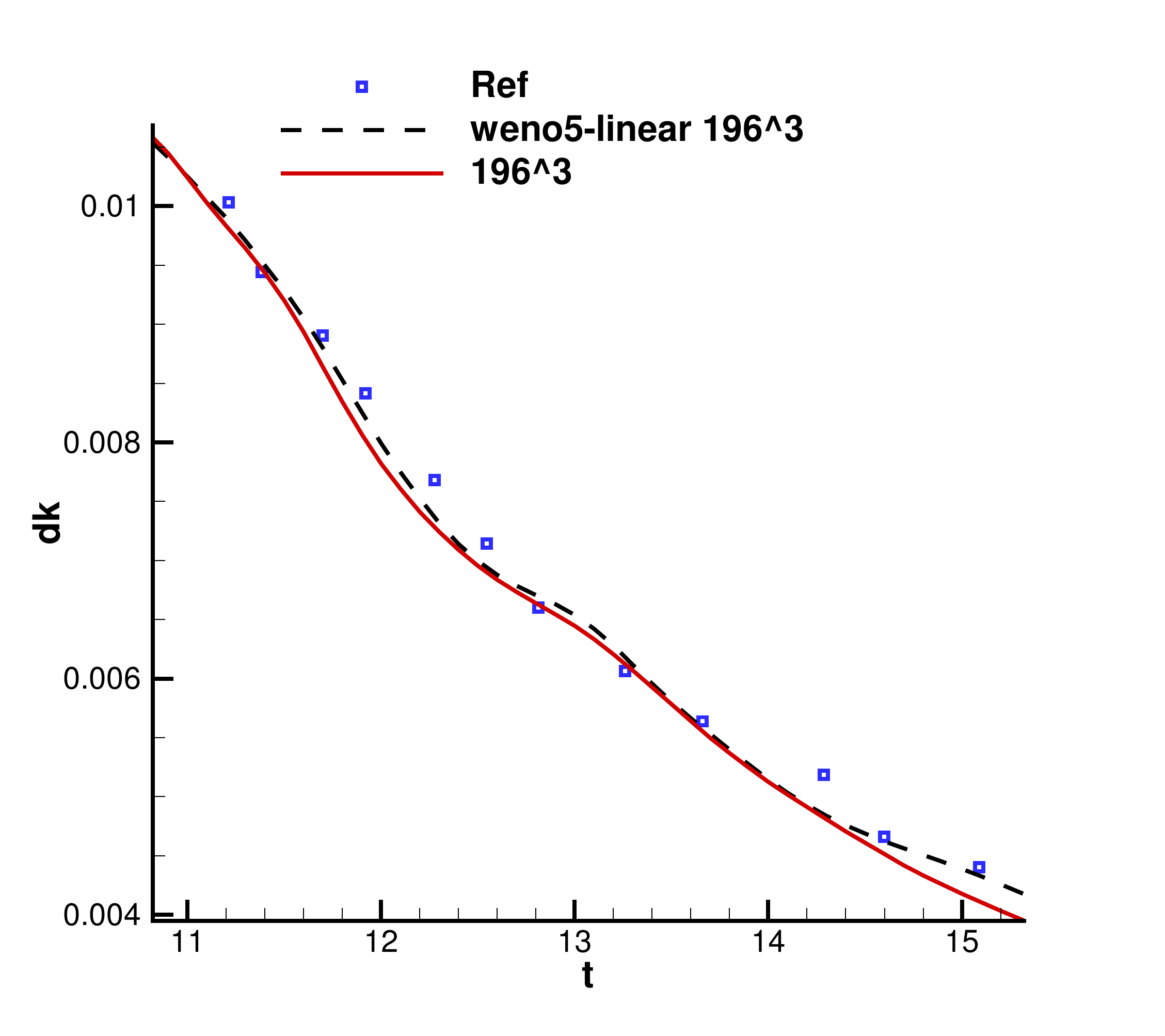}}
	\caption{Taylor-Green vortex: Re=1600. The local enlargement of the time history of kinetic energy dissipation rate. }
	\label{tg-re1600-local}
\end{figure}

For $Re=280$, the quantitative result agrees nicely with the reference data under a coarse mesh $96^3$, as shown in Fig.\ref{tg-re280}.
The new third-order compact scheme even shows better ability against the traditional fifth-order GKS with the same mesh, seen as Fig.\ref{tg-re280-dk-local}.
 The time history of the normalized volume-averaged kinetic energy and dissipation rate under $Re=1600$ with $128^3$ and $196^3$ mesh points are presented in Fig.\ref{tg-re1600}. A detailed zoom-in plot is given in Fig.\ref{tg-re1600-local}.
The compact scheme is capable to capture the complicated vortex structure as the non-compact one under the same mesh.

\subsection{One dimensional Riemann problem}
\begin{figure}[htp]	
	\centering
		\subfigure[]{
		\label{sod-1d-a}
  	    \includegraphics[width=0.45\textwidth]{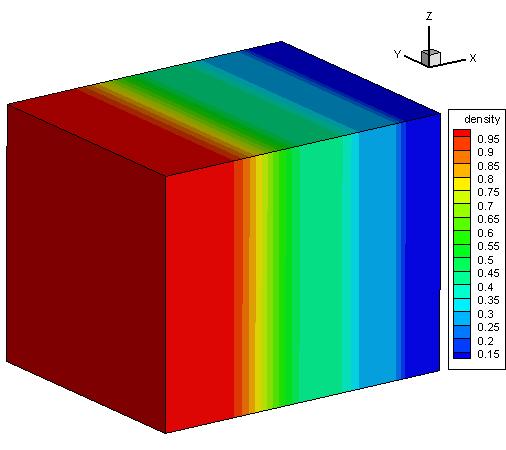}
        }
		\subfigure[]{
		\label{sod-1d-b}
	    \includegraphics[width=0.45\textwidth]{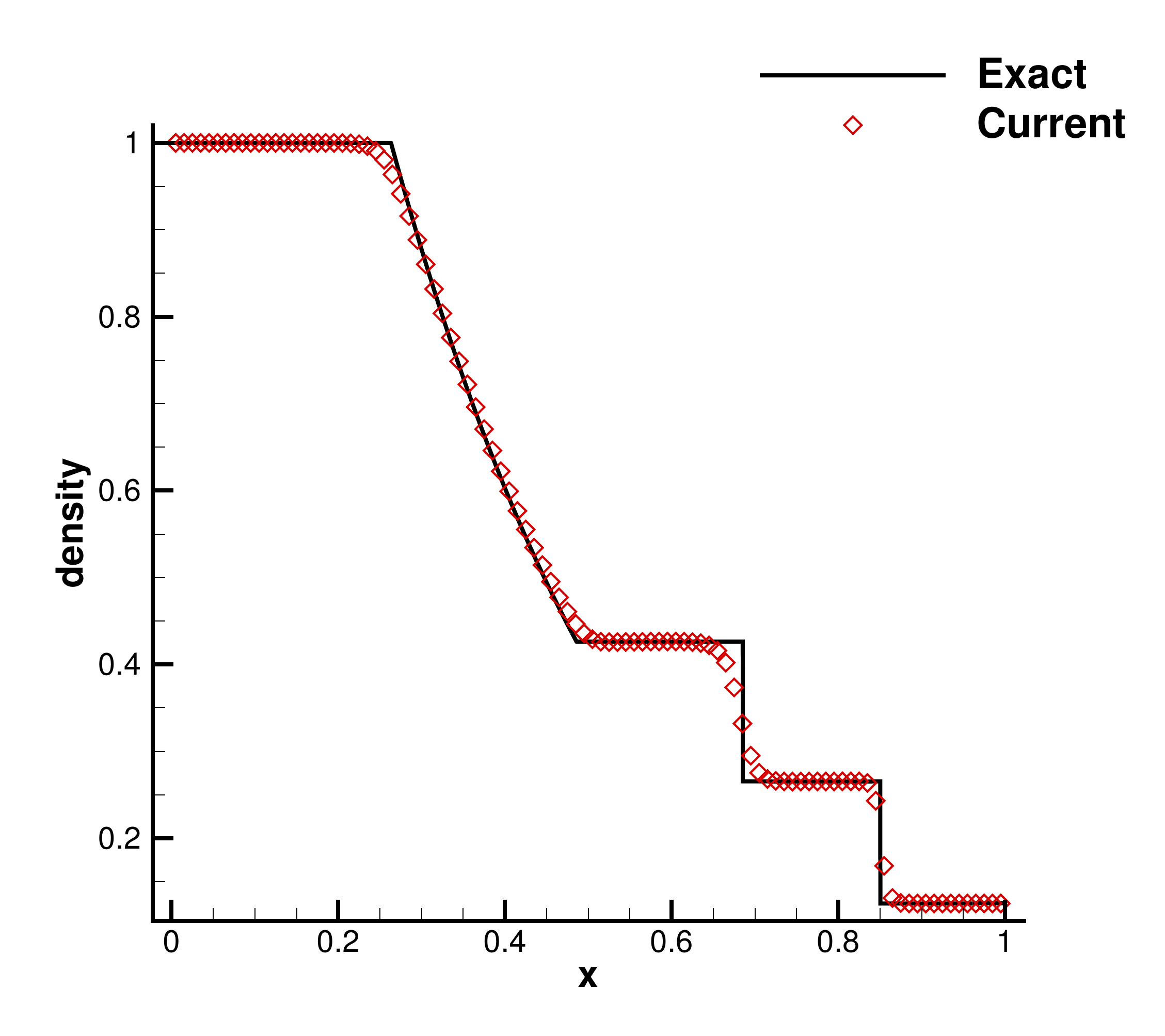}
        }
	\caption{Sod problem: 3-D view of density distribution and its extraction along the center x-horizontal line. The mesh size along the x-direction is $1/100$. }
	\label{sod-1d}
\end{figure}

The one-dimensional Sod test case is performed in three-dimensional simulation.
The initial condition is given by
\begin{equation*}
(\rho,U,V,W,p)=\left\{\begin{aligned}
&(1, 0,0,0, 1), 0<x<0.5,\\
&(0.125,0,0,0,0.1),  0.5 \leq x<1,
\end{aligned} \right.
\end{equation*}
where $100 \times 5 \times 5$ uniform cells are used in the computational domain of $[0,1]\times[0,1] \times [0,1]$.
The solutions are presented at $t=0.2$.
Non-reflection boundary condition is adopted at the left and right boundaries of the computational domain,
and periodic boundary condition is adopted at the rest of the boundaries.
The 3-D plot of density distribution in Fig.\ref{sod-1d-a} shows the uniformity in the flow distributions along $x$ direction.
The density distribution at the center horizontal line is also extracted, as shown in Fig.\ref{sod-1d-b}.
For this mild case, the numerical result agrees well with the exact solution.
However, the contact discontinuity is not as sharp as the traditional fifth-order WENO-GKS. It is consistent with Remark \ref{rmk-reconstruction}, which shows the scheme will recover to a third-order non-compact GKS under 1-D smooth case.

\subsection{Three-dimensional explosion test problem}

\begin{figure}[htp]	
	\centering
	\includegraphics[width=0.32\textwidth]{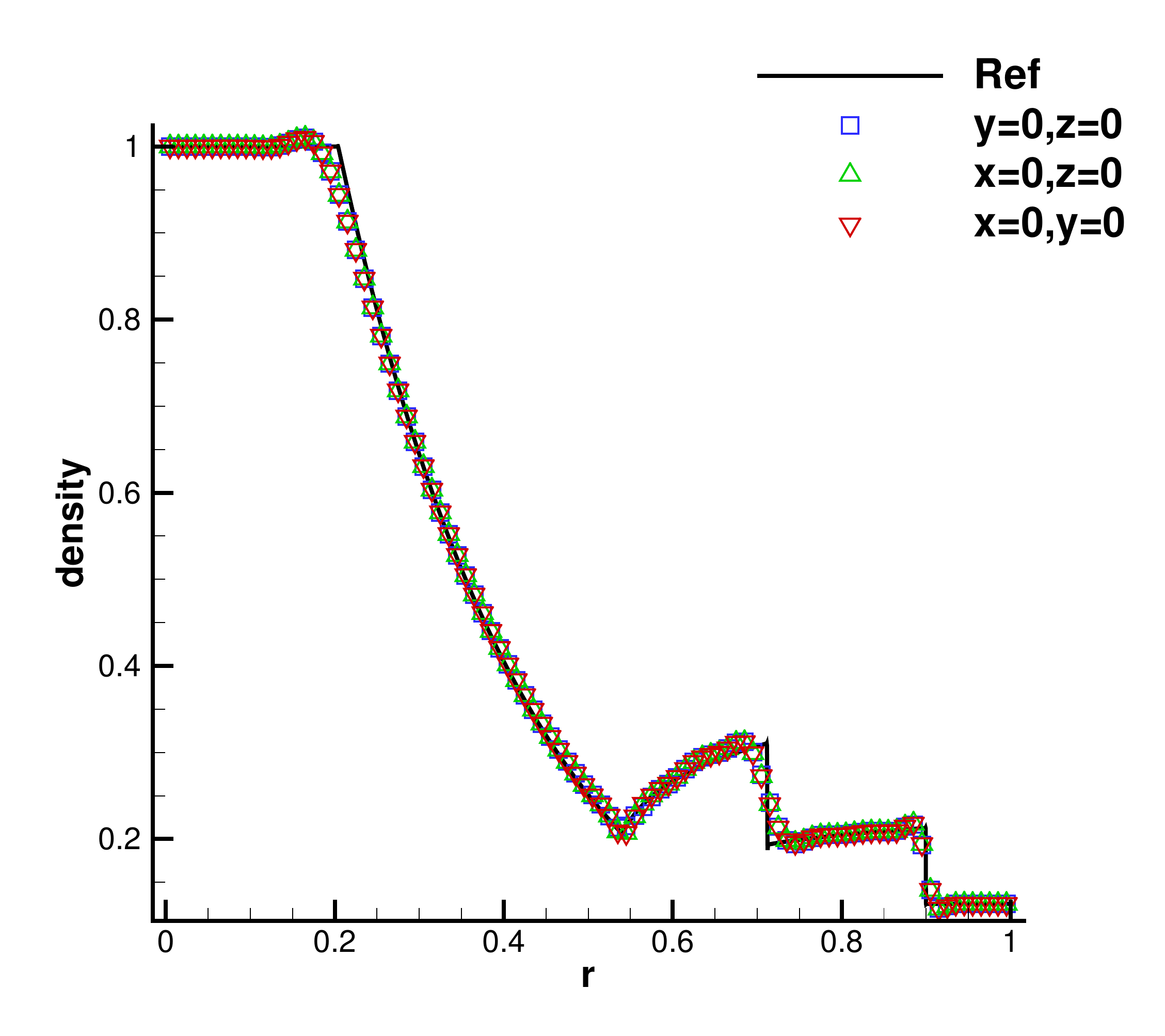}
	\includegraphics[width=0.32\textwidth]{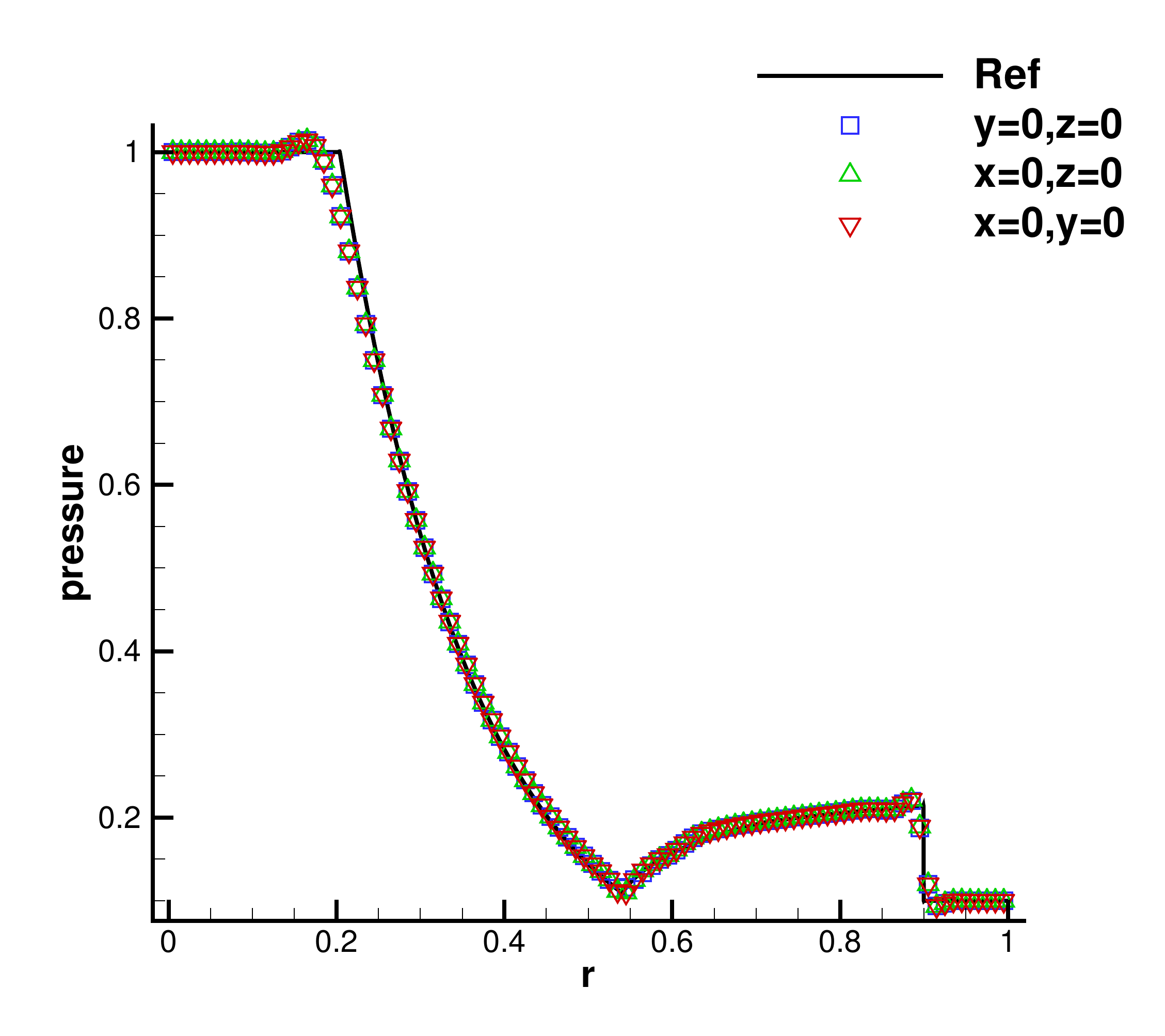}
	\includegraphics[width=0.32\textwidth]{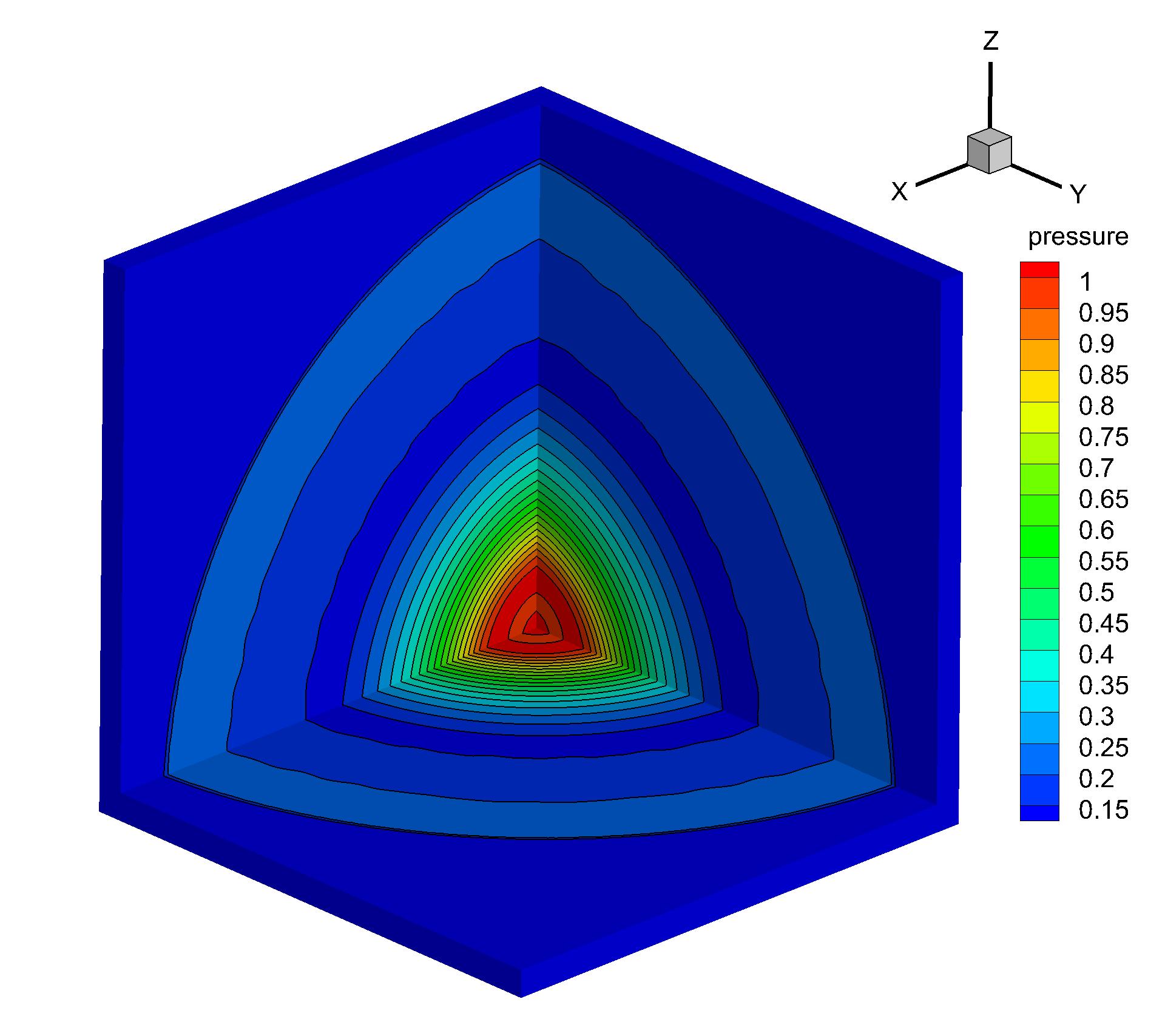}		
	\caption{Three-dimensional explosion problem with $100^3$ cells. Left and middle: the density and pressure profiles extracted along the $y=0,z=0$, $x=0,z=0$, and $x=0,y=0$ lines. Right: 3-D view of pressure distribution.}
	\label{sod-3d}
\end{figure}

As an extension of the Sod problem, the spherical explosion test problem is considered.
The initial conditions are given by
\begin{equation*}
(\rho,U,V,W,p)=\left\{\begin{aligned}
&(1, 0,0,0, 1), 0<r<0.5,\\
&(0.125,0,0,0,0.1),  0.5 \leq r<1,
\end{aligned} \right.
\end{equation*}
where $r=\sqrt{(x^2+y^2+z^2)}$.
The solution contains a spherical shock wave and a contact surface traveling away from the
center and a spherical rarefaction wave moving towards the origin $(0,0,0)$.
To save the computational cost, the computational domain is $(x,y,z) \in [0,1] \times [0,1] \times [0,1] $ and the uniform grid with $dx=dy=dz=1/100$ is used.
The symmetric boundary conditions are imposed on the planes $x=0$, $y=0$, and $z=0$,
while the outflow boundary conditions are imposed on the planes $x=1$, $y=1$, $z=1$.
The density and pressure profiles along different radial directions at $t = 0.25$ are given in Fig.\ref{sod-3d}.
The compact scheme resolves the wave profiles crisply.
Slightly overshoot can be observed at the front of the rarefaction wave.

\subsection{High-speed inviscid flow passing through a sphere}

\begin{figure}[htp]	
	\centering
	\includegraphics[width=0.32\textwidth]
	{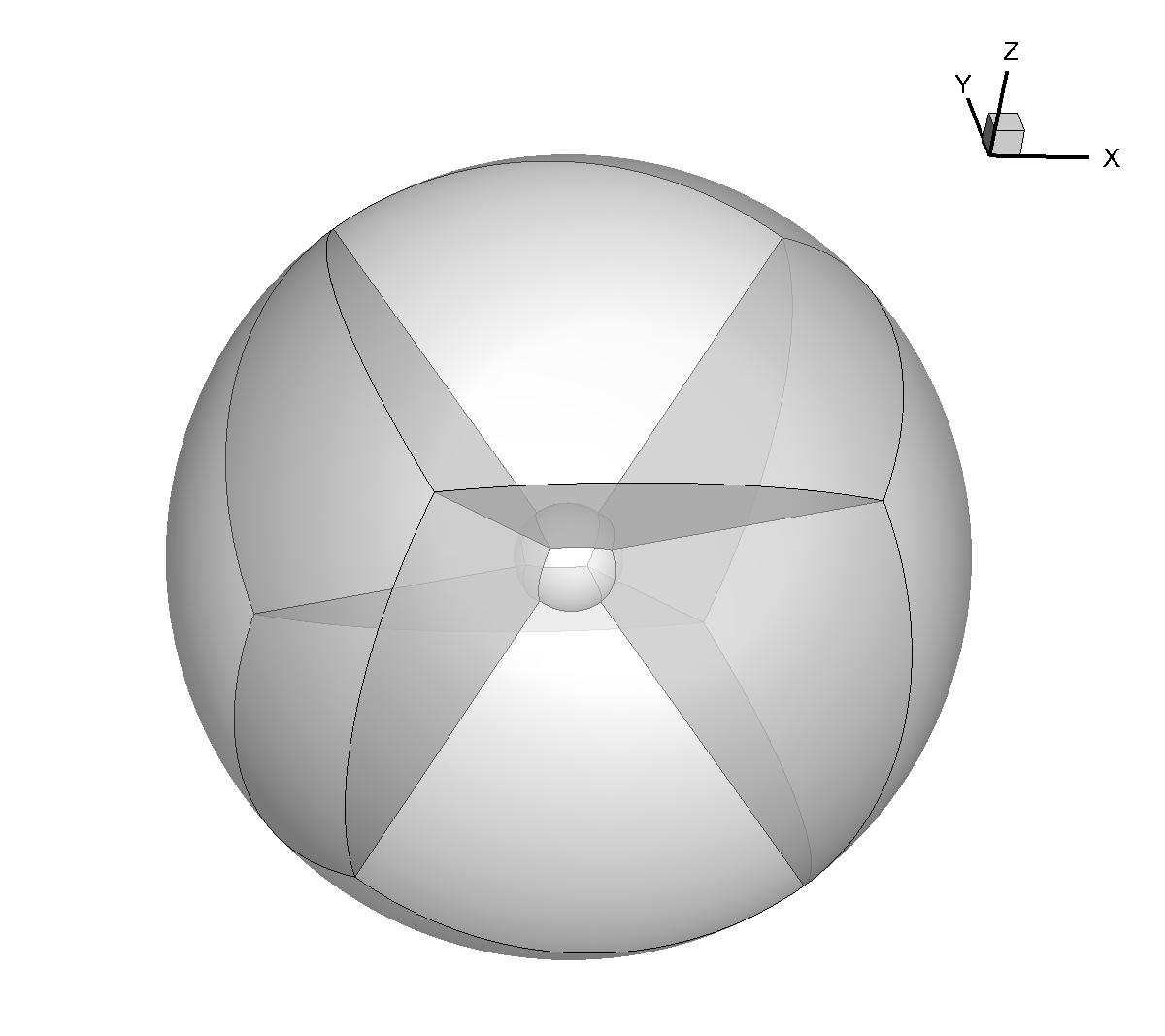}
	\includegraphics[width=0.32\textwidth]
	{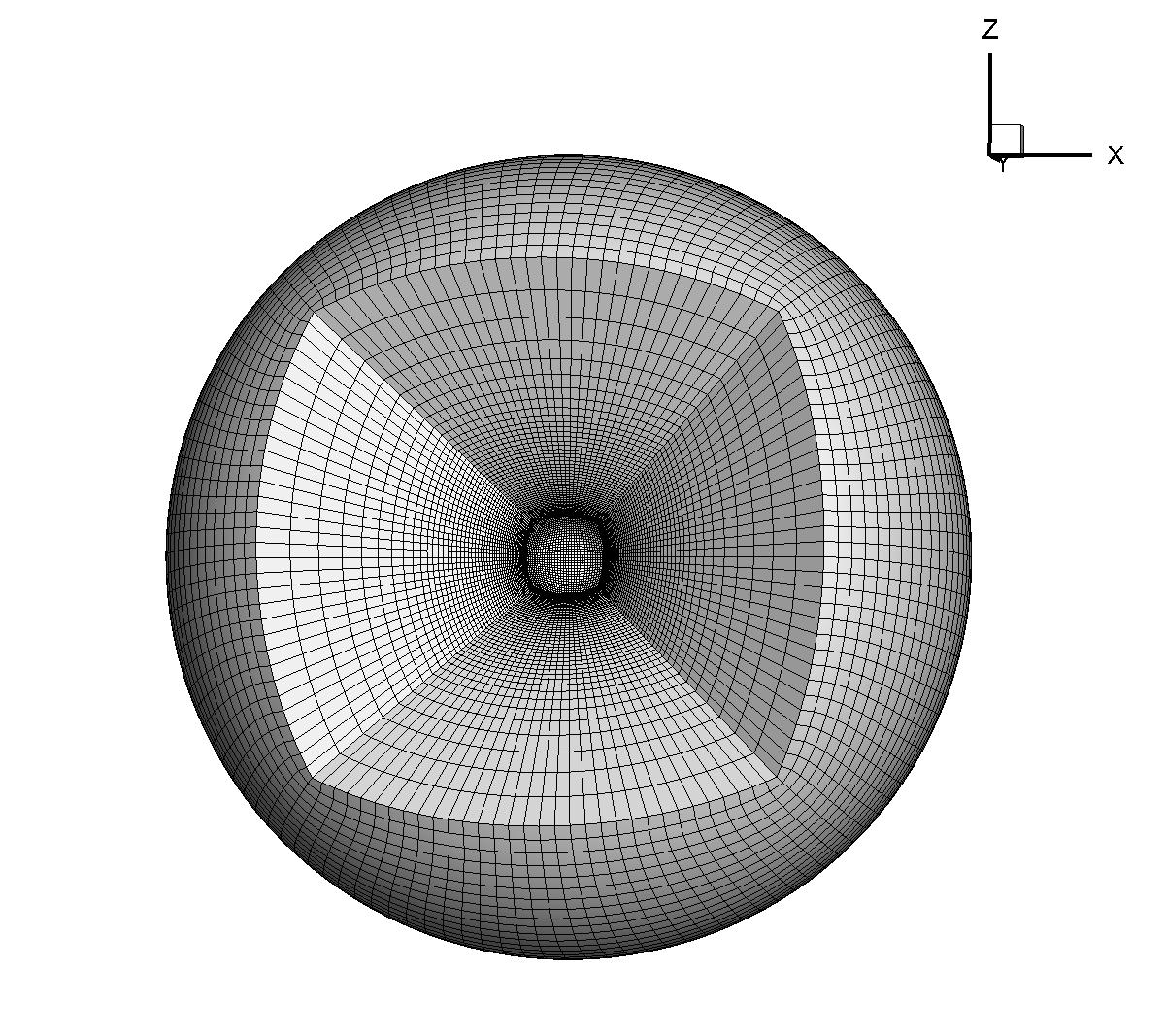}
	\includegraphics[width=0.32\textwidth]
	{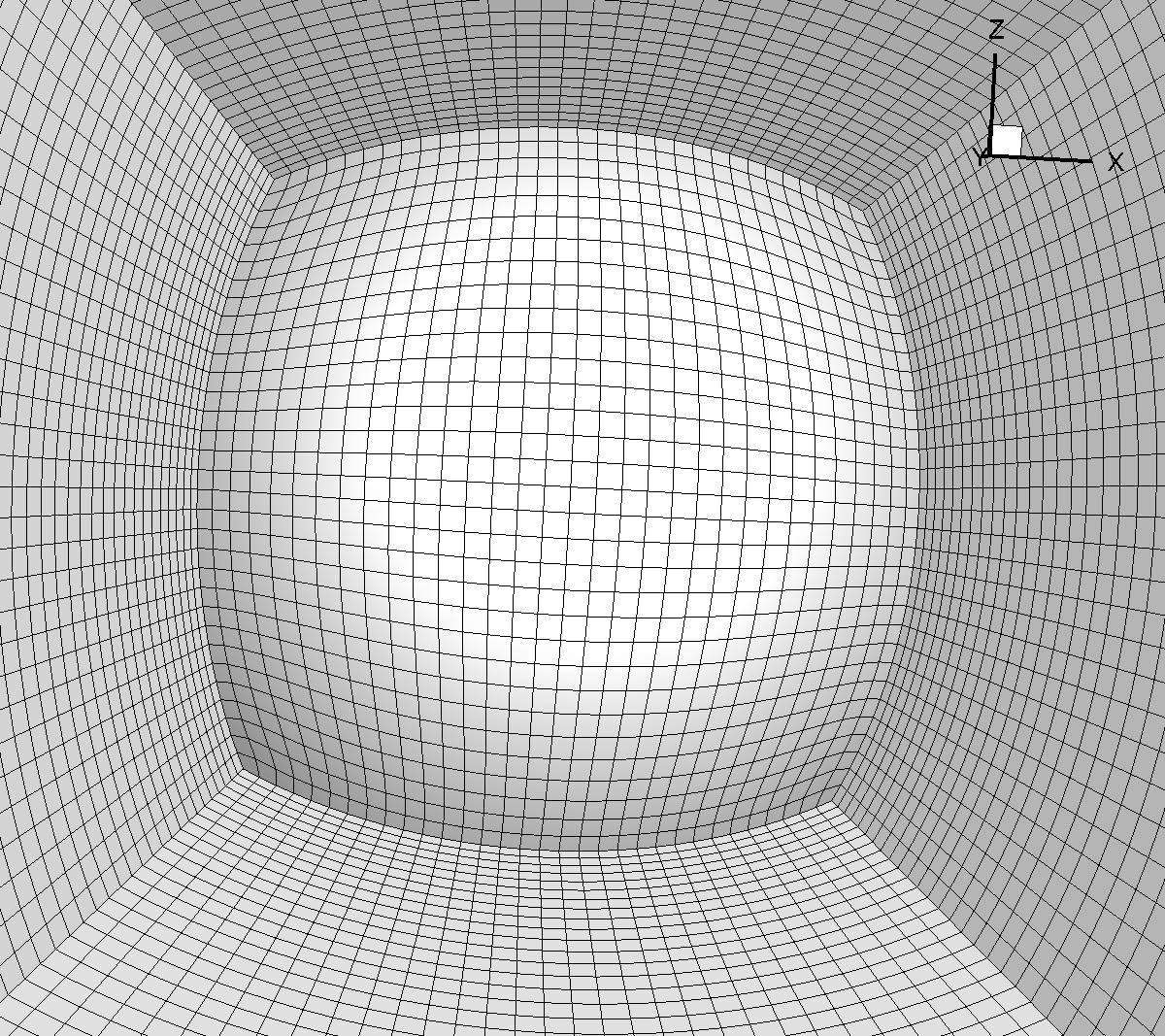}
	\caption{The mesh for inviscid flow passing through a sphere with $32\times 32 \times 50 \times 6 $ cells.}
	\label{inviscid-sphere-mesh}
\end{figure}

\begin{figure}[htp]	
	\centering
	\includegraphics[height=0.32\textwidth]
	{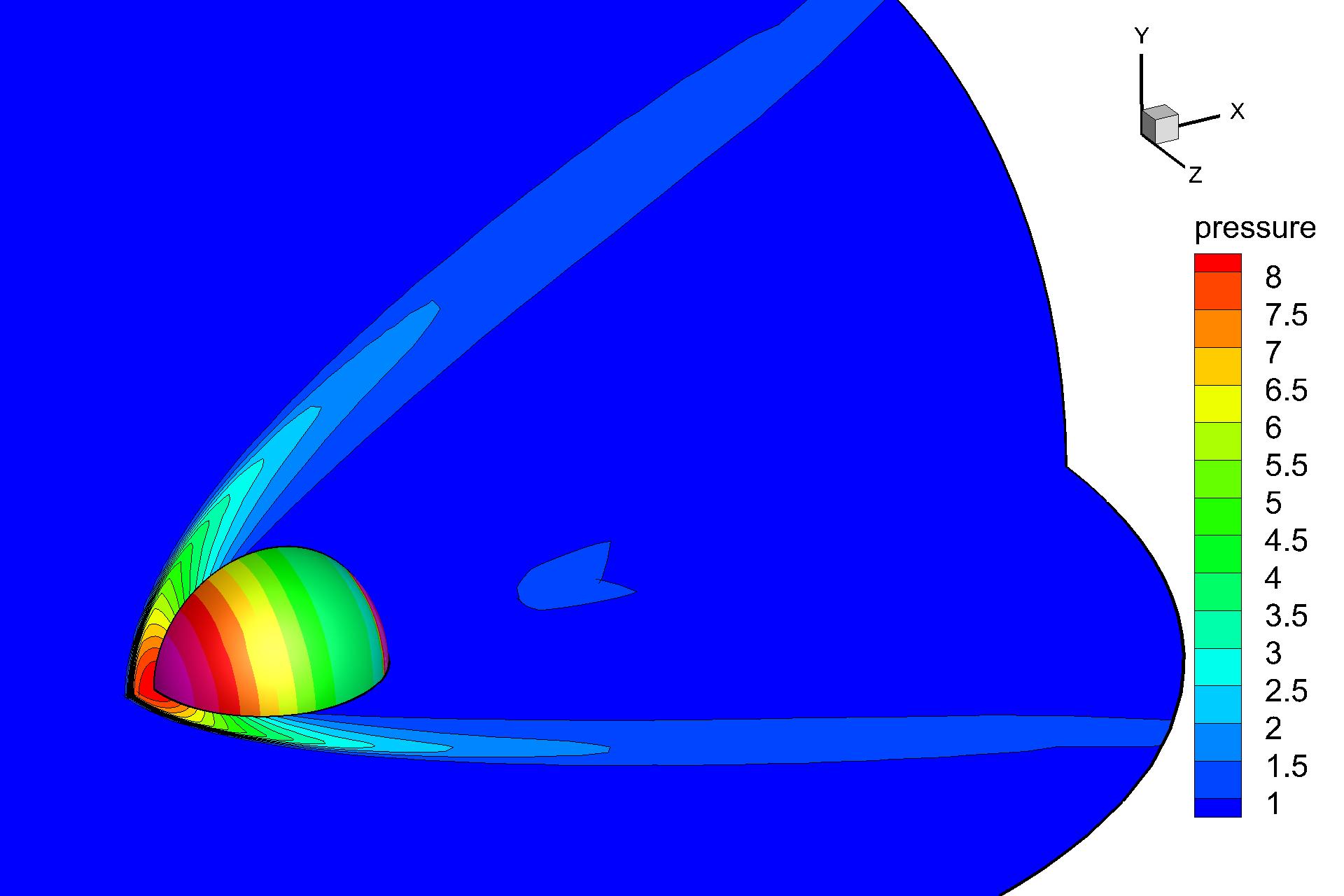}	
	\includegraphics[height=0.32\textwidth]
	{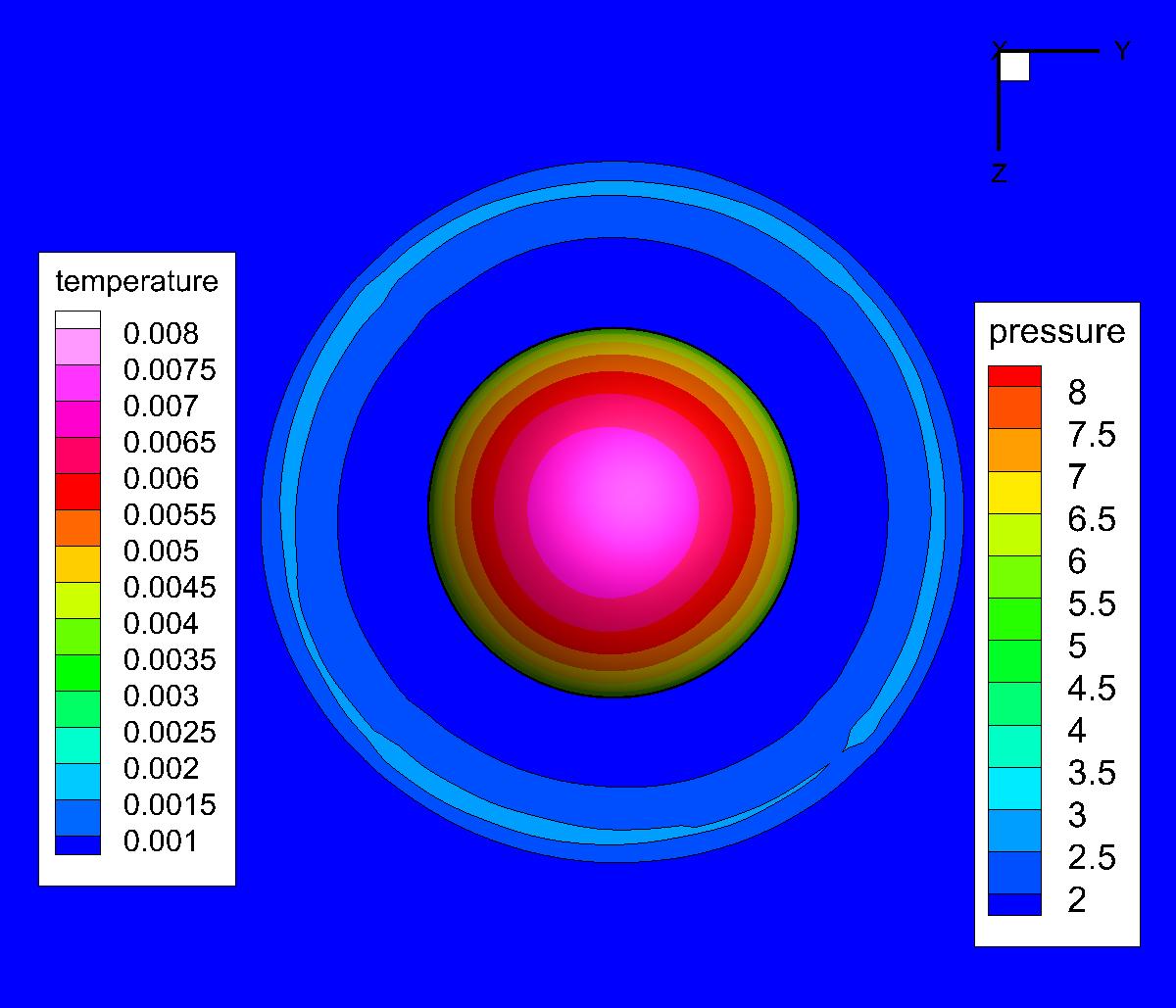}	
	\includegraphics[height=0.32\textwidth]
	{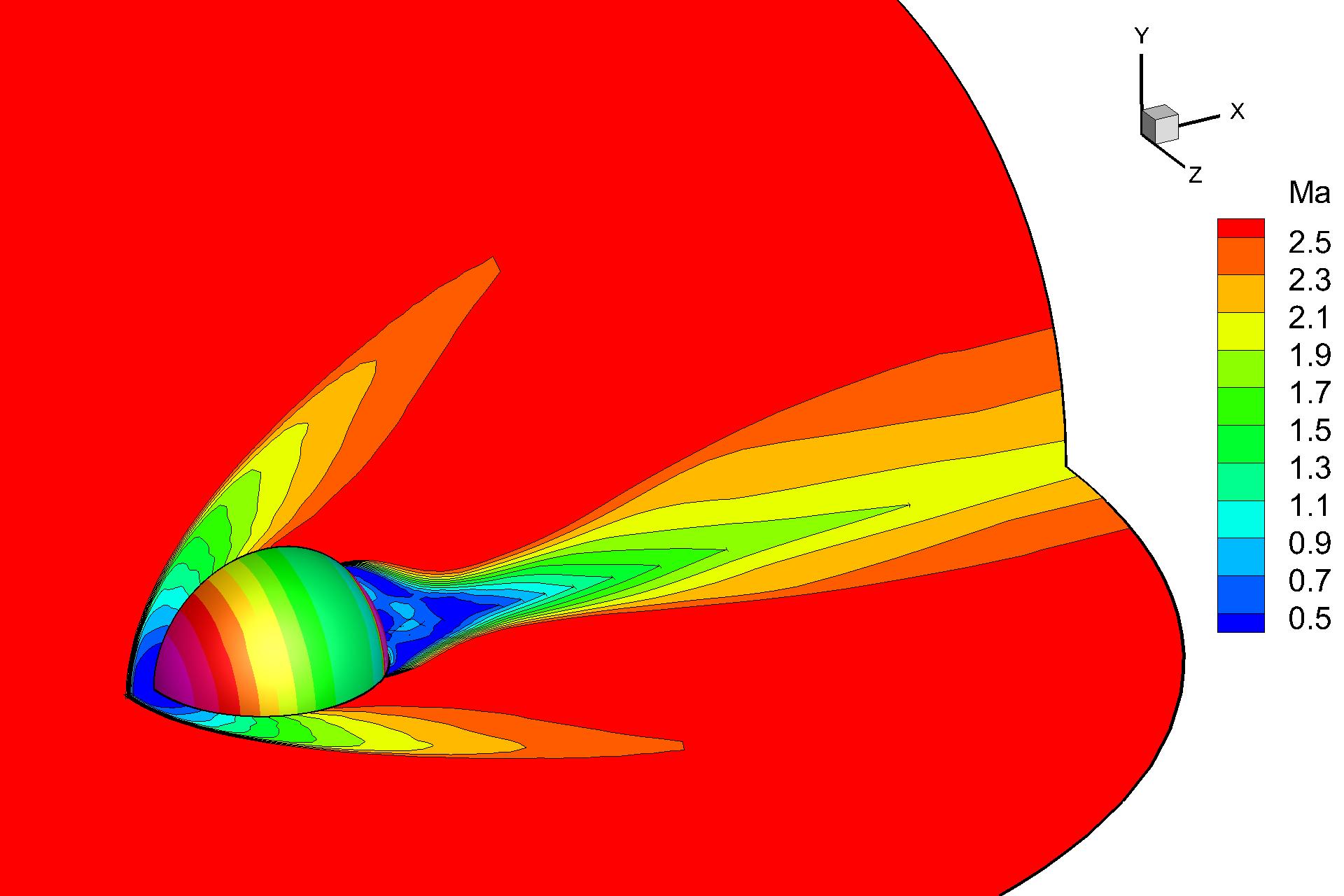}
	\includegraphics[height=0.32\textwidth]
	{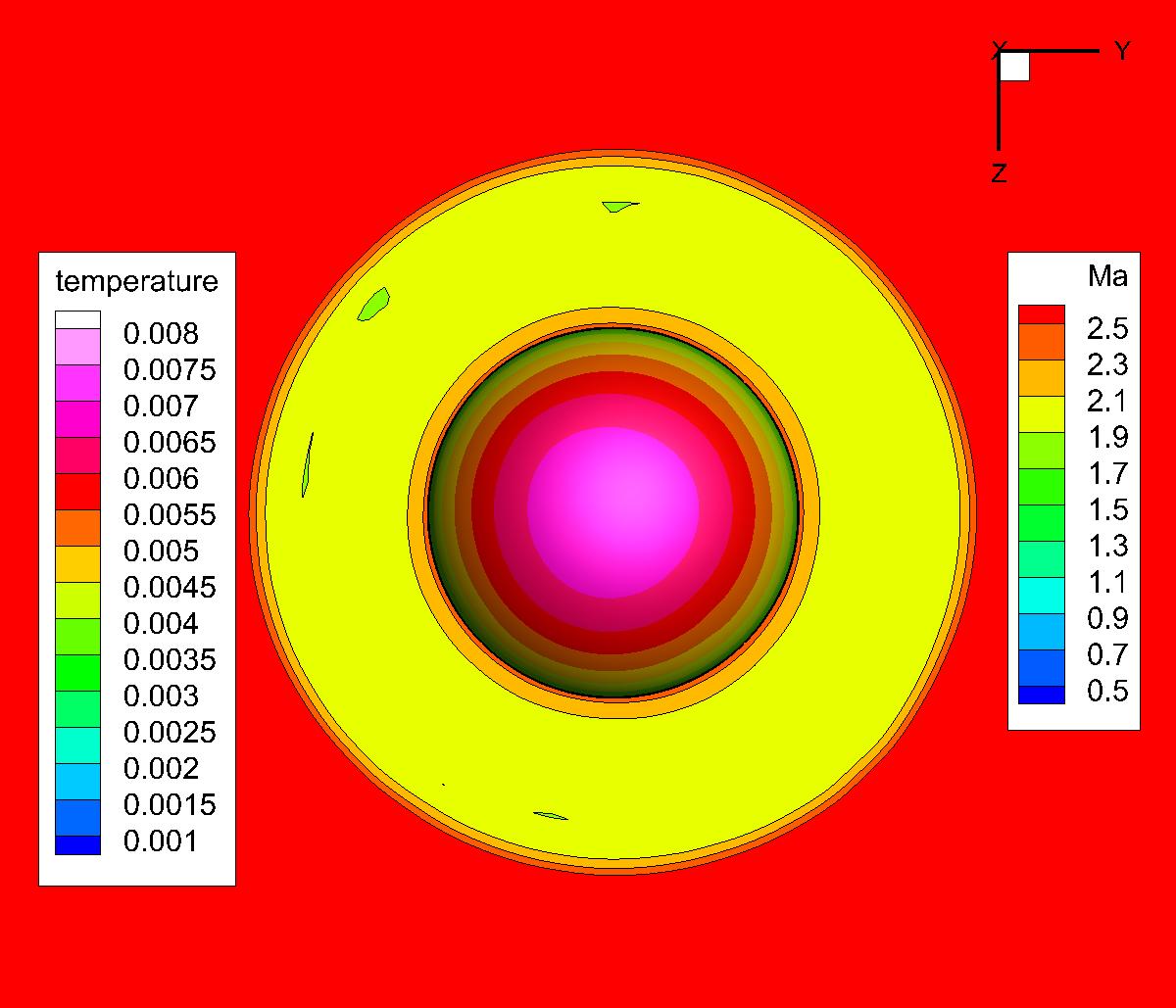}
	\caption{Supersonic inviscid flow passing through a sphere: Ma=3. Left: the pressure and Mach distributions for the x-z and y-z planes. Right: the pressure and Mach distributions for the y-z plane. The sphere is colored by the temperature. CFL=0.5. The maximum Mach number $Ma_{\text{Max}} \approx 5.5$ in the whole domain. }
	\label{inviscid-sphere-contour-ma3}
\end{figure}

\begin{figure}[htp]	
	\centering
	\includegraphics[height=0.32\textwidth]
	{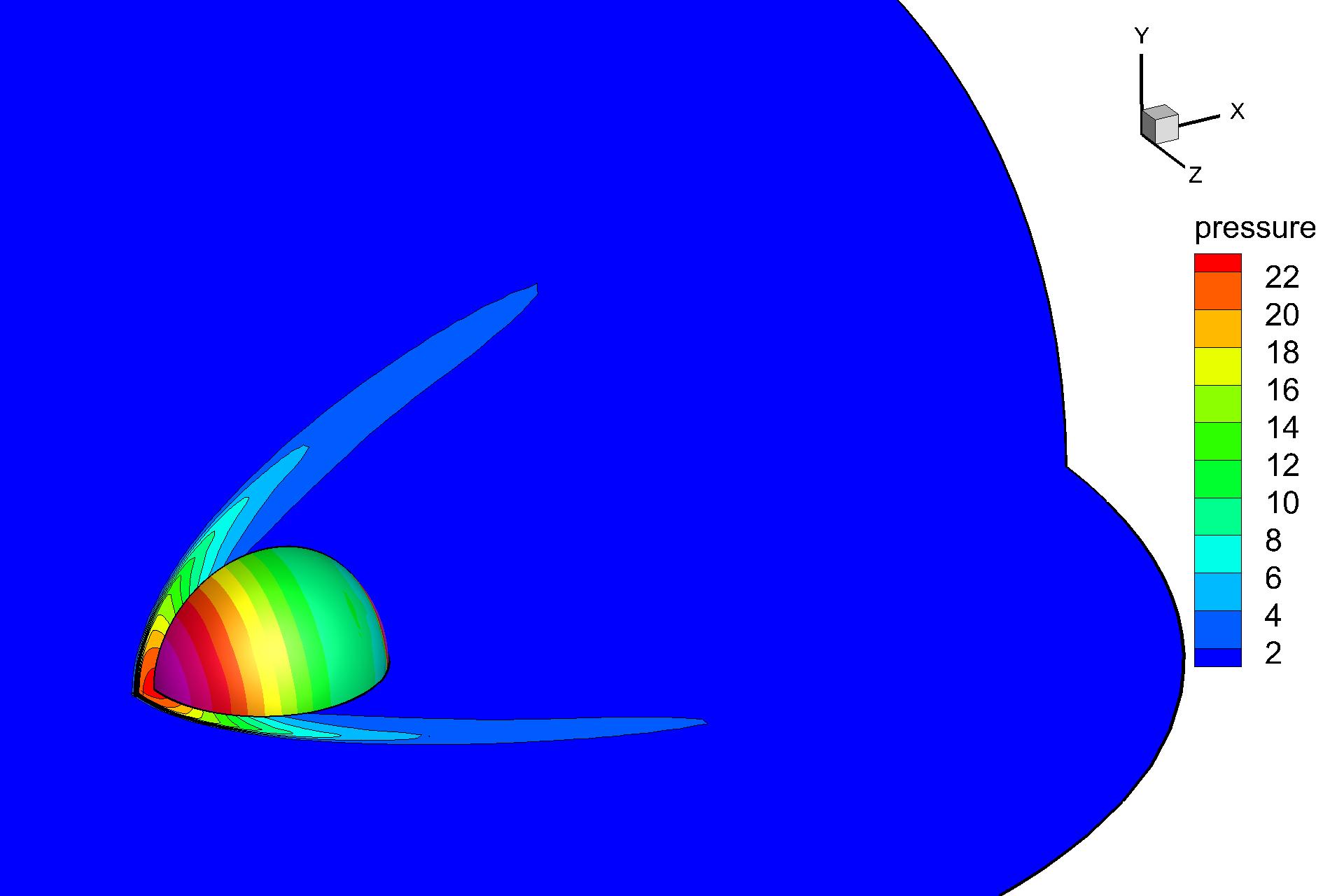}	
	\includegraphics[height=0.32\textwidth]
	{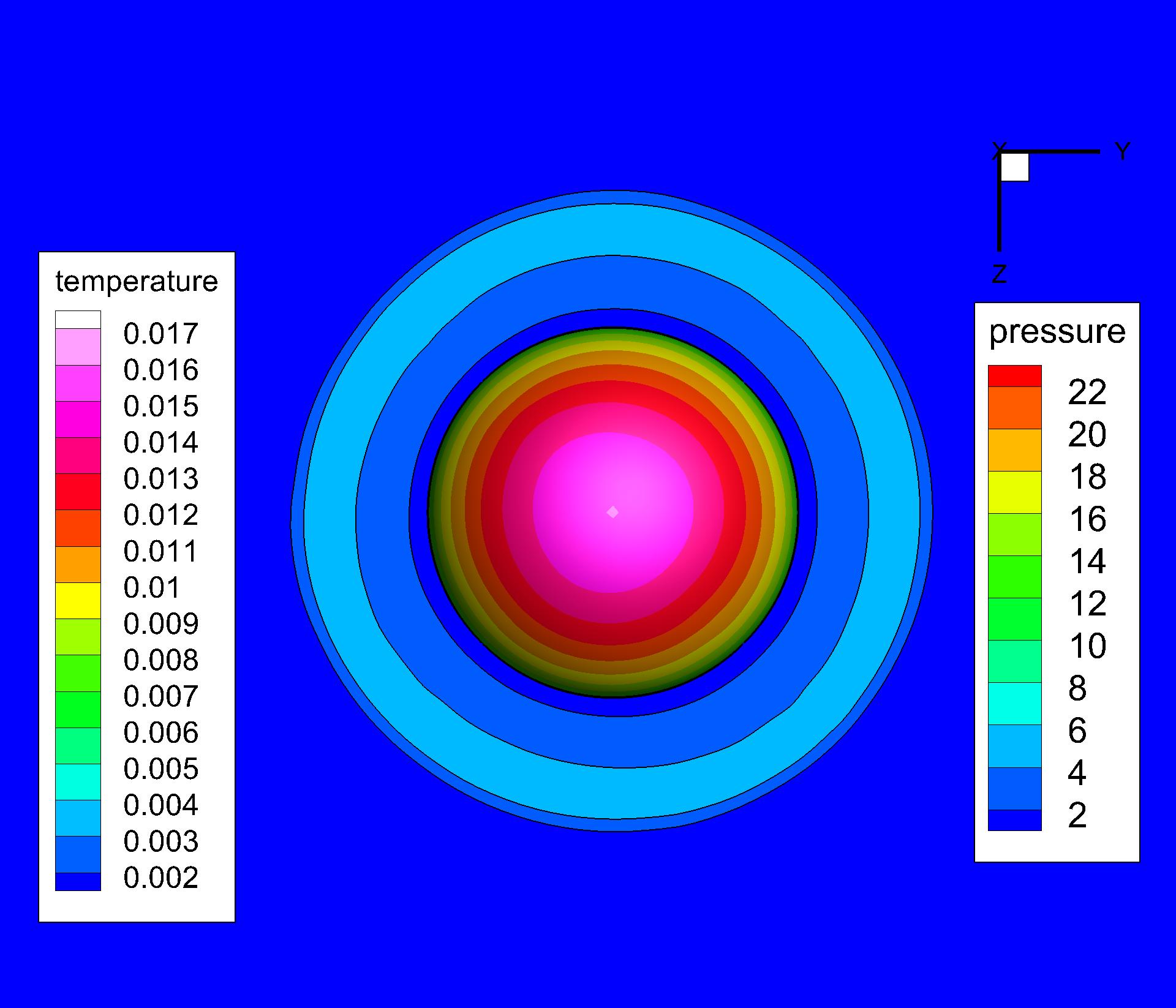}	
	\includegraphics[height=0.32\textwidth]
	{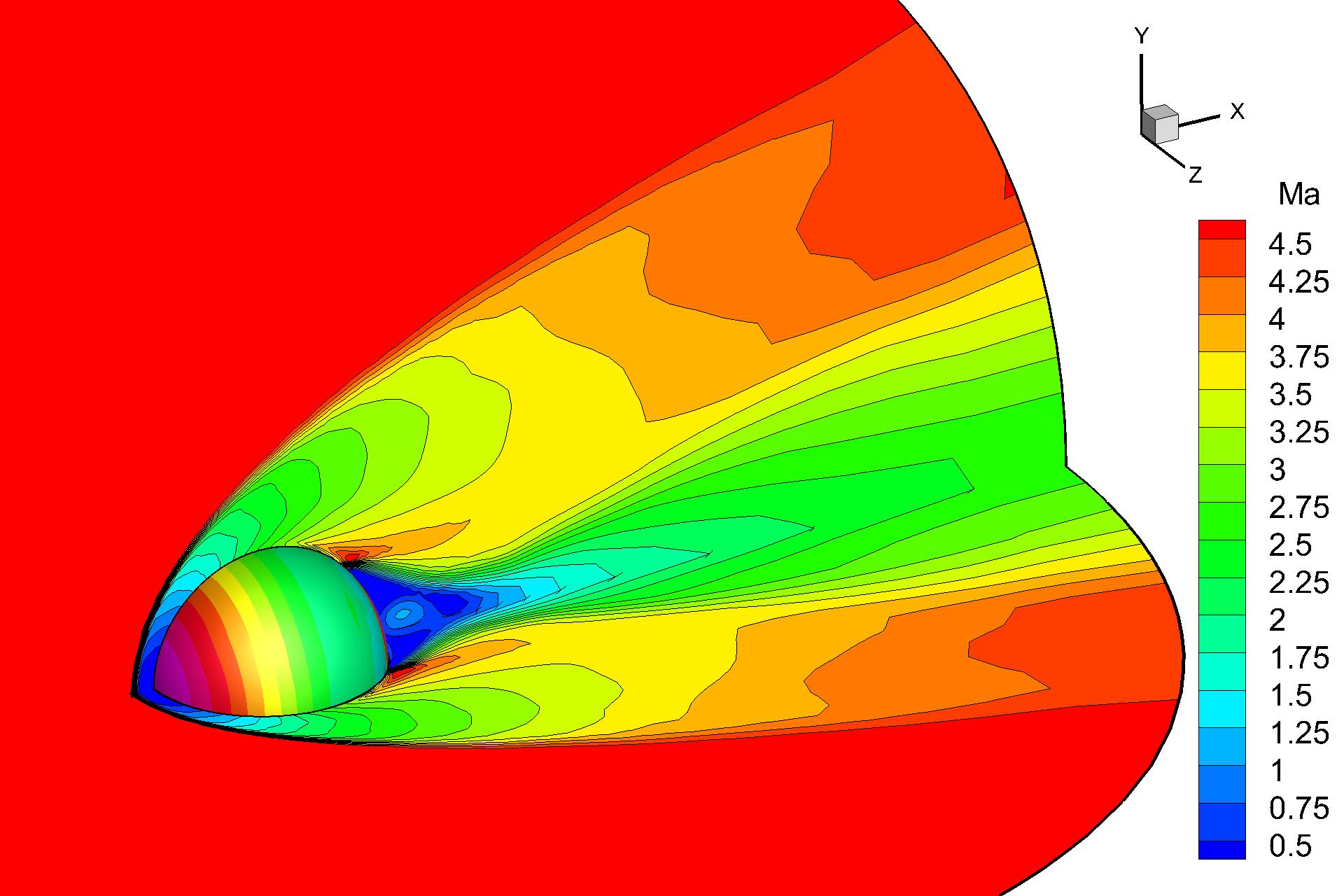}
	\includegraphics[height=0.32\textwidth]
	{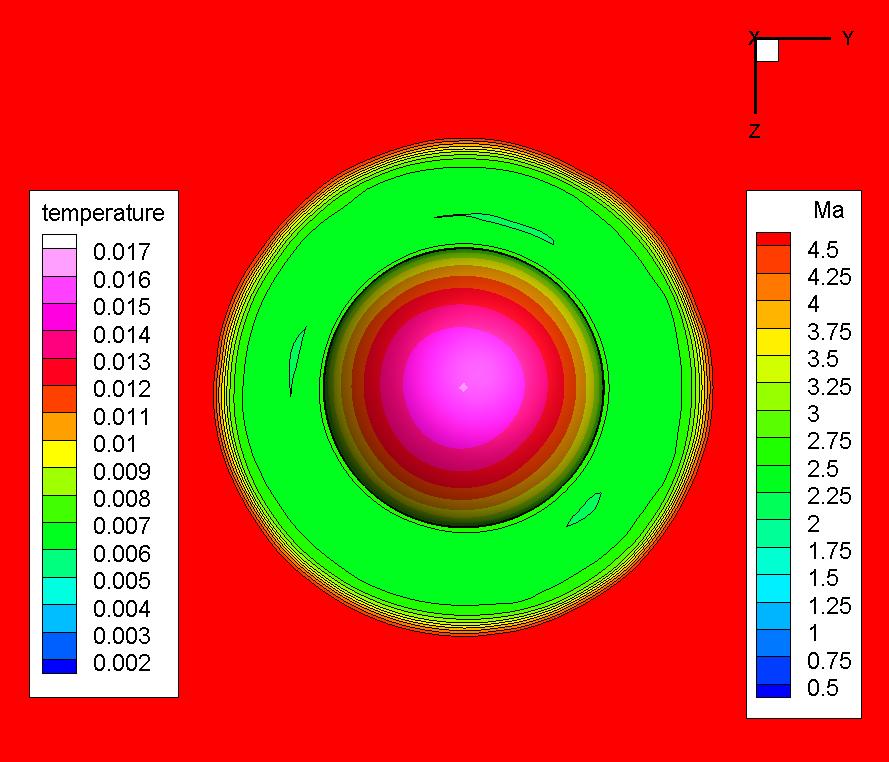}
	\caption{Hypersonic inviscid flow passing through a sphere: Ma=5. Left: the pressure and Mach distributions for the x-z and y-z planes. Right: the pressure and Mach distributions for the y-z plane. The sphere is colored by the temperature. CFL=0.5. The maximum Mach number $Ma_{\text{Max}} \approx 7.3$ in the whole domain.}
	\label{inviscid-sphere-contour-ma5}
\end{figure}

To validate the robustness of the current scheme with non-coplanar meshes,
a high-speed inviscid flow pasting through a sphere is tested.
The structured grids with six blocks are used in the computation, as shown in Fig.\ref{inviscid-sphere-mesh}.
The diameter of the sphere is $D=1$ and the first grid off the wall is $1 \times 10^{-2} D$.
The slip boundary condition is imposed on the surface of the sphere.
The outer domain is around $3.75D$ in the radial direction.
The supersonic inlet/outlet is adopted on the outside boundary,
which is set according to the angle between the outer-pointing normal vector of each boundary interface and the incoming velocity.
A supersonic flow with $Ma=3$ is tested first.
The simulation starts with the free stream flow condition.
The computation is directly started.
The pressure and Mach distributions at steady state are shown in Fig.\ref{inviscid-sphere-contour-ma3}.
Then a hypersonic flow with $Ma=5$ is tested.
A primary flow field calculated by the first-order kinetic method \cite{GKS-lecture} is used as the initial field.
The numerical results are shown in Fig.\ref{inviscid-sphere-contour-ma5}.
The shock is captured sharply  and the carbuncle phenomenon does not appear in both cases.
The result is essentially axis-symmetric.
The asymmetric pattern can be observed at the leeward side of the sphere.
A nearly vacuum state forms in this region and the HWENO reconstruction will easily give a negative density or pressure and then reduce to a low-order reconstruction.

\subsection{Compressible isentropic turbulence}

A decaying homogeneous isotropic compressible turbulence is computed within a square box defined as $-\pi \leq x, y, z\leq \pi$, and the periodic boundary conditions are used in all directions \cite{samtaney2001direct}.
Given spectrum with a specified root mean square $U'$
\begin{align*}
U'=\ll\frac{\textbf{U}\cdot \textbf{U}}{3}\gg^{1/2},
\end{align*}
a divergence-free random velocity field $\textbf{U}_0$ is initialized,
where $\ll...\gg$ is a volume average over the whole computational domain.
The specified spectrum for velocity is given by
\begin{align*}
E(k)=A_0k^4\exp(-2k^2/k_0^2),
\end{align*}
where $A_0$ is a constant to set initial kinetic energy, $k$ is the wavenumber, $k_0$ is the wavenumber at spectrum peaks.
The initial volume averaged turbulent kinetic energy $K_0$  and the initial large-eddy-turnover time $\tau_0$ are given by
\begin{align*}
K_0=\frac{3A_0}{64}\sqrt{2\pi}k_0^5,~~\tau_0=\sqrt{\frac{32}{A_0}}(2\pi)^{1/4}k_0^{-7/2}.
\end{align*}
The Taylor micro-scale $\lambda$ and corresponding Reynolds number $Re_\lambda$ and  are given by
\begin{align*}
\lambda^2=\frac{(U')^2}{\ll(\partial_1 U_1)^2\gg},
Re_\lambda=\frac{\ll\rho\gg U'\lambda}{\ll\mu\gg}=\frac{(2\pi)^{1/4}}{4}\frac{\rho_0}{\mu_0}\sqrt{2A_0}k_0^{3/2}.
\end{align*}
The turbulence Mach number $Ma_t$ is defined as
\begin{align*}
Ma_t=\frac{\sqrt{3}U'}{\ll c_s\gg}=\frac{\sqrt{3}U'}{\sqrt{\gamma T_0}}.
\end{align*}

\begin{figure}[htp]	
	\centering
	\includegraphics[width=0.48\textwidth]{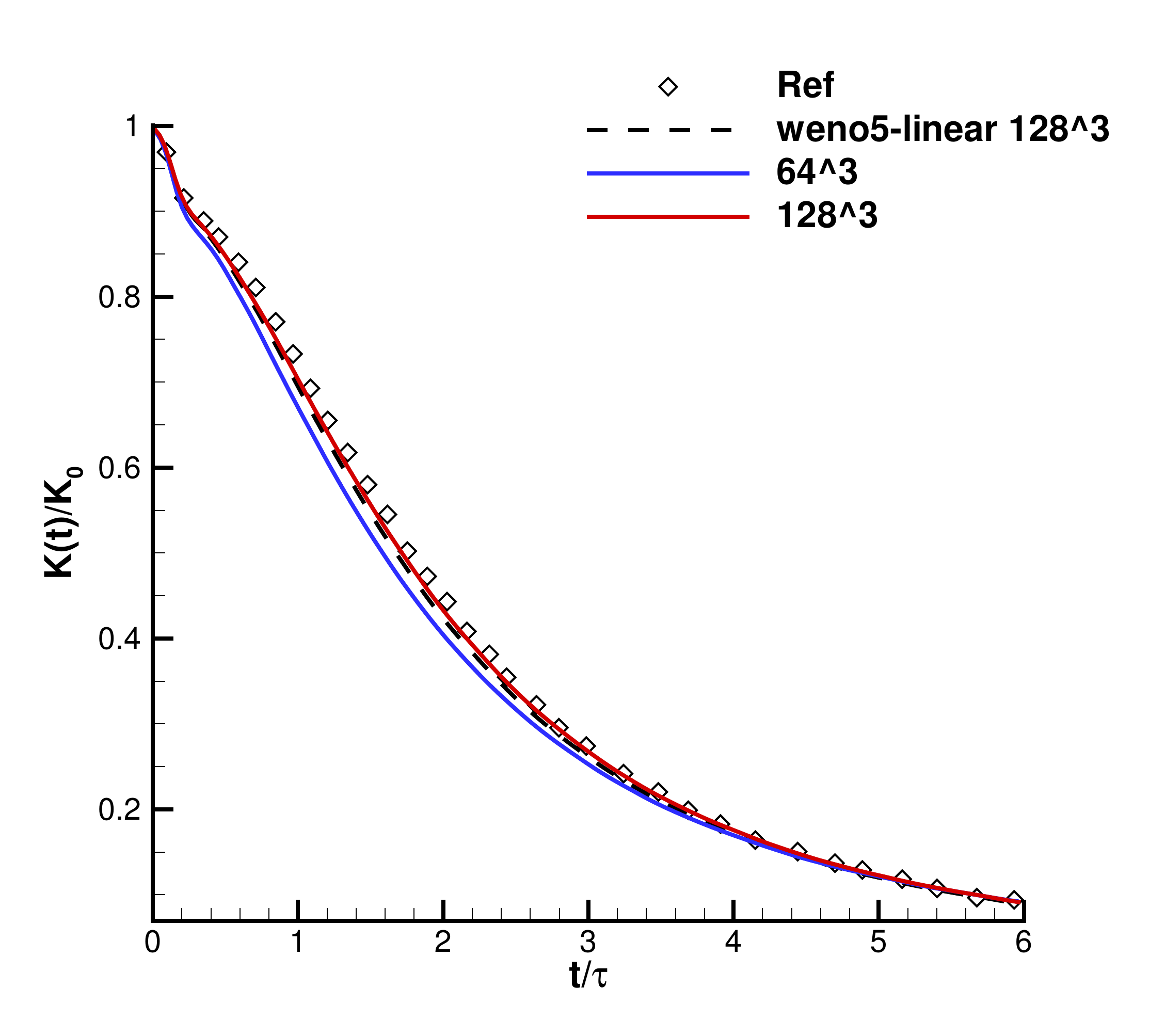}
	\includegraphics[width=0.48\textwidth]{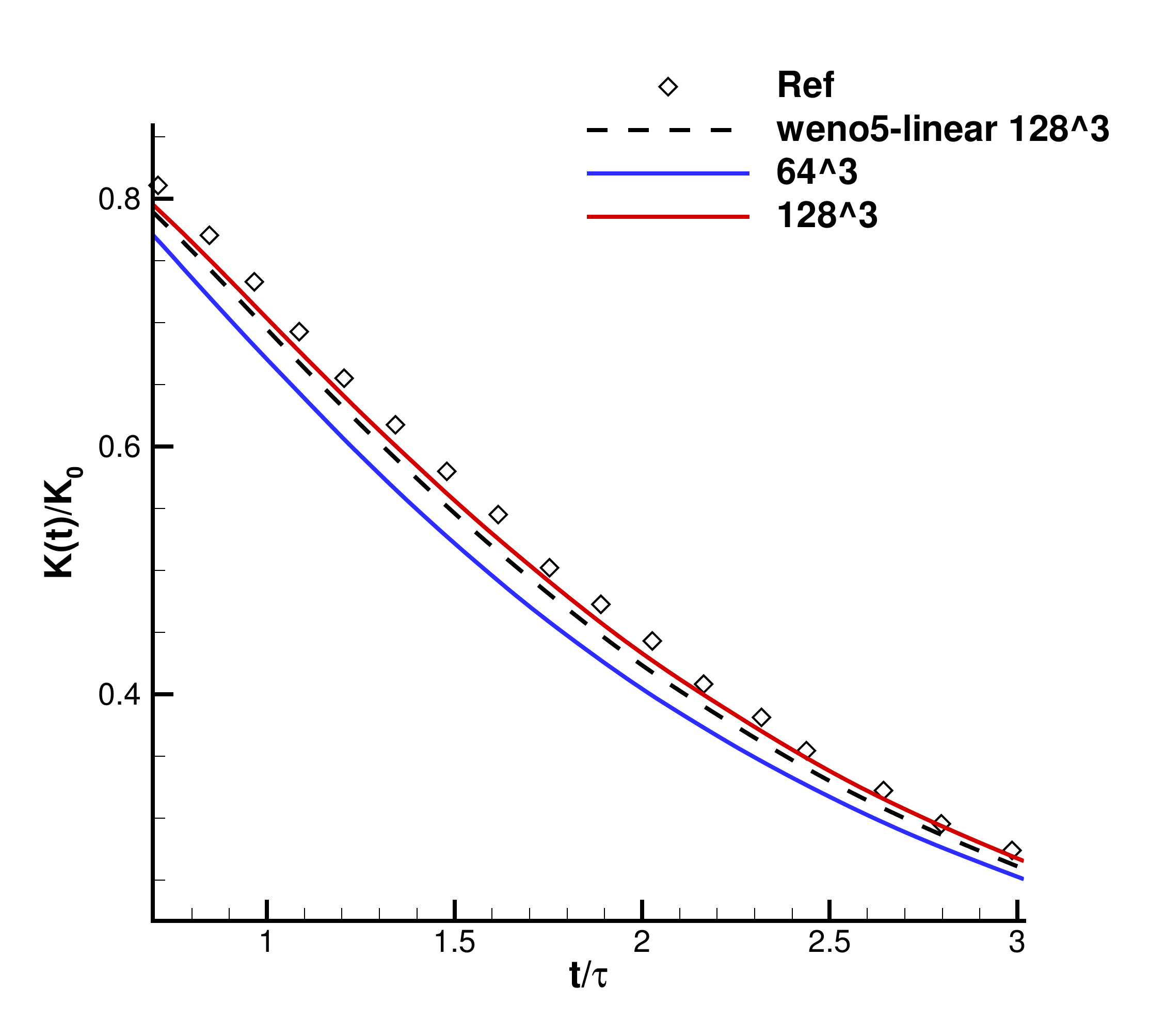}
	\caption{Compressible isotropic turbulence: $K(t)/K_0$. $Re_{\lambda}=72$, $Ma_t=0.5$. }
	\label{cit-ma05-ke}
\end{figure}

\begin{figure}[htp]	
	\centering
	\includegraphics[width=0.48\textwidth]{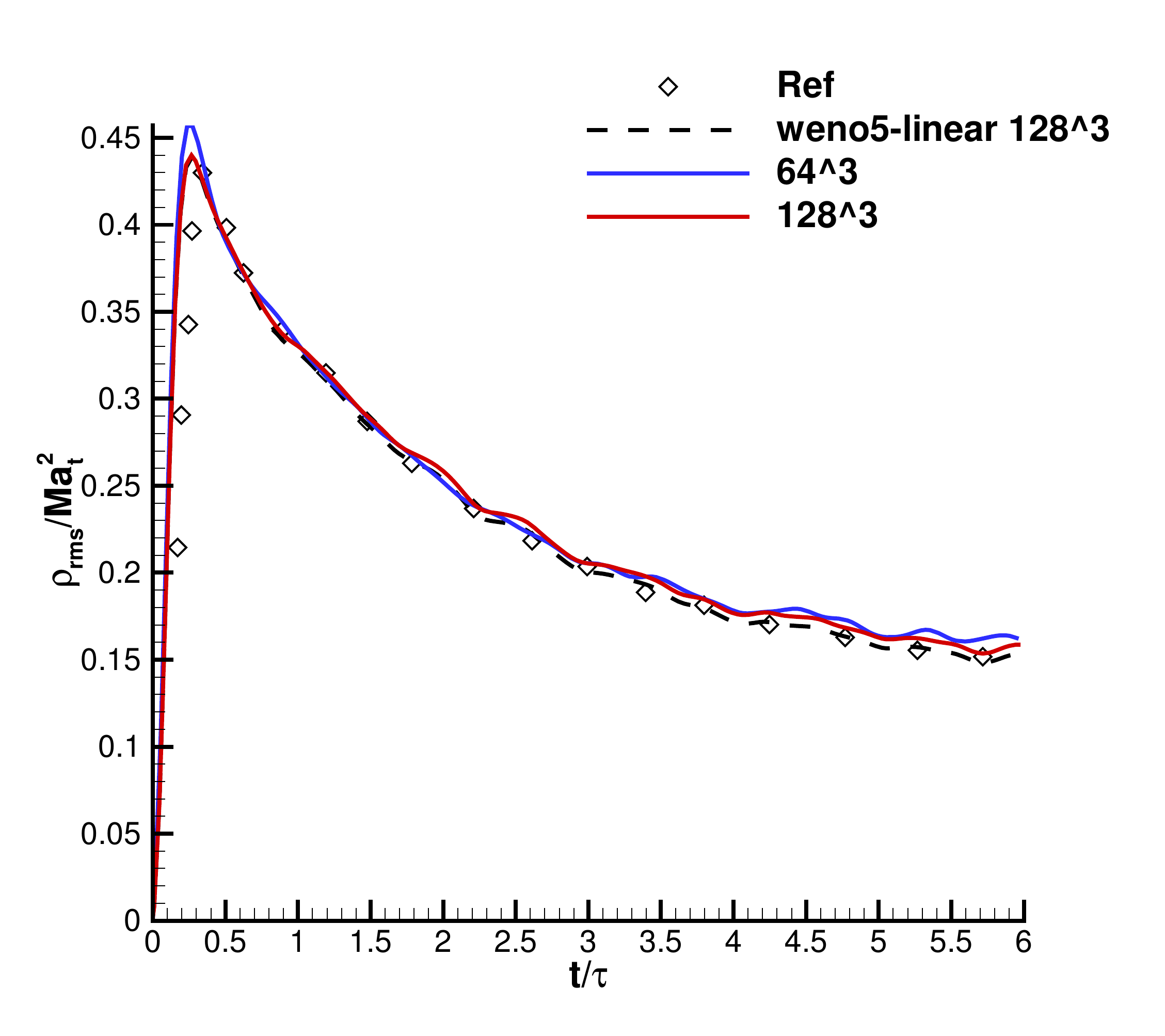}
	\includegraphics[width=0.48\textwidth]{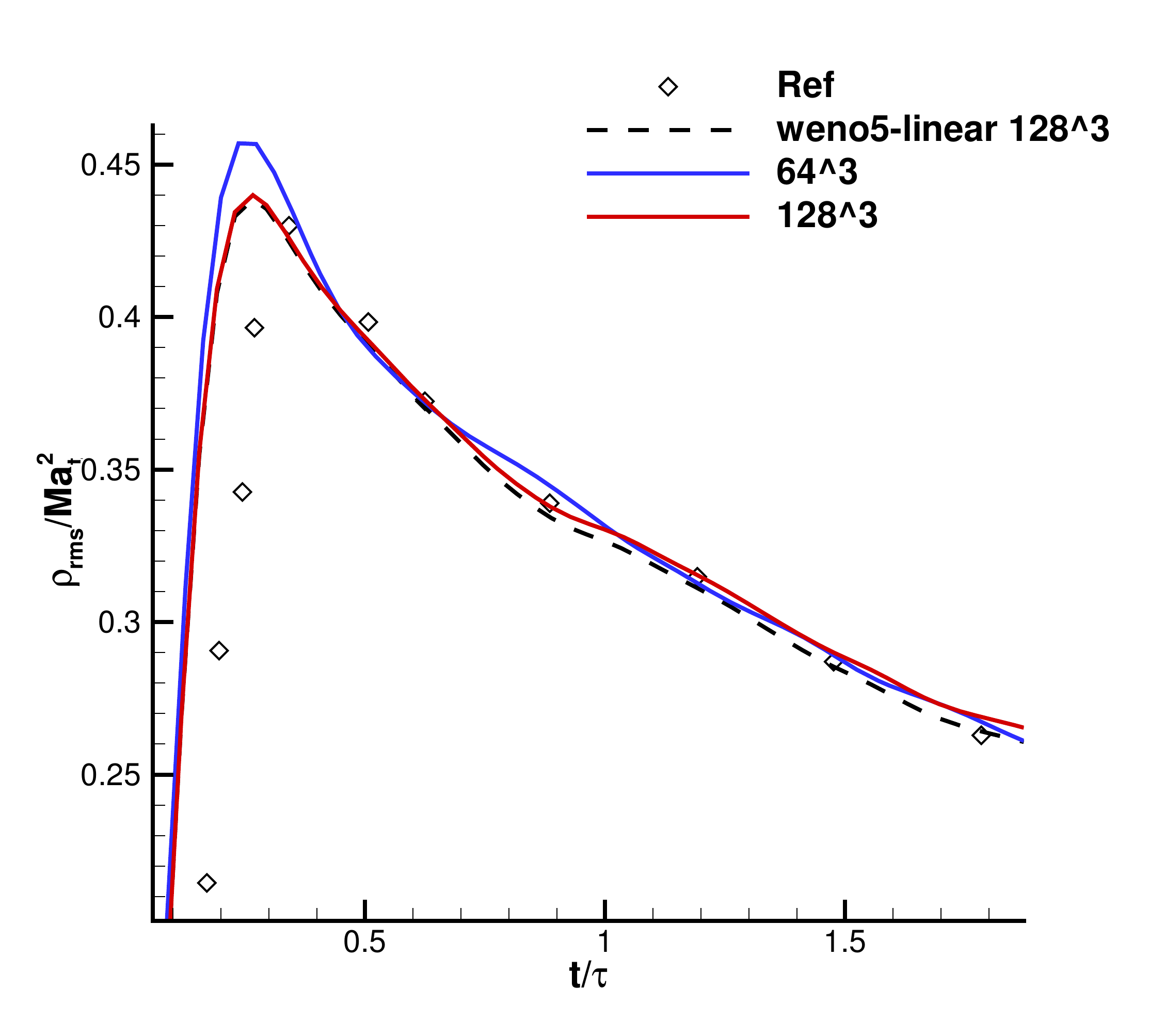}
	\caption{Compressible isotropic turbulence: $\rho_{rms}(t)/Ma_t^2$. $Re_{\lambda}=72$, $Ma_t=0.5$. }
	\label{cit-ma05-rms}
\end{figure}

\begin{figure}[htp]	
	\centering
	\includegraphics[width=0.48\textwidth]{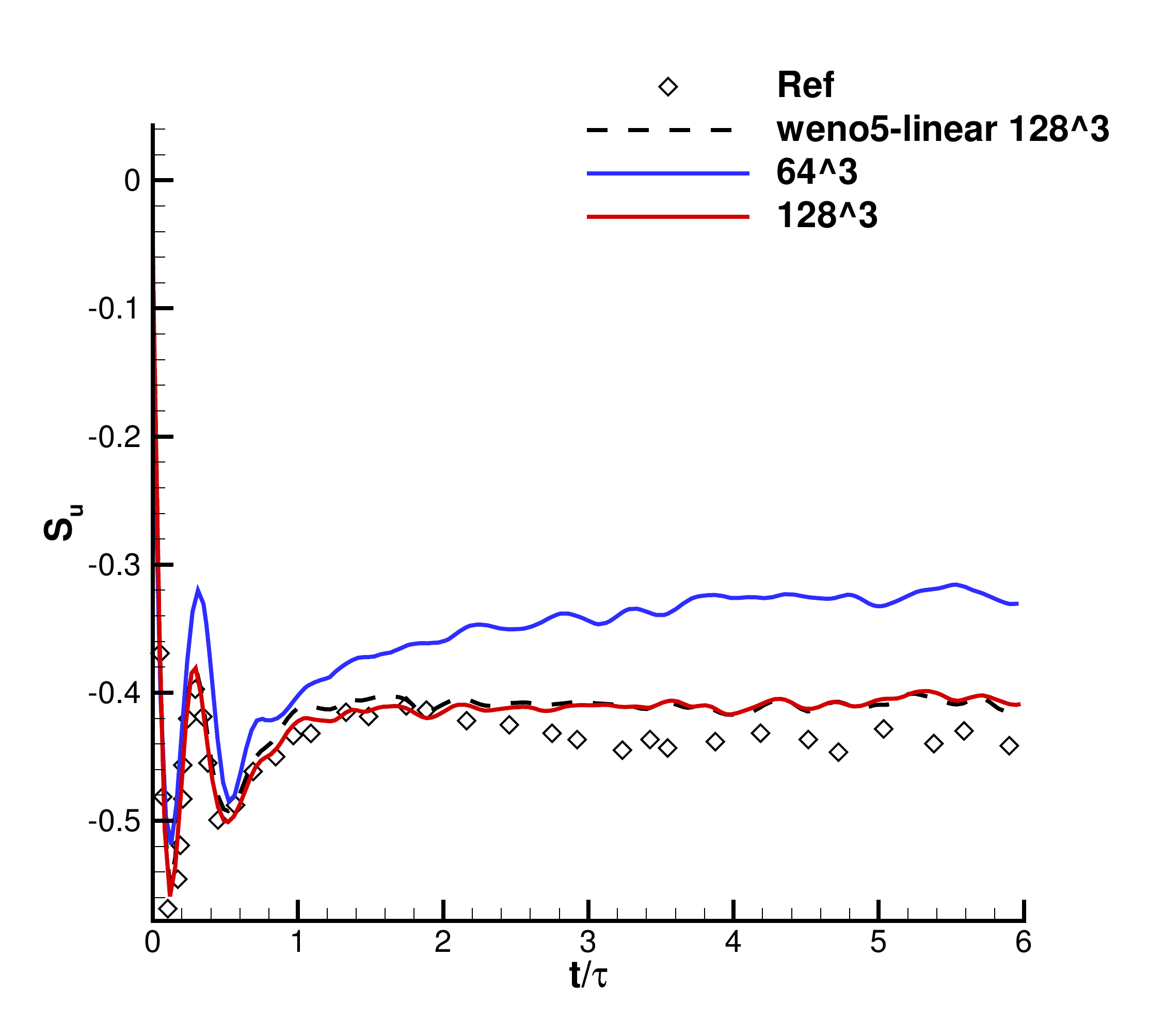}
	\includegraphics[width=0.48\textwidth]{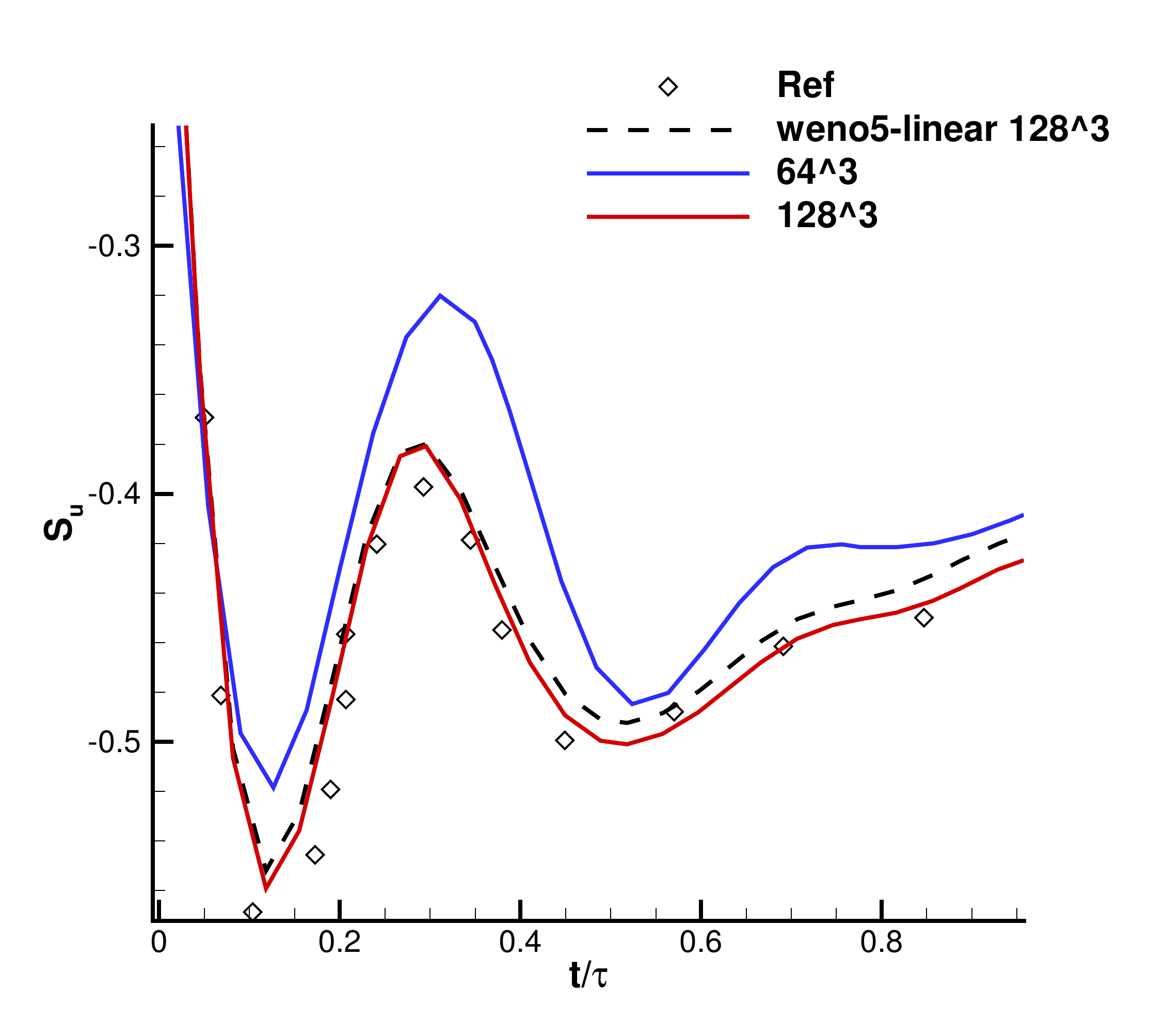}
	\caption{Compressible isotropic turbulence: $S_u(t)$. $Re_{\lambda}=72$, $Ma_t=0.5$. }
	\label{cit-ma05-skew}
\end{figure}

The dynamic viscosity is determined by the power law
\begin{align*}
\mu=\mu_0\big(\frac{T}{T_0}\big)^{0.76},
\end{align*}
where $\mu_0$ and $T_0$ can be determined from $Re_\lambda$ and $Ma_t$ with initialized $U'$ and $\rho_0=1$.
A fixed $Re_\lambda=72$ is investigated in the current work.
The other parameters, i.e., $A_0=1.3\times10^{-4}, k_0=8$, are chosen according to \cite{cao2019three}.
The random initial flow field evolves complex local structures, as shown in the direct numerical simulations by the conventional WENO-GKS \cite{cao2019three}.
When $Ma_t=0.5$, the flow is initially transonic, since the maximum Mach number in the flow filed is about three times of the initial turbulent Mach number.
The pure smooth GKS solver and the WENO-AO reconstruction with linear weights can be used under such mild case to achieve a higher resolution.
Uniform meshes with $64^3$ and $128^3$ cells are used in the simulations.
Initially the cell-averaged slopes are constructed automatically by setting the first explicit time step $\Delta t_1 =0$.
The kinetic energy, root-mean-square of density fluctuation, and the skew factor are calculated, which are given by
\begin{align*}
K(t)=\frac{1}{2}\ll\rho \textbf{U}\cdot \textbf{U}\gg,~~\rho_{rms}(t)=\sqrt{\ll(\rho-\overline{\rho})^2\gg},~S_u(t)=&\sum_i\frac{\ll(\partial_i u_i)^3\gg}{\ll(\partial_i u_i)^2\gg^{3/2}}.
\end{align*}
Since the cell-averaged $\overline{\textbf{W}}_{x_i}$ is stored in each cell,
the $\partial_i u_i$ can be calculated conveniently through the chain rule.
The time history of normalized kinetic energy $K(t)/K_0$, normalized root-mean-square of density fluctuation $\rho_{rms}(t)/Ma_t^2$ agree well with the reference data and the traditional WENO-GKS with $64^3$ cells, as shown in Fig.\ref{cit-ma05-ke}, and Fig.\ref{cit-ma05-rms}.
For the higher-order moment, the skew factor needs finer meshes to resolve, as shown in Fig.\ref{cit-ma05-skew}.

\begin{figure}[htp]	
	\centering
	\includegraphics[width=0.48\textwidth]{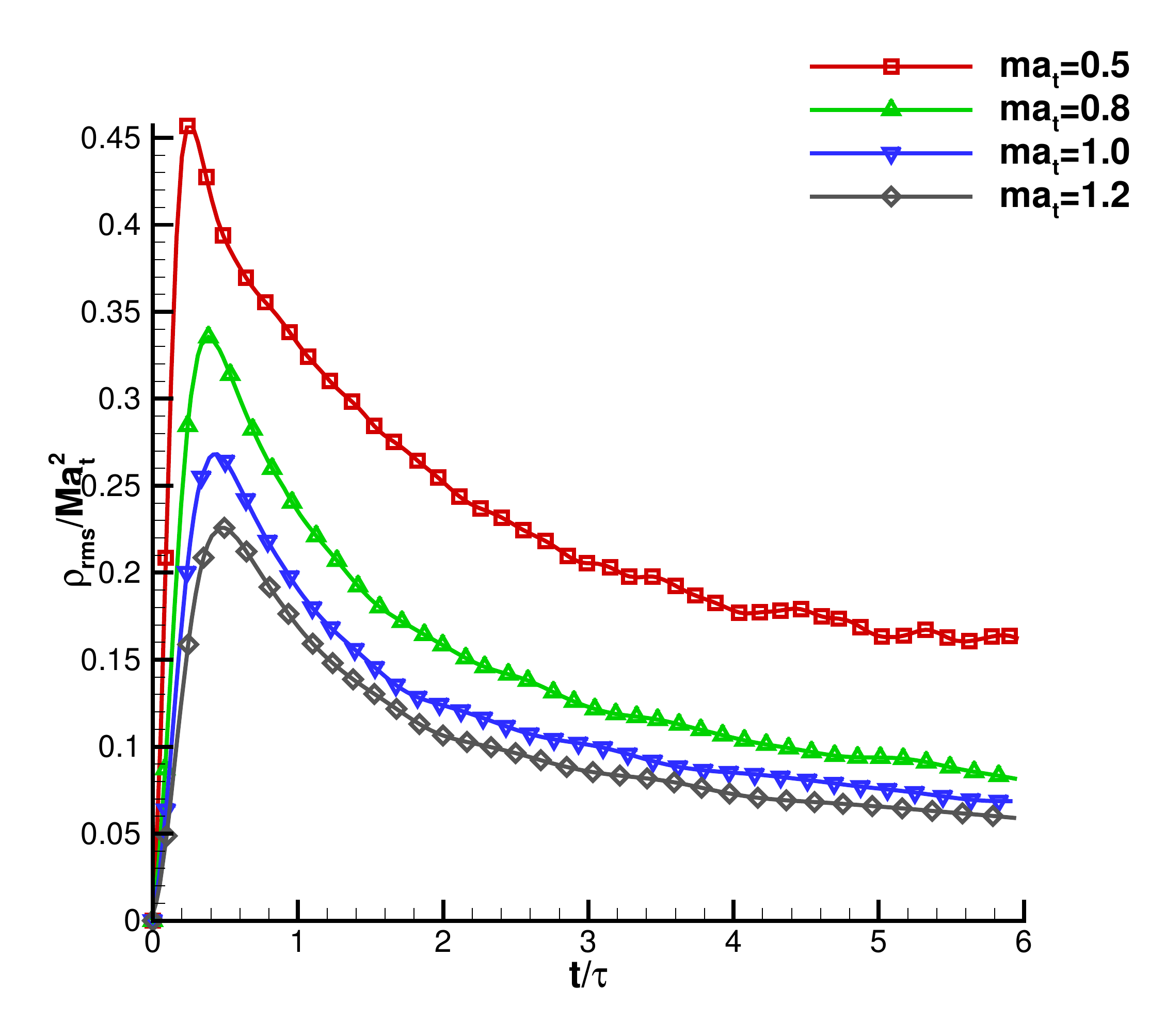}
	\includegraphics[width=0.48\textwidth]{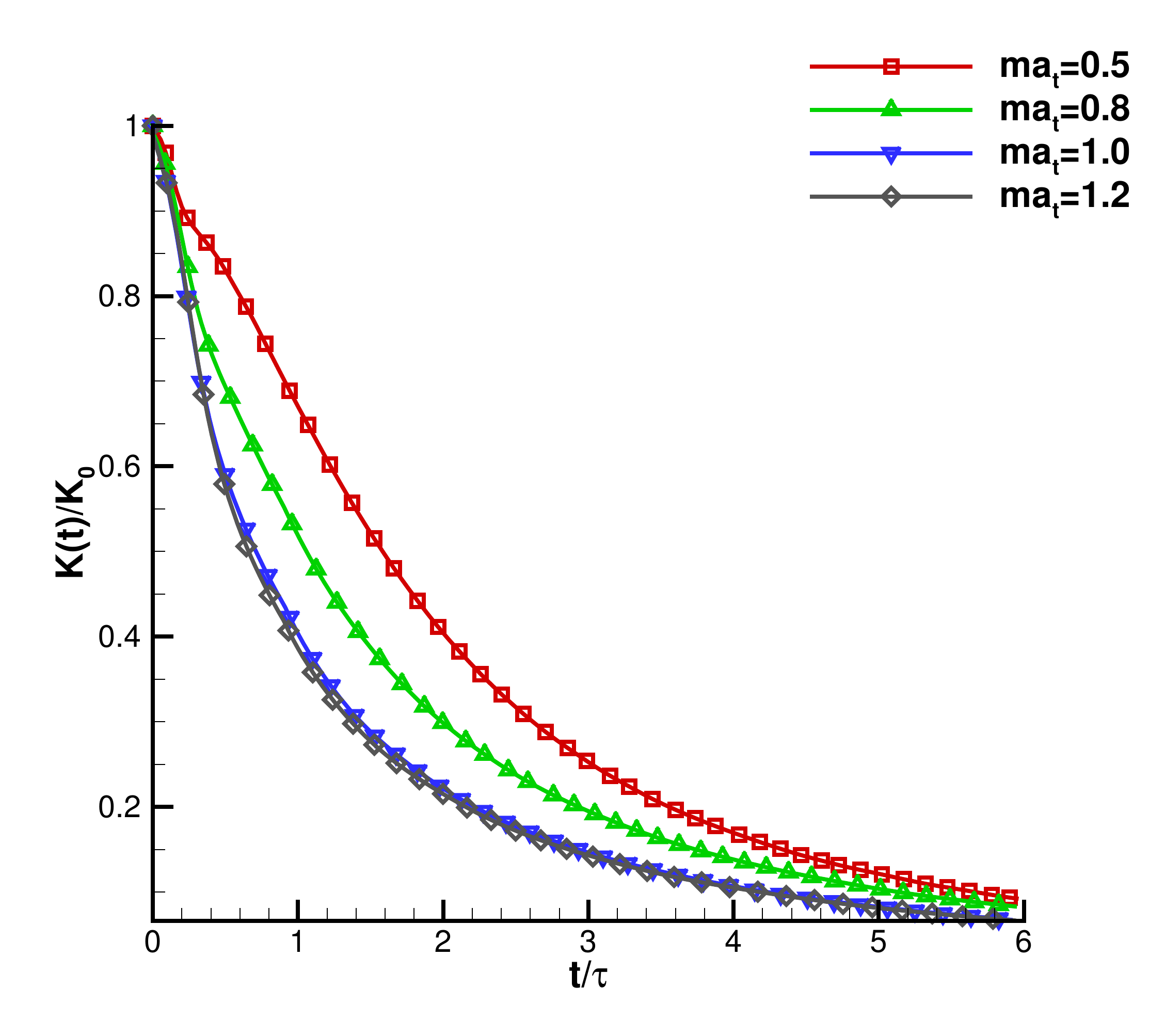}
	\caption{ Compressible isotropic turbulence: comparison with different $Ma_t$ numbers by the new compact GKS. Mesh: $64^3$. CFL=0.5.}
	\label{cit-ma-compare}
\end{figure}

\begin{figure}[htbp]
	\centering
	\includegraphics[width=0.48\textwidth]{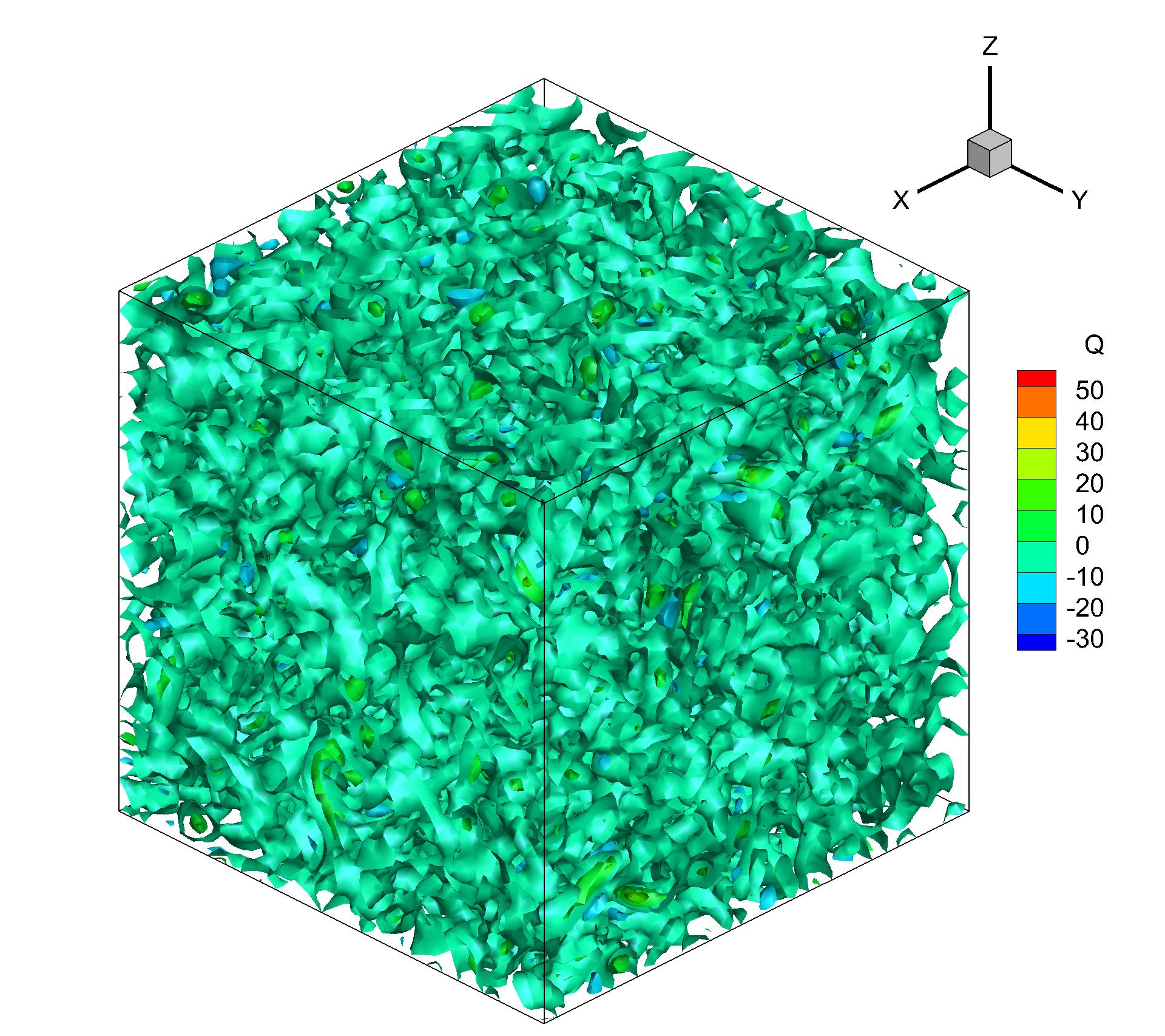}
	\includegraphics[width=0.48\textwidth]{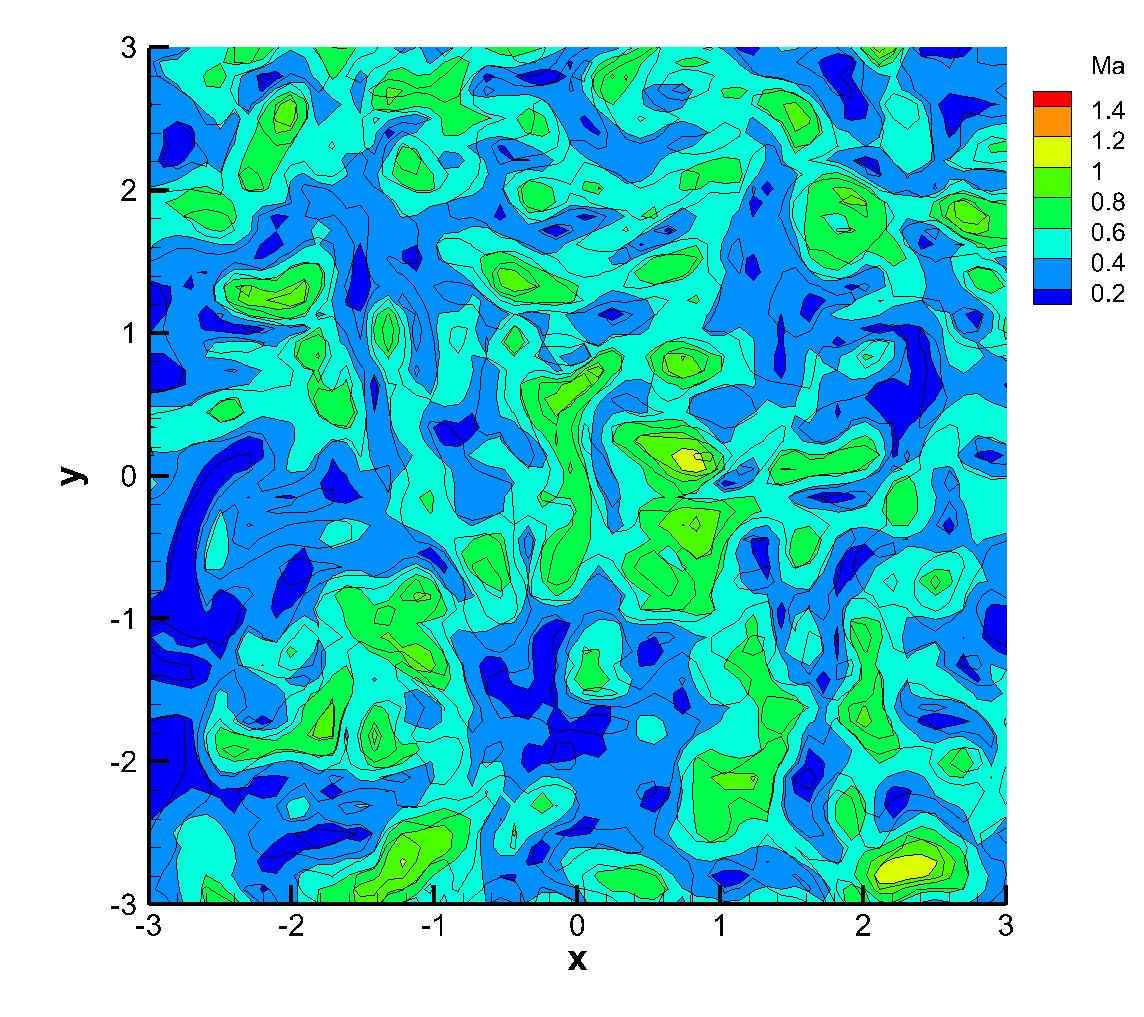}
	\caption{\label{cit-ma1-contour} Compressible homogeneous turbulence with $Ma_t=1$ at time $t/\tau_0=1$.
		Left: iso-surfaces of Q criterion colored by Mach number; right: the Mach number distribution with $z=-\pi$. Mesh: $64^3$}
\end{figure}

When the Mach number gets higher, the flow becomes supersonic, and the stronger shocklets are generated, followed by complex shock-vortex interactions.
It is nontrivial for high-order methods to survive under $Ma_t \geq 1$.
In addition, it becomes more challenging under a coarse mesh, since the discontinuities become stronger due to the limitation of the mesh resolution.
Thus, a series of turbulent Mach numbers have been chosen to test the robustness of the current scheme with the mesh $64^3$.
The full GKS solver and the non-linear HWENO reconstruction are used.
Considering the large velocity jump in the initial field,  a modified $\tau$ is taken as
$\tau=\frac{\mu}{p} +\sum_{1}^{5}\delta Q \Delta t$, where $Q$ means all five primitive variables, operator $\delta Q=\frac{|Q^l-Q^r|}{|Q^l|+|Q^r| + 1e^{-10}}$.
The statistical quantities with respect to different Mach numbers are presented in Fig.\ref{cit-ma-compare}.
With the increase of $Ma_t$, the kinetic energy gets dissipated more rapidly.	
The visualized results, i.e., the iso-surfaces of Q criterion and the selected surface slice of Mach number distribution at $z=-\pi$ are plotted in Fig.\ref{cit-ma1-contour}.
The complex vortexes and widespread shocklets can be observed clearly.

\section{Conclusion}
In this paper, a compact high-order gas-kinetic scheme for three-dimensional flow simulation is presented.
The distinguishable feature of the scheme is that the high-order GKS evolution model at a cell interface provides not only the fluxes, but also
the time accurate flow variables. As a result, based on the cell interface values the first-order spatial derivatives of flow variables
inside each control volume can be directly obtained through Gauss's theorem at the next time level.
The way for the updates of gradients in GKS is different from the weak formulation for the updates of similar degree of freedom in the compact DG-type methods. Therefore, equipped with the cell-averaged values and their gradients,
a new HWENO reconstruction with less sub-stencils and all positive weights has been designed for the initial data reconstruction in the scheme.
At the same time, the multi-stage and multi-derivative technique is used as a time marching strategy in the scheme, which subsequently leads to a
high efficiency in comparison with the traditional Runge-Kutta method for the same temporal accuracy.
The compact third-order scheme in 3D can use a relative large CFL number in the computations, and shows
similar resolution as the fifth-order gas-kinetic scheme with  the non-compact WENO reconstruction.
Overall, the time accurate evolution model, the HWENO reconstruction, and the MSMD time marching technique make the final scheme accurate, robust, and efficient for the compressible flow simulations with smooth and discontinuous solutions.

Although the current scheme is constructed on structured mesh, it can be directly extended to unstructured one.
The compact third-order and fourth-order HGKS  on three-dimensional unstructured mesh is on the development, which further enlarges the applicable regime of the high-order gas-kinetic schemes for flow computation with complex geometry.
The development of high-order compact scheme with implicit and other acceleration techniques is on the  investigation as well for the steady state solution.

\section*{Acknowledgment}
The authors would like to thank Dr. Liang Pan for helpful discussion.
The current research is supported by National Numerical Windtunnel project, Hong Kong research grant council 16206617, and  National Science Foundation of China 11772281, 91852114.

\section*{References}
\bibliographystyle{plain}%
\bibliography{jixingbib}

\begin{thebibliography}{10}

\bibitem{balsara2016efficient}
Dinshaw~S Balsara, Sudip Garain, and Chi-Wang Shu.
\newblock An efficient class of {WENO} schemes with adaptive order.
\newblock {\em Journal of Computational Physics}, 326:780--804, 2016.

\bibitem{BGK}
Prabhu~Lal Bhatnagar, Eugene~P Gross, and Max Krook.
\newblock {A model for collision processes in gases. I. {Small} amplitude
  processes in charged and neutral one-component systems}.
\newblock {\em Physical Review}, 94(3):511, 1954.

\bibitem{cao2018physical}
Guiyu Cao, Hualin Liu, and Kun Xu.
\newblock Physical modeling and numerical studies of three-dimensional
  non-equilibrium multi-temperature flows.
\newblock {\em Physics of Fluids}, 30(12):126104, 2018.

\bibitem{cao2019three}
Guiyu Cao, Liang Pan, and Kun Xu.
\newblock Three dimensional high-order gas-kinetic scheme for supersonic
  isotropic turbulence i: criterion for direct numerical simulation.
\newblock {\em Computers \& Fluids}, 192:104273, 2019.

\bibitem{cheng2017parallel}
Jian Cheng, Xiaodong Liu, Tiegang Liu, and Hong Luo.
\newblock A parallel, high-order direct discontinuous {Galerkin method for the
  Navier-Stokes equations on 3D hybrid grids}.
\newblock {\em Communications in Computational Physics}, 21(5):1231--1257,
  2017.

\bibitem{debonis2013solutions}
James DeBonis.
\newblock {Solutions of the Taylor-Green vortex problem using high-resolution
  explicit finite difference methods}.
\newblock In {\em 51st AIAA Aerospace Sciences Meeting including the New
  Horizons Forum and Aerospace Exposition}, page 382, 2013.

\bibitem{du2018hermite}
Zhifang Du and Jiequan Li.
\newblock {A Hermite WENO reconstruction for fourth order temporal accurate
  schemes based on the GRP solver for hyperbolic conservation laws}.
\newblock {\em Journal of Computational Physics}, 355:385--396, 2018.

\bibitem{RK-advantage2}
Sigal Gottlieb, Chi-Wang Shu, and Eitan Tadmor.
\newblock Strong stability-preserving high-order time discretization methods.
\newblock {\em SIAM review}, 43(1):89--112, 2001.

\bibitem{hu1999weighted}
Changqing Hu and Chi-Wang Shu.
\newblock Weighted essentially non-oscillatory schemes on triangular meshes.
\newblock {\em Journal of Computational Physics}, 150(1):97--127, 1999.

\bibitem{Huynh2007FR}
Ht~T Huynh.
\newblock A flux reconstruction approach to high-order schemes including
  discontinuous {Galerkin} methods.
\newblock {\em AIAA paper}, 4079:2007, 2007.

\bibitem{ji2019high}
Xing Ji.
\newblock {\em High-order non-compact and compact gas-kinetic schemes}.
\newblock PhD thesis, Hong Kong Univeristy of Science and Technology, 2019.

\bibitem{ji2018compact}
Xing Ji, Liang Pan, Wei Shyy, and Kun Xu.
\newblock {A compact fourth-order gas-kinetic scheme for the Euler and
  Navier-Stokes equations}.
\newblock {\em Journal of Computational Physics}, 372:446 -- 472, 2018.

\bibitem{ji2020performance}
Xing Ji and Kun Xu.
\newblock Performance enhancement for high-order gas-kinetic scheme based on
  {WENO}-adaptive-order reconstruction.
\newblock {\em Communications in Computational Physics}, 28(2):539--590, 2020.

\bibitem{ji2018family}
Xing Ji, Fengxiang Zhao, Wei Shyy, and Kun Xu.
\newblock {A family of high-order gas-kinetic schemes and its comparison with
  Riemann solver based high-order methods}.
\newblock {\em Journal of Computational Physics}, 356:150--173, 2018.

\bibitem{ji2020hweno}
Xing Ji, Fengxiang Zhao, Wei Shyy, and Kun Xu.
\newblock A {HWENO} reconstruction based high-order compact gas-kinetic scheme
  on unstructured mesh.
\newblock {\em Journal of Computational Physics}, page 109367, 2020.

\bibitem{pan2018high}
PAN Jianhua, WANG Qian, Yusi Zhang, and REN Yuxin.
\newblock High-order compact finite volume methods on unstructured grids with
  adaptive mesh refinement for solving inviscid and viscous flows.
\newblock {\em Chinese Journal of Aeronautics}, 31(9):1829--1841, 2018.

\bibitem{krivodonova2006high}
Lilia Krivodonova and Marsha Berger.
\newblock High-order accurate implementation of solid wall boundary conditions
  in curved geometries.
\newblock {\em Journal of computational physics}, 211(2):492--512, 2006.

\bibitem{li2016twostage}
Jiequan Li and Zhifang Du.
\newblock A two-stage fourth order time-accurate discretization for
  {Lax--Wendroff} type flow solvers {I.} hyperbolic conservation laws.
\newblock {\em SIAM Journal on Scientific Computing}, 38(5):A3046--A3069, 2016.

\bibitem{li2014efficient}
Wanai Li.
\newblock {\em Efficient implementation of high-order accurate numerical
  methods on unstructured grids}.
\newblock Berlin, Heidelberg: Springer, 2014.

\bibitem{luo2008computation}
Hong Luo, Joseph~D Baum, and Rainald L{\"o}hner.
\newblock On the computation of steady-state compressible flows using a
  discontinuous galerkin method.
\newblock {\em International Journal for Numerical Methods in Engineering},
  73(5):597--623, 2008.

\bibitem{pan2017two-multicomponent}
Liang Pan, Junxia Cheng, Shuanghu Wang, and Kun Xu.
\newblock A two-stage fourth-order gas-kinetic scheme for compressible
  multicomponent flows.
\newblock {\em Communications in Computational Physics}, 22(4):1123--1149,
  2017.

\bibitem{pan2016unstructuredcompact}
Liang Pan and Kun Xu.
\newblock A third-order compact gas-kinetic scheme on unstructured meshes for
  compressible {Navier--Stokes} solutions.
\newblock {\em Journal of Computational Physics}, 318:327--348, 2016.

\bibitem{pan2018two}
Liang Pan and Kun Xu.
\newblock Two-stage fourth-order gas-kinetic scheme for three-dimensional
  {Euler} and {Navier-Stokes} solutions.
\newblock {\em International Journal of Computational Fluid Dynamics},
  32:395--411, 2018.

\bibitem{Pan2016twostage}
Liang Pan, Kun Xu, Qibing Li, and Jiequan Li.
\newblock {An efficient and accurate two-stage fourth-order gas-kinetic scheme
  for the Euler and Navier--Stokes equations}.
\newblock {\em Journal of Computational Physics}, 326:197--221, 2016.

\bibitem{qiu2004hermite}
Jianxian Qiu and Chi-Wang Shu.
\newblock {Hermite WENO schemes and their application as limiters for
  Runge--Kutta discontinuous Galerkin method: one-dimensional case}.
\newblock {\em Journal of Computational Physics}, 193(1):115--135, 2004.

\bibitem{qiu2016stability-limter-dg-book}
Jianxian Qiu and Qiang Zhang.
\newblock Stability, error estimate and limiters of discontinuous galerkin
  methods.
\newblock In {\em Handbook of Numerical Analysis}, volume~17, pages 147--171.
  Elsevier, 2016.

\bibitem{samtaney2001direct}
Ravi Samtaney, Dale~I Pullin, and Branko Kosovi{\'c}.
\newblock Direct numerical simulation of decaying compressible turbulence and
  shocklet statistics.
\newblock {\em Physics of Fluids}, 13(5):1415--1430, 2001.

\bibitem{shu2009weno-review}
Chi-Wang Shu.
\newblock High order weighted essentially nonoscillatory schemes for convection
  dominated problems.
\newblock {\em SIAM review}, 51(1):82--126, 2009.

\bibitem{shu2016weno-dg-review}
Chi-Wang Shu.
\newblock High order weno and dg methods for time-dependent
  convection-dominated pdes: A brief survey of several recent developments.
\newblock {\em Journal of Computational Physics}, 316:598--613, 2016.

\bibitem{sun2007sd}
Y.~Sun, Z.~J. Wang, and Y.~Liu.
\newblock High-order multidomain spectral difference method for the
  {Navier-Stokes} equations on unstructured hexahedral grids.
\newblock {\em Communications in Computational Physics}, 2(2):310--333, 2007.

\bibitem{taneda1956experimental}
Sadatoshi Taneda.
\newblock Experimental investigation of the wakes behind cylinders and plates
  at low {Reynolds} numbers.
\newblock {\em Journal of the Physical Society of Japan}, 11(3):302--307, 1956.

\bibitem{wang2017thesis}
Qian Wang.
\newblock {\em Compact High-Order Finite Volume Method on Unstructured Grids}.
\newblock PhD thesis, Tsinghua University, 6 2017.

\bibitem{wang2016compact}
Qian Wang, Yu-Xin Ren, and Wanai Li.
\newblock Compact high order finite volume method on unstructured grids {II:
  Extension to two-dimensional Euler equations}.
\newblock {\em Journal of Computational Physics}, 314:883--908, 2016.

\bibitem{high-order-review}
Zhijian Wang, Krzysztof Fidkowski, R{\'e}mi Abgrall, Francesco Bassi, Doru
  Caraeni, Andrew Cary, Herman Deconinck, Ralf Hartmann, Koen Hillewaert,
  Hung~T Huynh, et~al.
\newblock High-order {CFD} methods: current status and perspective.
\newblock {\em International Journal for Numerical Methods in Fluids},
  72(8):811--845, 2013.

\bibitem{wang2017towards}
ZJ~Wang, Y~Li, F~Jia, GM~Laskowski, J~Kopriva, U~Paliath, and R~Bhaskaran.
\newblock Towards industrial large eddy simulation using the {FR/CPR} method.
\newblock {\em Computers \& Fluids}, 156:579--589, 2017.

\bibitem{xie2017hybrid-compact-mcv-2d-euler}
Bin Xie, Xi~Deng, Ziyao Sun, and Feng Xiao.
\newblock A hybrid pressure--density-based mach uniform algorithm for 2d euler
  equations on unstructured grids by using multi-moment finite volume method.
\newblock {\em Journal of Computational Physics}, 335:637--663, 2017.

\bibitem{GKS-lecture}
Kun Xu.
\newblock Gas-kinetic schemes for unsteady compressible flow simulations.
\newblock {\em Lecture series-van Kareman Institute for fluid dynamics},
  3:C1--C202, 1998.

\bibitem{GKS-2001}
Kun Xu.
\newblock A gas-kinetic {BGK} scheme for the {Navier--Stokes} equations and its
  connection with artificial dissipation and {Godunov} method.
\newblock {\em Journal of Computational Physics}, 171(1):289--335, 2001.

\bibitem{xu2014direct}
Kun Xu.
\newblock {\em Direct modeling for computational fluid dynamics: construction
  and application of unified gas-kinetic schemes}.
\newblock World Scientific, 2014.

\bibitem{xu2005multidimensional-implicitGKS}
Kun Xu, Meiliang Mao, and Lei Tang.
\newblock {A multidimensional gas-kinetic BGK scheme for hypersonic viscous
  flow}.
\newblock {\em Journal of Computational Physics}, 203(2):405--421, 2005.

\bibitem{yang2018reconstructed}
Xiaoquan Yang, Jian Cheng, Hong Luo, and Qijun Zhao.
\newblock A reconstructed direct discontinuous galerkin method for simulating
  the compressible laminar and turbulent flows on hybrid grids.
\newblock {\em Computers \& Fluids}, 168:216--231, 2018.

\bibitem{yang2019robust}
Xiaoquan Yang, Jian Cheng, Hong Luo, and Qijun Zhao.
\newblock Robust implicit direct discontinuous galerkin method for simulating
  the compressible turbulent flows.
\newblock {\em AIAA Journal}, 57(3):1113--1132, 2019.

\bibitem{yu2014accuracy}
Meilin Yu, Zhijian Wang, and Yen Liu.
\newblock {On the accuracy and efficiency of discontinuous Galerkin, spectral
  difference and correction procedure via reconstruction methods}.
\newblock {\em Journal of Computational Physics}, 259:70--95, 2014.

\bibitem{zhao2019acoustic}
Fengxiang Zhao, Xing Ji, Wei Shyy, and Kun Xu.
\newblock An acoustic and shock wave capturing compact high-order gas-kinetic
  scheme with spectral-like resolution.
\newblock {\em arXiv preprint arXiv:2001.01570}, 2019.

\bibitem{zhao2019compact}
Fengxiang Zhao, Xing Ji, Wei Shyy, and Kun Xu.
\newblock Compact higher-order gas-kinetic schemes with spectral-like
  resolution for compressible flow simulations.
\newblock {\em Advances in Aerodynamics}, 1(1):13, 2019.

\bibitem{zhao2017weighted}
Fengxiang Zhao, Liang Pan, and Shuanghu Wang.
\newblock Weighted essentially non-oscillatory scheme on unstructured
  quadrilateral and triangular meshes for hyperbolic conservation laws.
\newblock {\em Journal of Computational Physics}, in press, 2018.

\bibitem{zhu2009hermite}
Jun Zhu and Jianxian Qiu.
\newblock {Hermite WENO schemes and their application as limiters for
  Runge-Kutta discontinuous Galerkin method, III: unstructured meshes}.
\newblock {\em Journal of Scientific Computing}, 39(2):293--321, 2009.

\bibitem{zhu2018new}
Jun Zhu and Jianxian Qiu.
\newblock New finite volume weighted essentially nonoscillatory schemes on
  triangular meshes.
\newblock {\em SIAM Journal on Scientific Computing}, 40(2):A903--A928, 2018.

\bibitem{zhu2020new}
Jun Zhu and Chi-Wang Shu.
\newblock A new type of third-order finite volume multi-resolution weno schemes
  on tetrahedral meshes.
\newblock {\em Journal of Computational Physics}, 406:109212, 2020.

\end{thebibliography}
\end{document}